# The hidden solar cost of urban unfamiliarity

Lijia Liu[1], Yimeng Song[1,2*]

[1] *Department of Urban Planning and Design, The University of Hong Kong, Hong Kong SAR, China*

[2] *Urban Systems Institute, The University of Hong Kong, Hong Kong SAR, China*

* Correspondence to Yimeng Song: ymsong@hku.hk

**Abstract**

Unfamiliar walkers may follow the shortest route and miss nearby shade, indoor passages and multilevel connections. We define the resulting hidden solar cost as the direct-sun exposure avoidable through shade-informed routing without substantial detours. We developed a time-specific three-dimensional model of Hong Kong Island that combines indoor and outdoor walking networks with building geometry, vegetation mapped by light detection and ranging (LiDAR), and solar position. For 500 modelled journeys, we compared shortest and shade-informed routes at four daytime snapshots and five extra-distance limits. At noon, shade-informed routes reduced modelled direct-sun exposure by 13.5% relative to shortest routes under a 10% extra-distance limit. Route length increased by 2.85% on average. A 5% extra-distance limit delivered 77% of the reduction obtained at 10%. In a controlled comparison, outdoor shade information accounted for 86% of the reduction and indoor-protection information for 14%. Removing multilevel connections left only 191 of 500 journeys feasible within the original route-length limit. These route comparisons identify an avoidable solar cost when existing protection is missing from the information used to choose a route. Making protected routes visible and preserving their connections can reduce sun exposure without large detours.

## 1. Introduction

A city can contain abundant shade and still expose pedestrians to avoidable sun. For a pedestrian, protection is experienced as a sequence of shaded, sheltered and exposed segments that must form a connected route to a destination. This distinction matters as urban heat extremes intensify and cities seek thermally safer walking environments (IPCC, 2023; Tuholske et al., 2021; Abuwaer et al., 2025). Buildings, vegetation and indoor passages can sharply alter direct-sun exposure and outdoor thermal conditions (Middel et al., 2016; Middel et al., 2021). Protection becomes usable when it can be reached without an unacceptable detour and recognised in time to shape route choice. The practical climate value of urban shade therefore depends on the relation between physical provision, pedestrian information and network connectivity.

The scientific gap lies in the connection between physical protection and the information used to reach it. Exposure maps quantify where shade is supplied (Buo et al., 2023; Gu et al., 2026), while routing models identify paths that can reduce sun exposure (Feng et al., 2024; Kolaxidis et al., 2025; Wolf et al., 2025). Studies of pedestrian decisions show that walkers trade distance against sun and that familiarity shapes the information used in route choice (Melnikov et al., 2022; Tong and Bode, 2022; Basu et al., 2024). A route can be physically available yet effectively absent from the choices of a pedestrian who does not know a shaded street, an indoor passage or a multilevel connection. Existing maps and route averages do not establish the solar cost of this missing local information or the detour needed to recover it. They also leave unresolved whether the opportunity arises from better information or from the physical network itself. Quantifying that cost is the central scientific problem of this study.

Resolving this gap requires treating protection as a journey resource. Mapping this resource across the full walking network and day captures changes in building shade as solar position shifts, alongside shelter from vegetation and indoor space. Route benefit is measured through both total exposure and the continuity of exposed sections. The same total can occur as one long sunlit section or several shorter sections (Tomasi et al., 2024b; Zhao et al., 2026). Walking cost is evaluated across extra-distance limits because pedestrians trade detour against exposure and marginal benefits may decline (Basu et al., 2024). Protected segments also need to connect into practical routes. This condition is especially important on Hong Kong Island, where elevated, underground and indoor links change both pedestrian access and route choice (Sun et al., 2021; Wen et al., 2025). This setting allows nearby shade supply to be analysed separately from the journey-level benefit made possible by the network.

Here, we develop a time-specific three-dimensional model of walking exposure across Hong Kong Island. The model integrates outdoor and indoor walking networks, multilevel connections, building geometry, LiDAR-derived vegetation and solar position. For the same origins and destinations, we compare shortest routes with shade-informed routes that use building shade, vegetation shade and indoor protection within a defined extra-distance limit. The framework establishes how protection varies across the network and day, then quantifies avoidable total and continuous direct-sun exposure. It also tests how rapidly benefits diminish as detours increase and which informational, spatial and network conditions make protected routes usable. By separating local shade information from physical access, the analysis quantifies the hidden solar cost of urban unfamiliarity.

## 2. Methods

We modelled walking across Hong Kong Island as movement through a time-specific three-dimensional walking network. The model linked outdoor paths, indoor passages and multilevel connections with surrounding buildings and vegetation. We classified direct-sun exposure at four times of day and compared shortest and shade-informed routes under five extra-distance limits. We also tested the roles of protection information and network connections, and compared route gains with nearby protection. Known sun refers to path segments classified by the model as directly sunlit, and exposure was measured by their total length along each route.

### 2.1. Study area and spatial data

We built a connected walking network from the Hong Kong Lands Department's 3D Pedestrian Network and 3D Indoor Network (Lands Department, 2026c and 2026b). The network retained the original three-dimensional geometry, direction restrictions, elevation and lift connections. Walking distance followed the three-dimensional path geometry. Official shoreline and administrative boundary data defined the island-wide study area, which was divided into four administrative sampling areas (Lands Department, 2026a and Home Affairs Department, 2026). Coordinate systems and graph construction are described in Supplementary Methods 1.

Building surfaces came from the Lands Department's three-dimensional spatial data (Lands Department, accessed 2026). We reconstructed vegetation from airborne LiDAR returns acquired in 2020. The same product supplied terrain elevations (CEDD, 2020). Data coverage, acquisition dates and representative multilevel paths are reported in Supplementary Table S1 and Supplementary Figs. S1 and S2.

### 2.2. Estimating sun and shade along walking paths

Outdoor paths were sampled at 2.5 m intervals. From 1.1 m above each sample, we traced a ray towards the sun and tested whether buildings or vegetation blocked it. We modelled sun exposure on 15 July 2026 at 09:00, 12:00, 15:00 and 17:00 Hong Kong time (HKT). Solar direction was calculated with the Solar Position Algorithm in pvlib (Reda and Andreas, 2004 and Holmgren et al., 2018). LiDAR returns were grouped into 1 m voxels, with two or more returns treated as opaque. The 1.1 m ray origin corresponds to the abdomen height used in standing thermal environment measurements (ISO, 1998). Solar position remained fixed during each journey in the primary analysis.

Each sample was assigned to known sun (S), building shade (B), vegetation shade (T), indoor protection (I) or unresolved exposure (U). Known sun denotes positions classified as directly sunlit. Unresolved exposure was retained when the available geometry or data coverage could not establish sun or protection. Indoor sections were identified from network attributes. Samples were weighted by the path length they represented and summed to obtain route and network class lengths. Full classification rules and diagnostics are provided in Supplementary Methods 2, Supplementary Table S2 and Supplementary Figs. S3 to S7.

### 2.3. Comparing shortest and shade-informed routes

The main sample contained 250 spatially dispersed outdoor origins, each paired with two destinations. The resulting 500 journeys covered four sampling areas and shortest-route distances of 300 to 3,000 m. Endpoints were selected without using exposure outcomes. The shortest route minimised three-dimensional walking distance and was calculated with Dijkstra's algorithm (Dijkstra, 1959). Shade-informed routes used the building shade, vegetation shade and indoor protection classifications. Sampling and journey distributions are detailed in Supplementary Methods 3, Supplementary Table S3 and Supplementary Fig. S8.

We tested extra-distance limits of 0%, 5%, 10%, 20% and 30% at all four times. Noon with a 10% extra-distance limit was the primary condition. Each extra-distance limit set the maximum proportional increase in length relative to the shortest route. Candidate routes were generated by varying distance and exposure weights and then screened against this limit. Exposure cost was the combined length in known sun and unresolved exposure. A candidate qualified only when it reduced both known-sun distance and exposure cost. We selected the qualifying route with the lowest exposure cost, using route length to resolve ties. The shortest route was retained when no candidate qualified. Search procedures and diagnostics are reported in Supplementary Methods 4 and Supplementary Fig. S9.

Avoided known-sun distance was $\Delta S = S_0 - S$, where $S_0$ and S denote known-sun distance on the shortest and shade-informed routes. Additional walking distance was $\Delta L = L - L_0$. Proportional reduction in known sun was $R = \Delta S/S_0$ and was calculated when $S_0$ was greater than zero. We also measured the longest continuous sunlit section and the first tested extra-distance limit that reduced baseline sun by at least 50%. Route length did not include waiting time or climbing effort. Metric definitions are provided in Supplementary Methods 4 and Supplementary Table S4.

### 2.4. Examining shade information, walking connections and nearby protection

We compared three information conditions on the same network and common candidate routes. I0 selected the shortest route without protection information. I1 recognised outdoor building and vegetation shade but did not recognise indoor protection. I2 recognised outdoor shade and indoor protection. Indoor paths remained available in all conditions. The comparison included all 500 journeys at noon with 10% and 20% extra-distance limits. Condition definitions and the design using a common candidate-route set are described in Supplementary Methods 5, Supplementary Table S5 and Supplementary Fig. S10.

We tested walking connections by separately removing indoor paths or multilevel links. Shortest and shade-informed routes were recalculated for all 500 journeys at noon with a 10% extra-distance limit. Feasibility was assessed using the route-length limit derived from the corresponding full-network shortest route. Each journey was classified as feasible, disconnected or over the limit. Known-sun distance for feasible journeys was compared with the corresponding full-network route. Restriction rules and complete counts are reported in Supplementary Methods 5, Supplementary Table S6 and Supplementary Fig. S11.

Nearby protection was examined at 22 locations. Each location included 12 journeys spanning four directions and target distances of 250, 500 and 750 m. At noon with a 10% extra-distance limit, we compared mean avoided known-sun distance with the known protected share of the network reachable within 880 m of network distance. Protected length included building shade, vegetation shade and indoor protection, with overlapping physical sections counted once. Spearman rank correlations quantified the associations between nearby protection and route gains across locations. Sampling and nearby-protection calculations are documented in Supplementary Methods 6, Supplementary Table S7 and Supplementary Fig. S12.

### 2.5. Summarising outcomes and assessing uncertainty

Paired route differences were summarised with design-weighted means, medians and shares with $\Delta S > 0$. Aggregate proportional reduction was the design-weighted mean $\Delta S$ divided by the design-weighted mean $S_0$. Equal total weight was assigned to the 12 combinations of four sampling areas and three distance bands. We calculated 95% confidence intervals (CIs) from 1,000 bootstrap replicates that resampled origins within each area while keeping repeated observations from the same origin together (Efron, 1979). A spatial block bootstrap tested sensitivity to geographic clustering (Lahiri, 2003). Weighting and interval definitions are given in Supplementary Methods 7 and estimates in Supplementary Table S8.

Sensitivity analyses varied voxel size, occupancy threshold, grid alignment, path sampling, ray height, solar position during walking and terrain obstruction. Computational checks covered path connectivity, permitted travel directions, exposure totals and distance limits. Procedures, verification records and outcome definitions are provided in Supplementary Methods 8, Supplementary Methods 9 and Supplementary Methods 10. Sensitivity results are given in Supplementary Table S9 and Supplementary Figs. S13 and S14.

## 3. Results

### 3.1. Island-wide sun exposure and protection

The analysed walking network covered 1,323.4 km across Hong Kong Island (Fig. 1a). It connected footways and park paths with stairs, bridges, underpasses and indoor passages. At 12:00 HKT on 15 July 2026, 40.40% of the network was in known sun, equivalent to 534.6 km. Known protection covered 44.03% or 582.7 km. Vegetation shade accounted for 28.02%, building shade for 9.95% and indoor protection for 6.07%. Exposure remained unresolved along the other 15.57% of the network (Fig. 1b and Supplementary Table S2). Footways, footpaths and park walkways accounted for 83.96% of its total three-dimensional length (Fig. 1c).

Sun and protection changed through the day (Fig. 1b and Supplementary Table S2). Known sun reached its largest share at 12:00 and fell to 8.38% by 17:00, when building shade covered 43.34% of the network. Building shade covered 35.34% at 09:00 and 27.87% at 15:00. Vegetation shade remained between 26.16% and 28.02%, while indoor paths provided 80.3 km of protection at all four times. At 12:00, known protection ranged from 41.14% to 48.96% across the four administrative areas, while known sun ranged from 37.30% to 41.52% (Fig. 1d). Spatial classifications at all four times and complete area summaries are reported in Supplementary Fig. S5 and Supplementary Table S2.

### 3.2. Rerouting changes total and continuous sun exposure

To illustrate the route-level comparison, we examined two paths connecting the same endpoints (Fig. 2a,b). The shortest route minimised walking distance. The shade-informed route used information on building shade, vegetation shade and indoor passages to select an alternative within the permitted extra-distance limit. We report the proportional reduction in known sun as the primary exposure benefit and the proportional increase in route length as the walking cost.

At 12:00 HKT with a 10% extra-distance limit, the shade-informed route was 9.37% longer than the shortest route, increasing from 1,558.1 to 1,704.1 m (Fig. 2c,d). This additional 145.9 m of walking reduced known sun by 11.76%, from 529.3 to 467.0 m. The share of the route in known sun consequently fell from 33.97% to 27.41%. The route also shifted more of the journey indoors, with the indoor share increasing from 2.70% to 28.10%. Unresolved exposure decreased by 60.39%, from 323.5 to 128.2 m.

Total and continuous sun exposure did not always change in the same direction (Fig. 2c). In the illustrative journey, total known sun decreased by 11.76%, while the longest continuous sunlit section increased by 37.20%, from 82.2 to 112.7 m. Across the 269 journeys in which the shade-informed route reduced total known sun, this pattern was uncommon. Using a ±1 m threshold, the longest sunlit section shortened in a design-weighted 57.97% of journeys, was unchanged in 35.72% and lengthened in 6.31%. Its design-weighted mean length was 30.16% lower on the shade-informed routes, decreasing from 155.5 to 108.6 m. The mean absolute change was −46.9 m (95% CI, −55.7 to −39.0 m). Complete estimates are reported in Supplementary Table S10. Route geometry and exposure patterns also changed across the four evaluation times (Fig. 2e). Further examples are shown in Supplementary Figs. S15 and S16.

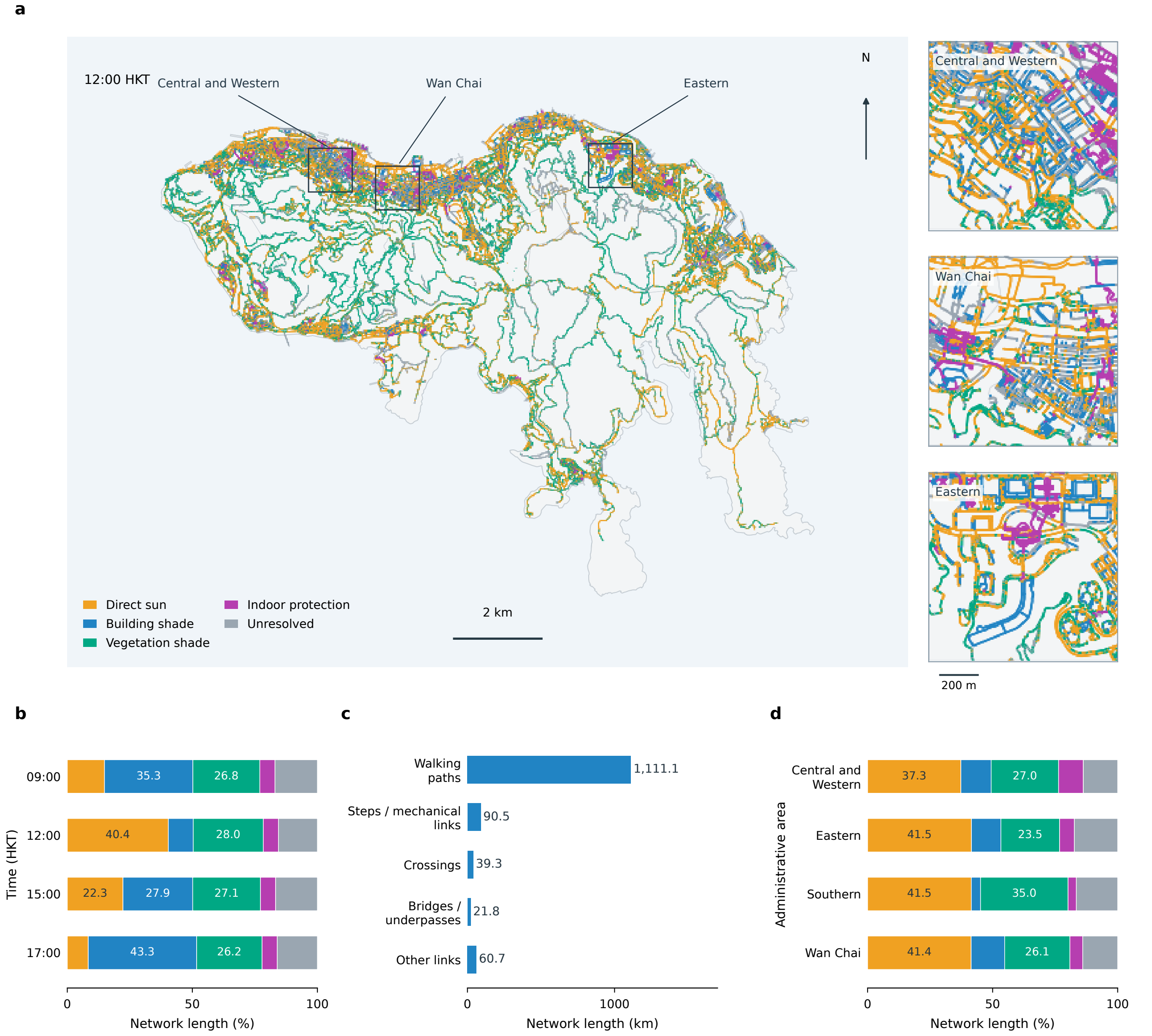

**Fig. 1. Sun exposure and protection across the walking network.** **a**, Exposure across Hong Kong Island at 12:00 HKT. Boxed areas are enlarged on the right at a common scale. **b**, Exposure-class fractions at four times. **c**, Network length by path type. **d**, Noon exposure fractions by administrative area. Fractions are weighted by three-dimensional network length, using the full network in b and each area's network in d. The network totals 1,323.4 km. Exposure colours in a also apply to the detail maps, b and d. All times refer to 15 July 2026.

### 3.3. Sun-avoidance gains vary across journeys, locations and times

The island-wide comparison showed that avoidable sun exposure was concentrated in a subset of journeys. The analysis included 500 journeys, with two destinations sampled for each of 250 origins. At 12:00 HKT with a 10% extra-distance limit, shade-informed routing reduced known sun by 13.54% across the design-weighted sample (95% CI, 11.98% to 15.02%). This corresponded to a design-weighted mean reduction of 74.23 m. After design weighting, 53.77% of journeys had less known sun on the shade-informed route than on the shortest route (95% CI, 48.98% to 58.26%). At the journey level, the median proportional reduction was 1.36%. The 10% of journeys with the largest gains contributed about half of the total

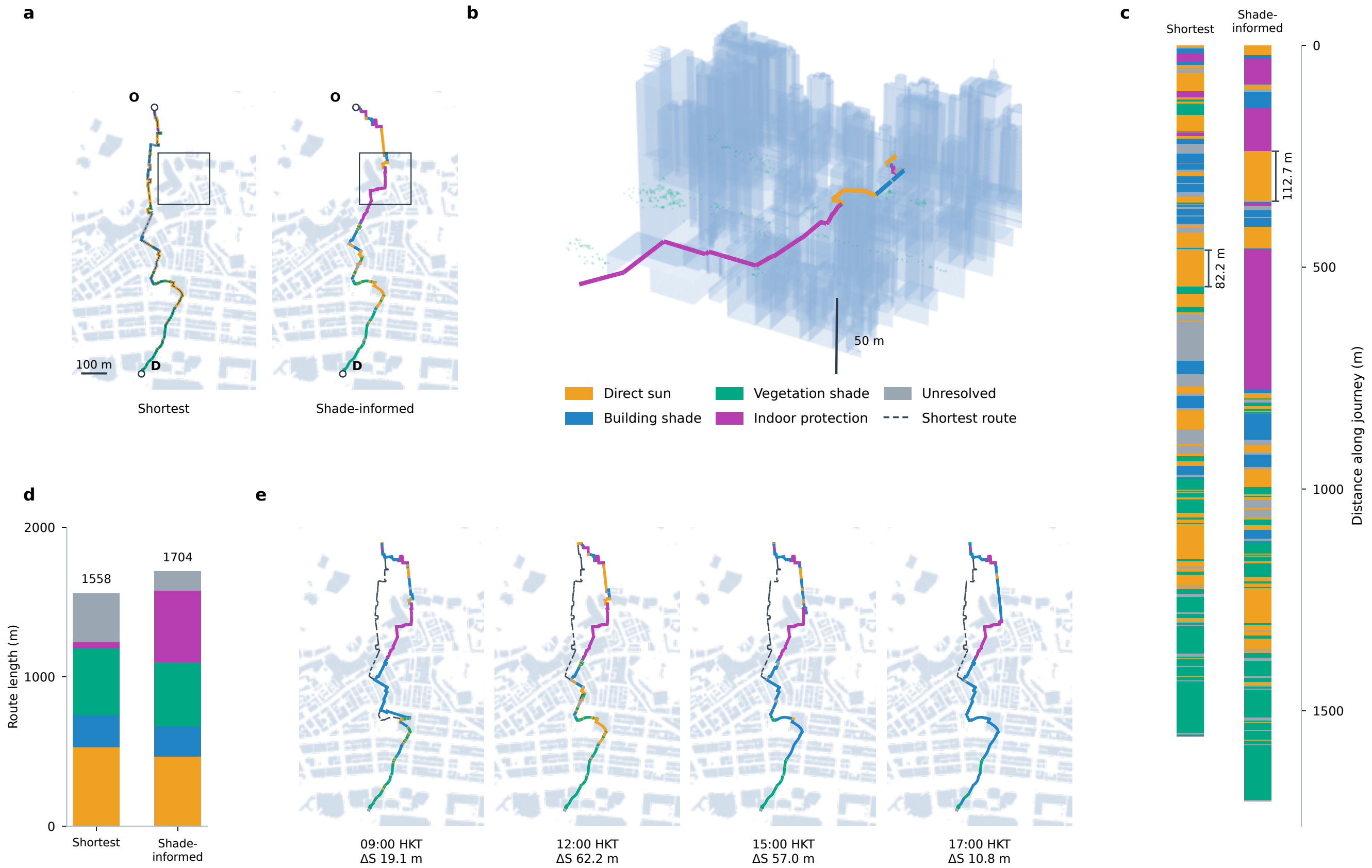

**Fig. 2. Sun exposure along shortest and shade-informed routes.** **a**, Routes connecting the same origin (O) and destination (D). The boxed area is enlarged in b. **b**, Buildings, vegetation and walking paths within a 220 × 220 m window, shown at equal scale in three dimensions. **c**, Exposure classes along the two complete routes, read from top to bottom in walking order. Brackets identify the longest continuous sunlit sections. **d**, Route lengths partitioned by exposure class. **e**, Routes at four times, with labels showing the reduction in known-sun distance (ΔS). Dashed lines identify shortest routes and colours indicate exposure classes on shade-informed routes. Panels a, b, c and d use 12:00 HKT. All panels use 15 July 2026 and a 10% extra-distance limit.

avoided sun (Fig. 3a,b). Full estimates are provided in Supplementary Tables S8 and S10.

Gains also varied among individual origins within the island-wide sample (Fig. 3c). Across the four areas, aggregate proportional reductions ranged from 6.81% to 18.89% under a common journey-length composition. Longer journeys were more likely to benefit (Fig. 3d). Positive-gain shares were 40.29%, 47.62% and 73.39% for shortest-route lengths of 300 to 800 m, 800 to 1,500 m and 1,500 to 3,000 m, respectively. Median proportional reduction was zero in the two shorter bands and 8.74% in the longest (95% CI, 6.01% to 12.17%). Median absolute reductions were also greater in the longest band (Fig. 3e). Journeys without an alternative route within the route-length limit were less common in the longest band. Details of zero-gain journeys and area and distance-band estimates are provided in Supplementary Table S8a,c.

The opportunity to avoid sun changed through the day even for the same endpoints. With the 10% extra-distance limit fixed, aggregate proportional reductions were 27.89% at 09:00, 13.54% at 12:00, 22.90% at 15:00 and 26.51% at 17:00. Mean avoided sun was greatest at 15:00 (76.57 m) and lowest at 17:00 (35.05 m) (Fig. 3f). The difference between the percentage and absolute patterns reflects changes in baseline sun exposure through the day. Relative to noon, individual journeys showed both increases and decreases in avoided sun at each comparison time (Fig. 3g). Across the four times, 190 of the 500 journeys switched between positive and zero gains. Full repeated-time results are provided in Supplementary Table S8b and Supplementary Fig. S17.

At noon, 195 of the 269 route pairs with positive reference gains remained positive across all 13 vegetation settings (Fig. 3h). Unresolved exposure was assigned to yield the smallest gain for each pair. Updating solar position during walking changed avoided sun by no more than 10 m in most of the 60 comparisons (Fig. 3i). All retained their positive, zero or negative gain classification under a 1 m tolerance (Supplementary Table S9 and Supplementary Figs. S13 and S14).

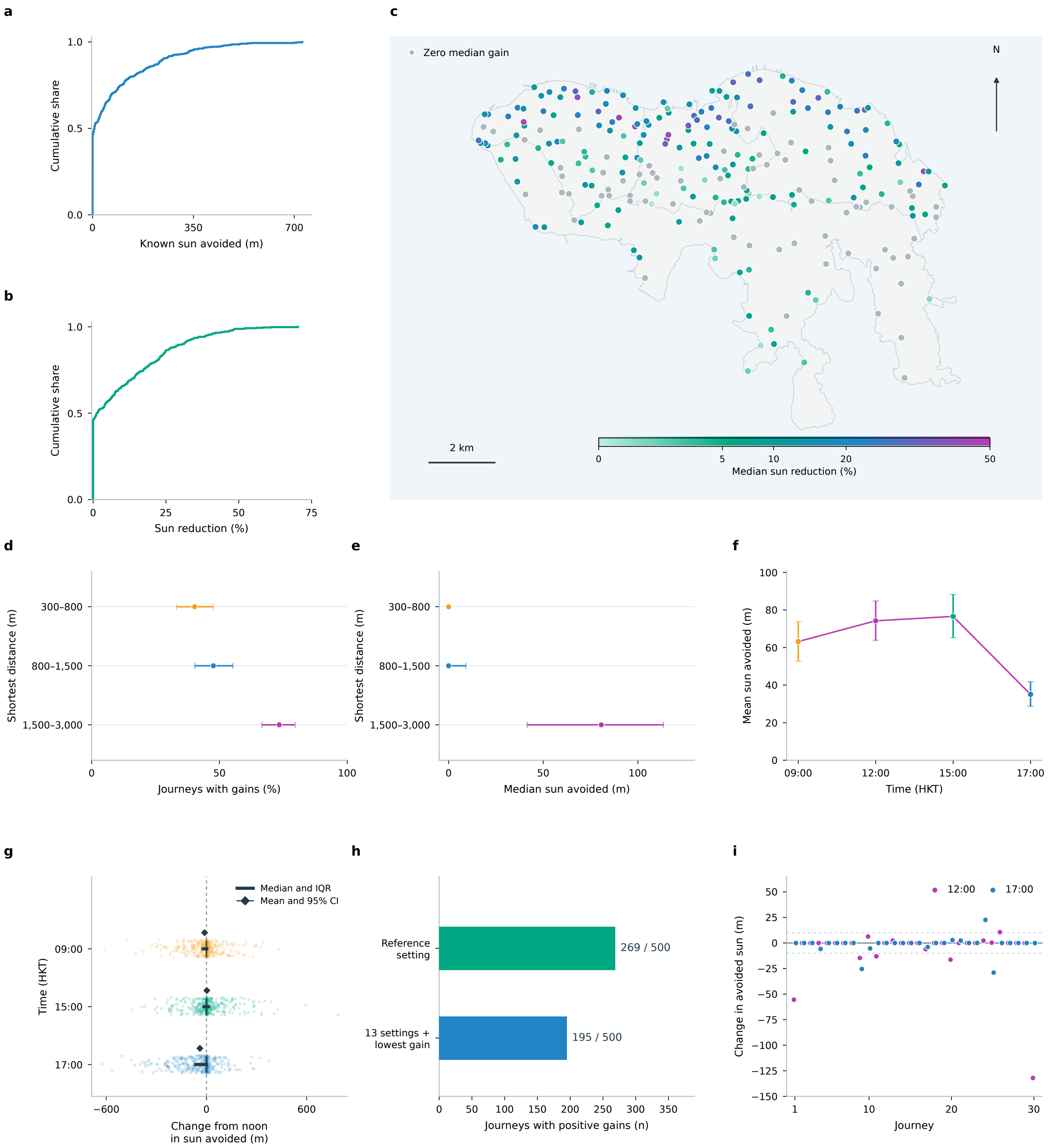

**Fig. 3. Variation and stability of sun-avoidance gains. a**, Cumulative avoided-sun distribution for 500 journeys. **b**, Proportional-reduction distribution for the 499 journeys with baseline sun. **c**, Median proportional reduction at 250 origins, each with two journeys. Grey indicates zero median reduction. **d**,**e**, Shares of journeys with reduced sun and median avoided sun across three shortest-route distance bands containing 166, 168 and 166 journeys. **f**, Mean avoided sun at four times. **g**, Paired changes from noon for 500 journeys at each comparison time. Points represent journeys, thick lines and ticks show interquartile ranges and medians, and diamonds show means. **h**, Unweighted counts of positive gains among 500 fixed route pairs in the reference setting and across all 13 vegetation settings. The latter assigns unresolved exposure to minimise each pair's gain. **i**, Changes in avoided sun after updating solar position during walking, relative to fixed departure-time exposure, for 30 journeys at 12:00 and 17:00 HKT. Dashed lines mark ±10 m. Colours identify distance bands in d,e and departure times in f,g,i. Estimates in a,b,d–g are design-weighted. Confidence intervals in d–g use 1,000 origin-cluster bootstrap replicates. All panels use 15 July 2026 and a 10% extra-distance limit. Panels a–e and h use 12:00 HKT.

### 3.4. Additional walking brought progressively smaller gains in sun avoidance

The practical cost of avoiding sun was the extra distance a person would need to walk. The extra-distance limit specified how much longer a shade-informed route could be than the shortest route. At noon, when up to 10% extra distance was allowed, the shade-informed routes were 2.85% longer on average (Fig. 4a). Under this setting, mean additional walking was about 47 m per journey (Supplementary Table S8). With up to 30% extra distance allowed, the average increase was 7.05%.

Much of the average sun avoidance could be obtained with short detours. Allowing up to 5% extra distance provided about three-quarters of the mean avoidance achieved with 10% extra distance (Fig. 4b). The opportunity for a much larger reduction depended on the individual journey. Among 499 journeys with known sun on the shortest route, 25 could cut that exposure by at least half under one of the tested extra-distance limits (Fig. 4c). Those 25 journeys first reached the target at extra-distance limits ranging from 5% to 30% extra distance.

The average benefit of further detours declined as the permitted extra distance increased (Fig. 4d). Across the sample, sun avoided per extra metre of walking fell from 2.58 m with 5% permitted extra distance to 0.85 m with 30% (Supplementary Table S10). Some journeys gained no further avoidance, and a few contained more known sun after the shade-informed route changed. Full route comparisons and tests allowing still longer detours are reported in Supplementary Tables S4 and S11 and Supplementary Fig. S18.

### 3.5. Shade information and walking connections shape the use of existing protection

Local shade information changed how routes made use of existing protection. All comparisons in this section evaluated exposure at noon and allowed routes no more than 10% longer than the original shortest route. Shade-informed routes used more building shade, vegetation shade and indoor passages, with building shade providing the largest added protected length (Fig. 5a). To isolate the contribution of information, we compared route choices from the same set of available paths. Recognising outdoor shade provided most of the mean reduction in known sun, and indoor protection information reduced exposure further (Fig. 5b). Complete estimates are reported in Supplementary Table S5 and Supplementary Table S8c.

Usable walking connections were also needed to reach these protected alternatives. In separate network tests, we removed indoor paths or links between different levels, such as stairs, footbridges and underpasses. Without indoor paths, 479 of 500 journeys remained possible within the same route-length limit, compared with 191 without multilevel links (Fig. 5c). The remaining journeys either required a longer detour or no longer had a connected route between their endpoints. Among the 190 journeys possible under both restrictions, removing either type of connection increased mean known sun on the shade-informed routes (Fig. 5d). Full feasibility counts and exposure changes are reported in Supplementary Table S6.

The amount of nearby protection was a weak guide to how much sun a journey could avoid (Fig. 5e). We compared 22 starting locations, each with 12 journeys to destinations in different directions and at different distances. The share of nearby paths in known protection ranged from 35.78% to 54.78%, but correlated only weakly with mean avoided sun (Spearman $\rho = -0.343$). Location-level results and tests using different surrounding walking distances are reported in Supplementary Table S7 and Supplementary Fig. S12.

Two locations with similar nearby protection illustrated how this opportunity depended on destination (Fig. 5f). Protected paths made up about 41% of the nearby network at both locations. At location B, journeys to the nearest targets gained no sun avoidance, whereas those to the furthest targets did. At location A, eastward journeys gained none at all three target distances. The representative journeys show coincident shortest and shade-informed routes at A, but an alternative with less known sun at B (Fig. 5g,h).

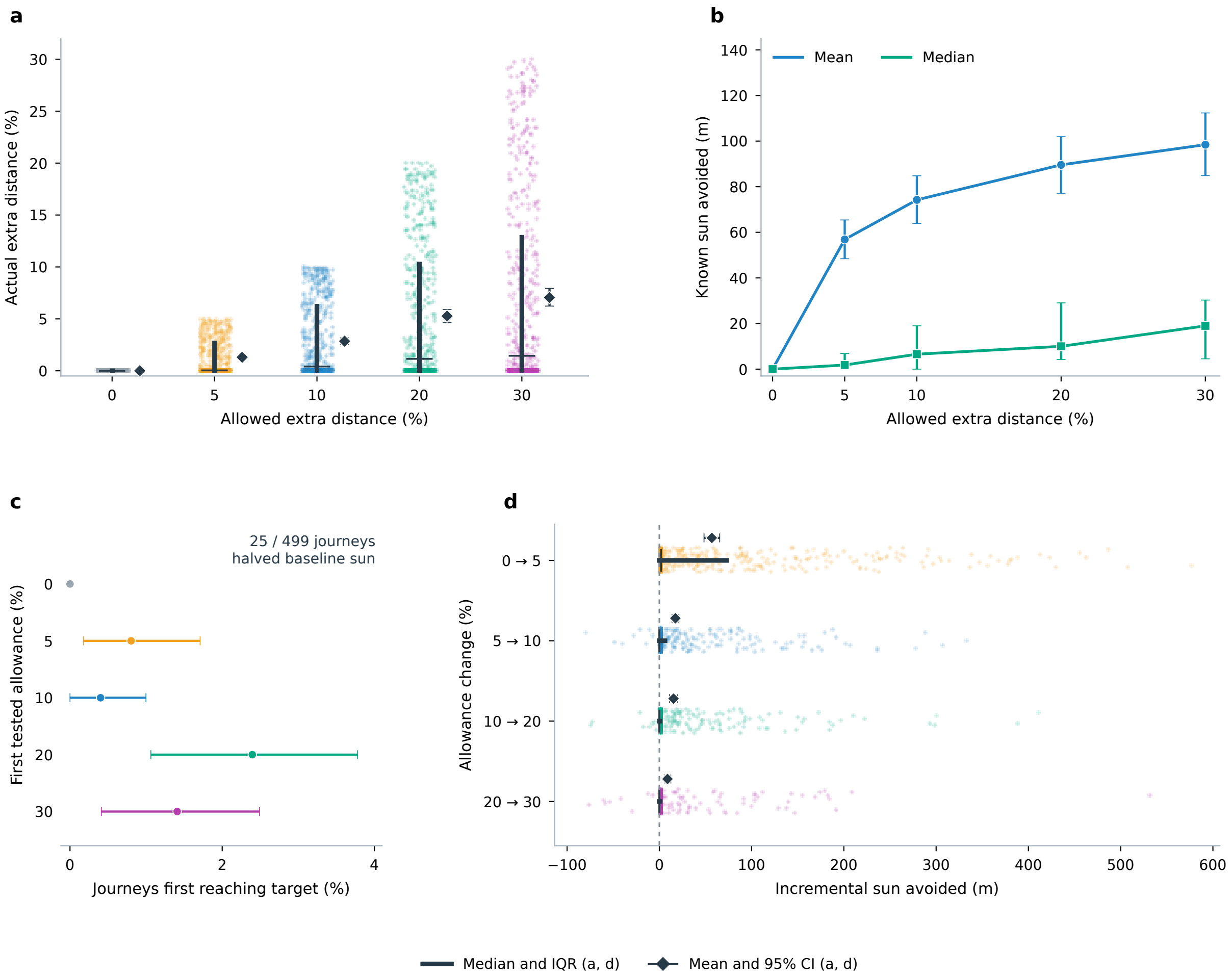

**Fig. 4. Sun avoidance and additional walking. a**, Actual additional walking as a percentage of shortest-route length for 500 journeys per extra-distance limit. **b**, Weighted mean and median avoided known sun across five extra-distance limits. **c**, Shares of journeys first halving baseline known sun at each extra-distance limit. The annotation gives the total reaching this target among the 499 journeys with baseline sun. **d**, Additional sun avoidance between adjacent extra-distance limits for the same 500 journeys. In a and d, points represent journeys, thick segments and ticks show interquartile ranges and medians, and diamonds with bars show means and 95% confidence intervals. Estimates are design-weighted and intervals use 1,000 origin-cluster bootstrap replicates. All panels use 12:00 HKT on 15 July 2026.

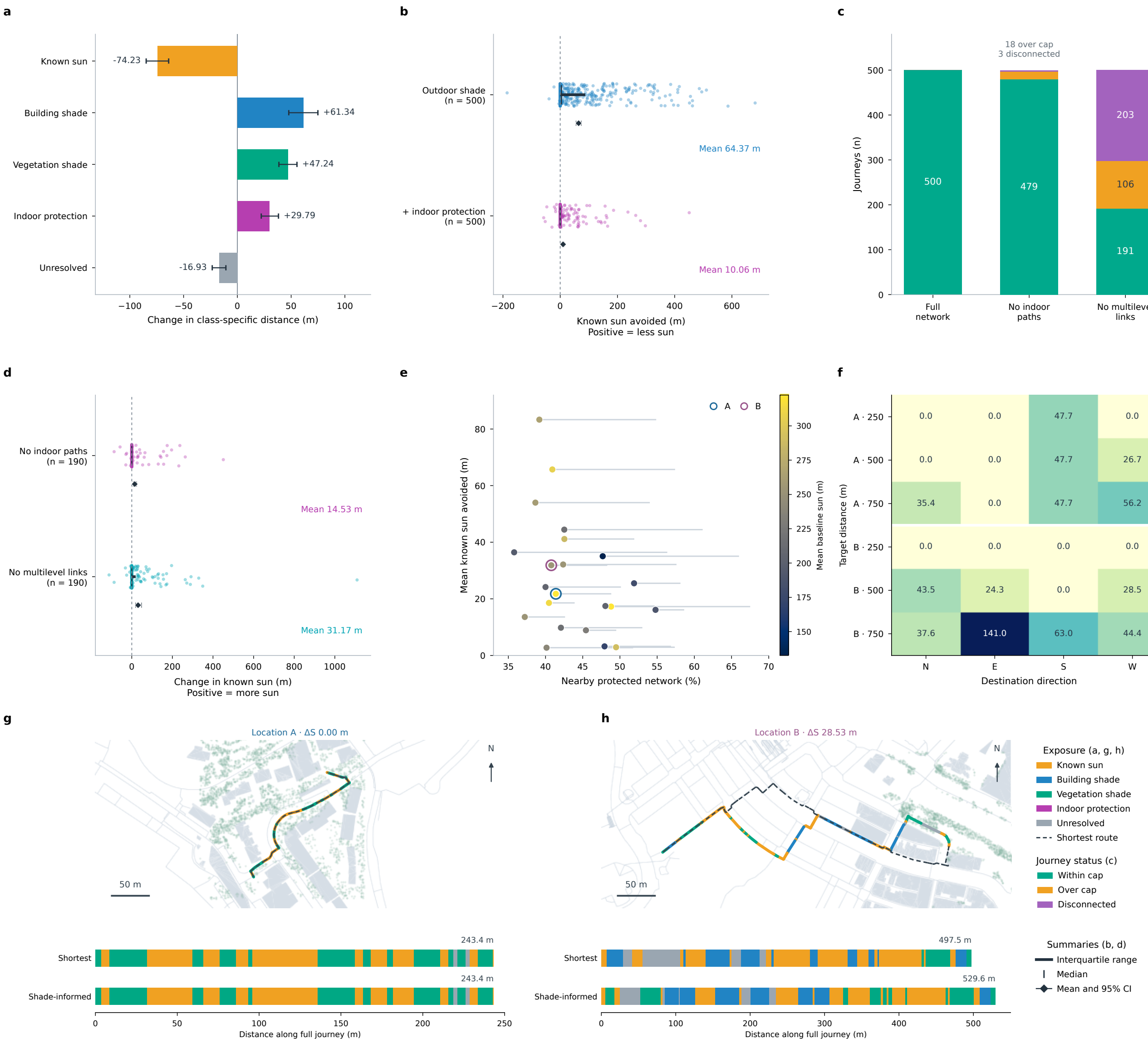

**Fig. 5. Shade information, walking connections and the use of existing protection.** **a**, Mean exposure-class length differences between shade-informed and shortest routes for 500 journeys. **b**, Additional known-sun reductions from recognising outdoor shade and then indoor protection, using a common candidate-route set with 500 journeys per comparison. **c**, Counts of journeys within or beyond the original route-length limit, or disconnected after removing indoor paths or multilevel links. **d**, Changes in known sun after removal for the 190 journeys feasible in all three networks. Positive values indicate less sun in b and more sun in d. In b,d, points represent journeys, thick segments and ticks show interquartile ranges and medians, and diamonds with bars show means and 95% confidence intervals. **e**, Nearby protected-network share and mean avoided sun at 22 locations, each with 12 journeys. Colours show mean shortest-route sun. Horizontal bars extend from known protection to the share including unresolved sections. Circles identify matched locations A and B. **f**, Avoided sun at these locations for four destination directions and three target distances (250, 500 and 750 m). **g**,**h**, Local route maps and full-journey exposure profiles for locations A and B, respectively. Exposure profiles use separate distance axes for the two representative journeys. Dashed lines identify shortest routes and colours indicate exposure classes on shade-informed routes. Means in a,b,d and quantiles in b,d are design-weighted. Confidence intervals use origin-cluster bootstrap resampling. All panels use 12:00 HKT on 15 July 2026 and a 10% extra-distance limit.

# 4. Discussion

This study links protection across an urban walking network to the sun exposure that can be avoided on individual journeys. We combined Hong Kong Island's three-dimensional walking network with building shade, vegetation and indoor passages, then compared shortest and shade-informed routes between identical endpoints. The findings show uneven opportunities to avoid direct sun within a limited additional walking distance. Their distribution depends on the alternatives available for each journey, the information used to recognise protection and the connections that make it reachable. Similar nearby protection can therefore support different journey benefits.

How long a walker remains in uninterrupted sun matters alongside total sun exposure. A reduction in total exposure can still leave a longer continuous sunlit section, as the route example illustrates. Network studies have examined time to shade for pedestrians with different walking speeds and stair-climbing abilities (Tomasi et al., 2024b). Whole-trip comfort research has likewise developed separate criteria for shaded proportions and maximum continuous sun-exposure duration (Zhao et al., 2026). The thermal conditions within protected sections also matter, as different shade types produce different radiant environments (Middel et al., 2021). A microclimate-routing study in Hong Kong found that shade-based and integrated thermal-discomfort objectives could favour different paths (Ma et al., 2025). Avoiding direct sun reduces one component of heat exposure, while thermal comfort also depends on the surrounding microclimate and the walker's response. Combining route exposure profiles with radiant-heat modelling would allow these differences to be evaluated along the journey (Lindberg et al., 2008).

The concentration of gains among journeys makes their distribution as important as their average magnitude. Unequal shade provision has been documented within cities even where overall coverage is high (Gu et al., 2026). Our route comparisons extend this spatial question to the journeys that can make use of that protection. Longer journeys allow larger absolute detours when extra distance is limited to a fixed percentage of the shortest route. They also more often had an alternative within that limit, consistent with their greater opportunity to avoid sun. Time of day changes the exposure of those alternatives and hence their value. Buildings cast longer shadows when the sun is lower in the morning and late afternoon, changing the protection available on alternative paths. Studies of experienced temperatures and GPS-recorded walking have also shown that activities and trip purposes affect heat exposure (Kuras et al., 2015, Ge et al., 2024). Combining these route opportunities with observed trip frequencies would help prioritise improvements for everyday walking.

The value of a detour depends on how much sun it avoids for the extra walking it requires. Pedestrians trade distance against sun exposure in controlled experiments, and shade influences choices during free walking (Melnikov et al., 2022, Azegami et al., 2023). Route-choice modelling further associates heat stress with perceived walking distance (Basu et al., 2024). The distance limit sets how much longer an alternative may be than the shortest route. Actual extra walking describes the cost of the shade-informed route. Multi-objective comparison can make different preferences for shorter walking and less exposure explicit (Marler and Arora, 2004). Presenting actual additional distance alongside total and continuous sun exposure would allow walkers to assess the compromise relevant to their journey.

The information comparison shows how local knowledge can change route choice on the same walking network. With the network and candidate routes fixed, additional protection information changes how existing alternatives are evaluated. Familiarity with a place may supply this knowledge, and route guidance can make it available to walkers with different levels of familiarity. This interpretation is consistent with research distinguishing information perception, integration and decision-making in pedestrian route choice (Tong and Bode, 2022). Navigation prototypes have implemented shade-based alternatives using user preferences (Feng et al., 2024). The use of shade guidance also depends on everyday travel constraints. Smartphone-based observations found limited changes in activity at heavily used outdoor locations during extreme heat, especially on weekdays (Derakhshan et al., 2025). Guidance would therefore need to describe usable entrances, connections and access times as well as expected exposure and additional distance. Comparing distance-based guidance with verified shade guidance on matched journeys would test whether the modelled information benefit changes the paths people use.

Walking connections determine which protected alternatives a pedestrian can reach. Studies of three-dimensional pedestrian systems show how elevated, underground and indoor paths change accessibility (Sun et al., 2021, Xu et al., 2022). Using actual sidewalk positions instead of road centrelines also changes estimated opportunities for shade-informed routes (Wolf et al., 2025). Indoor-network research identifies both shelter and shortcut contributions (Wen et al., 2025). A connection can therefore benefit a journey through its own protection and by enabling access to protected paths elsewhere. The connection-removal results support assessing shade along connected walking corridors, including entrances, exits and links to surrounding paths. For planning, this means asking which destinations remain within practical walking distance and how exposed the available routes are. Operating hours and step-free access would further determine who can use these corridors in practice.

Nearby protection and the sun a journey can avoid describe different features of the walking environment. A shortest route that is already well protected leaves less sun to avoid. Protected paths in other directions may increase local coverage without providing an alternative towards the journey's destination. This complements observations linking shade availability to shade use at individual sites (Levenson et al., 2025) by examining the additional benefit available through route choice. Walkability assessments that include thermal comfort can account for how people reach and use protection (Abuwaer et al., 2025). Iterative shade-design methods can test the position and performance of proposed protection (Tomasi et al., 2024a). Adding journey comparisons to those tests would help assess whether a proposed intervention closes an exposed gap or connects useful alternatives for the trips it serves.

The sensitivity tests identify routes whose benefits persist when estimated exposure changes. Unresolved sections traversed equally on both routes cancel from the exposure difference. Classification uncertainty therefore matters most on sections whose use differs between the routes. LiDAR-based tree modelling has demonstrated how detailed canopy geometry influences simulated light conditions and can be checked against measurements (Schindler et al., 2024). Shade can also move during a journey as the walker progresses. Updating solar position during walking evaluates exposure when each section is reached, complementing comparisons at fixed departure times. Field checks can focus on canopy boundaries and unresolved sections used differently by competing routes.

Passage entrances also warrant checks where access conditions could change the recommended route.

The design combines same-journey comparisons with a three-dimensional representation of access, allowing information use and physical connectivity to be examined separately. The main constraints concern the spatial data and how they represent shade and access. Spatial layers acquired in different years can reflect different stages of urban change. The shade model omits some overhead shelter, such as awnings and footbridge roofs, and terrain obstruction was examined separately. Treating vegetation as opaque also simplifies partial light transmission through foliage, particularly at canopy edges. The scenarios assume clear skies on one summer day and indoor passages open at each evaluation time. Applying routes under other weather or access conditions requires updated exposure and passage-access information. Updated spatial layers and observations of canopy transmission would refine these estimates.

## 5. Conclusion

Shade-informed routing can reduce direct-sun exposure by using protection already embedded in an urban walking network. For 500 modelled journeys on Hong Kong Island, shade-informed routes reduced known sun by 13.5% at noon under a 10% extra-distance limit. Route length increased by 2.85% on average. The benefit varied with the journey, time of day, extra-distance limit, and available information and walking connections. Further detours brought progressively smaller average gains. The findings can guide the communication of usable protected alternatives and maintenance of the walking connections that make them accessible. They can also inform assessments of additional protection where journeys retain long sunlit sections or offer few alternatives.

## Declarations

### Funding

This research received no grant funding.

### Competing interests

The authors declare no competing interests.

### Data and code availability

Source datasets and their access links are listed in Supplementary Table S1. The analysis code is not publicly available.

### Use of generative AI

OpenAI Codex was used to assist with drafting and editing the manuscript and preparing plotting code. The authors reviewed the text and figures and take responsibility for the final content.

# Supplementary Material
# The hidden solar cost of urban unfamiliarity

Lijia Liu[1], Yimeng Song[1,2*]

[1] Department of Urban Planning and Design, The University of Hong Kong, Hong Kong SAR, China

[2] Urban Systems Institute, The University of Hong Kong, Hong Kong SAR, China

* Correspondence to Yimeng Song: ymsong@hku.hk

## Supplementary Methods 1. Source data and graph construction

The network, building and administrative-boundary geometries were locally archived on 26 September 2026. Light detection and ranging (LiDAR) records reflected laser measurements as three-dimensional points called returns. These returns were acquired between December 2019 and February 2020; the 2020 product was downloaded for this research in September 2026. The shoreline metadata were revised on 29 May 2026 and retrieved on 10 October 2026. Supplementary Table S1 records product descriptions, local archive dates and source links. The modelled date, 15 July 2026, defines the evaluation scenario independently of the geometry acquisition dates.

The study-area land polygon was formed from closed shoreline lines (HYDRLINE features) with CLASS=SLN and TYPE=COA in the official 1:50,000 topographic product (Lands Department, 2026a). It retained the Hong Kong Island polygon by retaining physical segments whose complete horizontal geometry lay within that polygon. Nearby islands were outside the retained study polygon. The construction used the closed source features directly. Administrative boundaries supplied administrative area assignment (Home Affairs Department, 2026). Physical-segment administrative-area summaries use edge midpoints, with cross-area edges flagged separately.

Horizontal coordinates use the Hong Kong 1980 Grid (EPSG:2326); elevations use Hong Kong Principal Datum (HKPD, EPSG:5738), the source height reference. XYZ denotes two horizontal coordinates, X and Y, and elevation Z. The network is a directed graph, whose nodes represent path endpoints or connection states and whose arcs represent permitted travel directions. It retains the original three-dimensional paths, parallel links, lifts and boarding/exit state transitions. State transitions connect boarding and exit states; zero-length transitions add connectivity without adding walking distance. Rail and fare-related transfers are excluded from the walking graph. Projected crossings retain their source connection states. Route metrics sum the three-dimensional length of every physical-segment traversal, including repeated and reverse traversal. Network eligibility follows the archived source attributes. Supplementary Fig. S2 shows ordered facility profiles; its twofold vertical display scale leaves the computational XYZ geometry unchanged. Source coverage and three-dimensional profiles are shown in Supplementary Figs. S1–S2.

## Supplementary Methods 2. Exposure sampling and occlusion rules

Physical segments were sampled at 2.5 m three-dimensional arclength intervals with endpoints included. Each sample represented the interval between midpoints to its neighbours, with end intervals truncated at segment boundaries. These intervals partitioned each physical segment and summed to its length. Outdoor ray origins were path elevation plus 1.1 m. Geometric solar directions at the four evaluation times in Hong Kong time (HKT) were calculated with the pvlib solar-modelling library's Solar Position Algorithm (Reda and Andreas, 2004; Holmgren et al., 2018) and transformed to the projected grid. Solar elevation ranged from approximately 27.8° to 83.2° across the four times. Exposure classes were known sun (S), building shade (B), vegetation shade (T), indoor protection (I) and unresolved exposure (U). Known sun denotes path length classified as directly sunlit; U denotes positions where geometry or coverage did not establish sun or protection. Indoor segments were assigned I from source attributes; outdoor exposure used the same interval weights.

Building-ray tests used original triangular building faces from the finest available source geometry. Component checks used temporary closing faces to test open building-mesh boundaries. Building meshes with unaccepted open boundaries, incomplete coverage along rays and unresolved contacts near ray origins received U. Building shade took precedence over vegetation shade. For building-unblocked rays, vegetation tests used valid vegetation returns in LAS point-cloud files (classes 3–5), excluding points flagged as withheld from use. The reference occupancy grid used 1 m voxels, cubic spatial cells containing at least two returns, treated as opaque. Source voxel occupancy supplied the vegetation representation. Coverage along each full ray was checked against the source tiles and the native 0.5 m digital terrain model (DTM). The DTM records ground elevations. Unresolved coverage or occupied ray-origin cells received U. Primary occlusion included buildings and vegetation, with terrain evaluated in a separate sensitivity screen.

Five classes remain exhaustive and mutually exclusive at sample, physical-segment and route levels. Supplementary Figs. S3–S4 illustrate actual building and vegetation intersections. Supplementary Fig. S5 maps the four-time classification and its physical-length shares; Supplementary Figs. S6–S7 summarize unresolved exposure and its source reasons. The native terrain mask supplies coverage information, while the terrain-screen experiment tests solar obstruction. LiDAR vegetation classes describe low, medium and high vegetation under the source classification. Complete physical-network class lengths and unresolved-source categories are reported in Supplementary Table S2.

## Supplementary Methods 3. Journey sampling

Origins and destinations were sampled without using exposure outcomes, with random seed 20261011. Each origin-destination (OD) pair defined one journey. Origin quotas were 63 for Central and Western (CW), 63 for Wan Chai (WC), 62 for Eastern (EST) and 62 for Southern (STH). Each administrative-area and distance-band combination formed a sampling stratum, the group used to balance route comparisons. Nonempty 500 m administrative area cells and their eligible outdoor nodes were shuffled. Cells were visited in rounds, selecting at most one origin per cell per round with a minimum separation of 100 m. Two distinct destinations per origin followed rotating shortest-distance bands [300,800), [800,1500) and [1500,3000] m. Within a nonempty band, a destination cell and then an eligible node were sampled. The final band counts were 166, 168 and 166 OD. Endpoints are existing eligible network locations.

The 500 OD pairs were evaluated in 20 repeated combinations of time of day and extra-distance limit. Main and local journey samples were analysed separately. Precision checks evaluated administrative area and distance coverage, sample counts in combinations of administrative area and distance band and primary interval widths. The checked thresholds included 40 origins per administrative area, 100 OD per distance band, 30 OD and 20 origins per combination of administrative area and distance band, and interval widths of 10 percentage points for median proportional reduction in known sun (R) and positive-gain share. The final primary precision record is retained in Supplementary Table S3 and Fig. S8; intervals for avoided known-sun distance ($\Delta S$) are reported descriptively in Table S8. All 33 checks passed. The precision assessment was documented after the route outcome files had been generated.

## Supplementary Methods 4. Candidate search and route selection

Dijkstra searches on nonnegative directed arc costs supplied the length baseline and candidate routes (Dijkstra, 1959). Let $L_0$ denote shortest-route length and b the extra-distance limit as a fraction of that length. The route-length limit is $C=L_0(1+b)$. A directed connection (u,v) was retained in the search corridor, the network subset eligible for path searching, when $d(o,u)+\ell(u,v)+d(v,d)\leq C+10^{-7}$ m. Here, o and d are the origin and destination, u and v are the directed arc's endpoints. The function d gives shortest network distance, and $\ell$ is directed-connection length. Search costs were $L+\lambda V$, where L is route length and V is the exposure cost under the corresponding information condition. The parameter $\lambda$ controls the relative weight of exposure and distance. The final search expanded the eligible network corridors. It combined the original ten-$\lambda$ candidates with $\lambda=0$ and 41 positive logarithmically spaced values from 1/32 to 32 in five search corridors defined by extra-distance limits. Duplicate sequences of directed connections were removed to form a common candidate-route set for each OD and time of day, then screened for each extra-distance limit. Supplementary Fig. S9 reports candidate counts and the initial-to-final search comparison.

Under full information, the unprotected-length measure was S and exposure cost was $V=S+U$. An alternative was accepted when $L\leq C+10^{-7}$ m and both $S_0-S$ and $V_0-V$ exceeded $10^{-6}$ m. Selection minimised V, then L, then stable generation order. When no candidate met both exposure-improvement criteria, the shortest route was retained. The candidate-route set contains the alternatives available for route selection. A 50-OD noon search audit at 10% and 20% extra-distance limits identified six material reductions in exposure cost and prompted corridor expansion. Expansion was triggered when at least 5% of the 100 audited conditions reduced exposure cost by more than the larger of 1 m or 1% of baseline exposure cost. The main analysis comprised 10,000 repeated route-pair evaluations across OD pairs, evaluation times and extra-distance limits. Comparisons between adjacent extra-distance limits preserve signed changes in S, U and shade-informed route.

For a shade-informed route, L is total length and S, B, T, I and U are its exposure-class lengths, satisfying $L=S+B+T+I+U$. Subscript 0 denotes the corresponding shortest-route baseline. Avoided known-sun distance is $\Delta S=S_0-S$, additional walking distance is $\Delta L=L-L_0$, and proportional reduction in known sun is $R=\Delta S/S_0$. R is defined for $S_0>10^{-6}$ m; absolute gains include zero-baseline-sun routes. Continuous known-sun distances use ordered integration intervals. Current target attainment evaluates the shade-informed route at each extra-distance limit. First attainment records the earliest tested setting meeting the target. Its denominator includes valid R records that never reach the target; conditional first-cost summaries use the attained subset. Supplementary Table S4 retains both current and first-attainment measures.

## Supplementary Methods 5. Experiments on shade information and walking connections

The information experiment pooled routes generated with outdoor-only protection information (I1) and outdoor-plus-indoor protection information (I2) into a common candidate-route set on the original graph. I0 selected the shortest route. I1 counted S+I as unprotected length and S+I+U as exposure cost; I2 used S and S+U, respectively. The same route-length limit and exposure-improvement criteria were applied within each information condition, and final outcomes used full five-class labels. The experiment covered 500 OD at noon under 10% and 20% extra-distance limits. This separately generated common candidate-route set defines the information comparison. Its paired difference estimates the gain from changing recognised protection while keeping candidate routes available to every condition.

The indoor restriction removed source-designated indoor segments. The multilevel restriction removed stairs, escalators, ramps, lifts, stairlifts, footbridges and subways. These links totalled approximately 120 km, or 9.1% of network length. The indoor and multilevel attribute sets overlap and were evaluated in separate graphs. Each restricted graph supplied its own shortest-route selection baseline, while the original full-network route-length limit $L_0(1+b)$ determined feasibility. Full-sample results distinguish disconnected, over-limit and feasible OD. Signed exposure loss was Srestricted−Sfull. Comparisons among the 190 jointly feasible OD conditioned the original stratum weights on that subset. Supplementary Table S6 also records the complete 500-OD feasibility denominators and the separate common-set diagnostic. Information increments are reported in Supplementary Table S5 and Supplementary Fig. S10; connection restrictions are illustrated in Supplementary Fig. S11.

## Supplementary Methods 6. Local journey sampling and nearby protection

We screened 200 dispersed network nodes using connectivity and retained 22 study locations with reachable targets in all 12 direction–distance combinations. The target design combined four 90° sectors and distances of 250, 500 and 750 m. Each combination was a target cell, defined by one destination-direction sector and one target distance. Eligible shortest-distance intervals were [125,375), [375,625) and [625,800] m. Nodes on exact sector boundaries were excluded. Within each cell, selection minimised squared normalised deviations from target distance and bearing, with stable node-ID ties. Each cell received weight 1/12. The resulting 264 local OD were analysed at noon with a 10% extra-distance limit. The earlier 24-location coverage sample retained its 130 missing cells as a separate geographic diagnostic.

The surrounding walking network is reachable within 880 m of directed walking distance and covers the maximum 800 m local baseline plus its 10% extra-distance limit. Reachable physical-segment portions were merged across overlapping intervals and reverse directions of the same physical segments. The known protected share was P=(B+T+I)/Lphysical, where Lphysical is the length of reachable physical segments counted once. Sections with unresolved exposure remain in the denominator, yielding the classification interval [P,P+U/Lphysical]. Boundary portions inherit the full segment's exposure density; this is a separate approximation from the classification interval for unresolved exposure. Location means retained all 12 equal-weight cells. Mean R averages valid journey-level proportional reductions in known sun. The aggregate proportional reduction divides total avoided known-sun distance by total shortest-route known-sun distance. A fixed-route scale check used radii of 660, 880 and 1,100 m and 176 common short-distance OD.

Illustrative pairs of study locations were selected within the middle 20–80% of location mean R. Candidate location pairs were screened for similar nearby protection and baseline length before selecting a contrast close to the median response difference. The representative journey at each location had ΔS closest to the location median. The images for locations A (LC04) and B (LC14) therefore retain their actual zero and positive route gains alongside location means. Location-level results and radius comparisons are reported in Supplementary Table S7 and Supplementary Fig. S12.

## Supplementary Methods 7. Weighting and statistical uncertainty

Overall summaries assign total weight 1/12 to each stratum defined by administrative area and distance band and equal weight to OD within that stratum. Administrative-area summaries assign total weight 1/3 to each distance band; distance-band summaries assign total weight 1/4 to each administrative area. These weights standardise the study design. Weighted quantiles are the outcome values at which a specified cumulative share of sample weight is reached. Proportional summaries remove undefined R records before recalibrating the corresponding stratum weights. Connection-restriction subset summaries instead renormalise the original full-sample weights within the declared feasible subset. Local target cells use their separate equal-cell weights.

Origin-cluster bootstrap generated 1,000 replicates by resampling origins with replacement within administrative areas, using seed 2026101105. Each replicate retained both destinations and repeated conditions for each origin and recalibrated stratum weights (Efron, 1979). Pointwise intervals were the 2.5th and 97.5th percentiles. Spatial-block bootstrap used seed 2026101106 to resample occupied 2 km EPSG:2326 grid blocks, aligned to grid origin (0,0), with their origins; the main sample occupied 34 blocks (Lahiri, 2003). Replicates missing a required stratum were discarded and their counts recorded. Paired differences were calculated at OD level before summaries. For Fig. 3g, known-sun gains at 09:00, 15:00 and 17:00 were each differenced from noon within the same OD; their intervals used the same frozen bootstrap draws as the original summaries. Supplementary Table S8 reports both origin-cluster and spatial-block intervals. Descriptive local Spearman coefficients compare location-level nearby protection and route outcomes.

## Supplementary Methods 8. Sensitivity to exposure assumptions

Thirteen vegetation configurations comprise 0.5 m and 1 m voxels with one, two or three returns, plus seven nonzero combinations of 0/0.5 m grid shifts in x, y and z for 1 m voxels with two returns. Origin-occupancy conflicts in any configuration retain a shared unresolved exposure status. Fixed shortest and shade-informed routes for 500 OD at four times were evaluated under all configurations, yielding 26,000 route-pair rows. Five non-reference configurations were also rerouted for a 50-OD noon subset using the final search, yielding 250 comparisons. Fixed-route evaluation and rerouting remain separate analyses.

To calculate sun-avoidance bounds, unresolved exposure was assigned consistently on the same physical section in both routes. Let $d_e = n_{0,e} - n_{selected,e}$ be the difference in traversal counts of physical segment e. $U_e$ is its unresolved-exposure length per traversal. The lower and upper gain bounds are $\Delta S + \Sigma \min(d_e, 0) U_e$ and $\Delta S + \Sigma \max(d_e, 0) U_e$. Matching traversals cancel by physical identity and multiplicity. A route pair has a robust positive gain when its lower gain bound is positive in every vegetation configuration. These bounds describe allocation of unresolved exposure; bootstrap intervals describe variation from resampling origins.

A 30-OD subset selected without using gain outcomes was evaluated at noon and 17:00 on both shortest and shade-informed routes. Reference rays, 1 m sample spacing, 1.5 m ray-origin height and a walking-clock evaluation produced 480 route evaluations. The walking clock advanced time of day at 100 m $min^{-1}$ along the ordered fixed path and recomputed ray intersections. Changes greater than 10 m were counted as material; practical gain signs used a ±1 m tolerance. Native 0.5 m terrain elevations were summarised as envelopes spanning the minimum and maximum elevations within each 5 m cell for definite, possible and unresolved terrain obstruction. The separate extended route-search experiment evaluated a preselected 50-OD subset up to 50% and 100% extra-distance limits with an expanded candidate-route set. Fixed-path sensitivity results are reported in Supplementary Table S9 and Supplementary Figs. S13–S14. The separate extended route-search experiment is reported in Supplementary Table S11 and Supplementary Fig. S18.

## Supplementary Methods 9. Computational verification and reproducibility

Computational checks reconstructed ordered routes from network connections and three-dimensional coordinates. They verified endpoints, travel directions, permitted connection types, route lengths and distance limits. Exposure-class lengths were checked against total route length. A 50-journey, four-time benchmark provided 200 implementation comparisons. The reference-ray test reproduced all 120 archived route values for 30 journeys. The research archive includes input and output checksums, candidate and path records, and classification data.

The analysis used Python 3.12.14 on Windows 11, with NumPy 2.5.3, SciPy 1.18.1 and pvlib 0.15.2. The research archive includes an executable network example, route geometries and three local three-dimensional scenes. Full ray recomputation requires the original building and vegetation geometry. Public source products are listed in Supplementary Table S1.

## Supplementary Methods 10. Exposure and walking-cost measures

The summaries in Supplementary Table S10 were computed from the archived route tables without rerouting. The longest continuous sunlit section was derived from the ordered classified segments of each shortest and shade-informed route at noon. Consecutive known-sun segments were summed. Any other class, including unresolved exposure, ended the section. Changes were summarised for the 269 journeys with positive gains. Aggregate proportional reductions divide the design-weighted mean $\Delta S$ by the design-weighted mean $S_0$ at each time. The actual detour fraction is $\Delta L/L_0$. Walking efficiency divides the design-weighted mean $\Delta S$ by the design-weighted mean $\Delta L$ at each extra-distance limit. Baseline-exposure groups use the known-sun and protected shares of the shortest route. Endpoint groups use the path type of the first and last segments. Connection-restriction summaries condition the original weights on each restriction's own feasible set.

Intervals were obtained from 1,000 origin-cluster bootstrap replicates using the procedure in Supplementary Methods 7 and seed 20261107. Stratum weights were recalibrated in each replicate and renormalised within each subset. Applied to the primary noon outcomes, the procedure reproduced the reported median $\Delta S$ interval of 0 to 18.98 m.

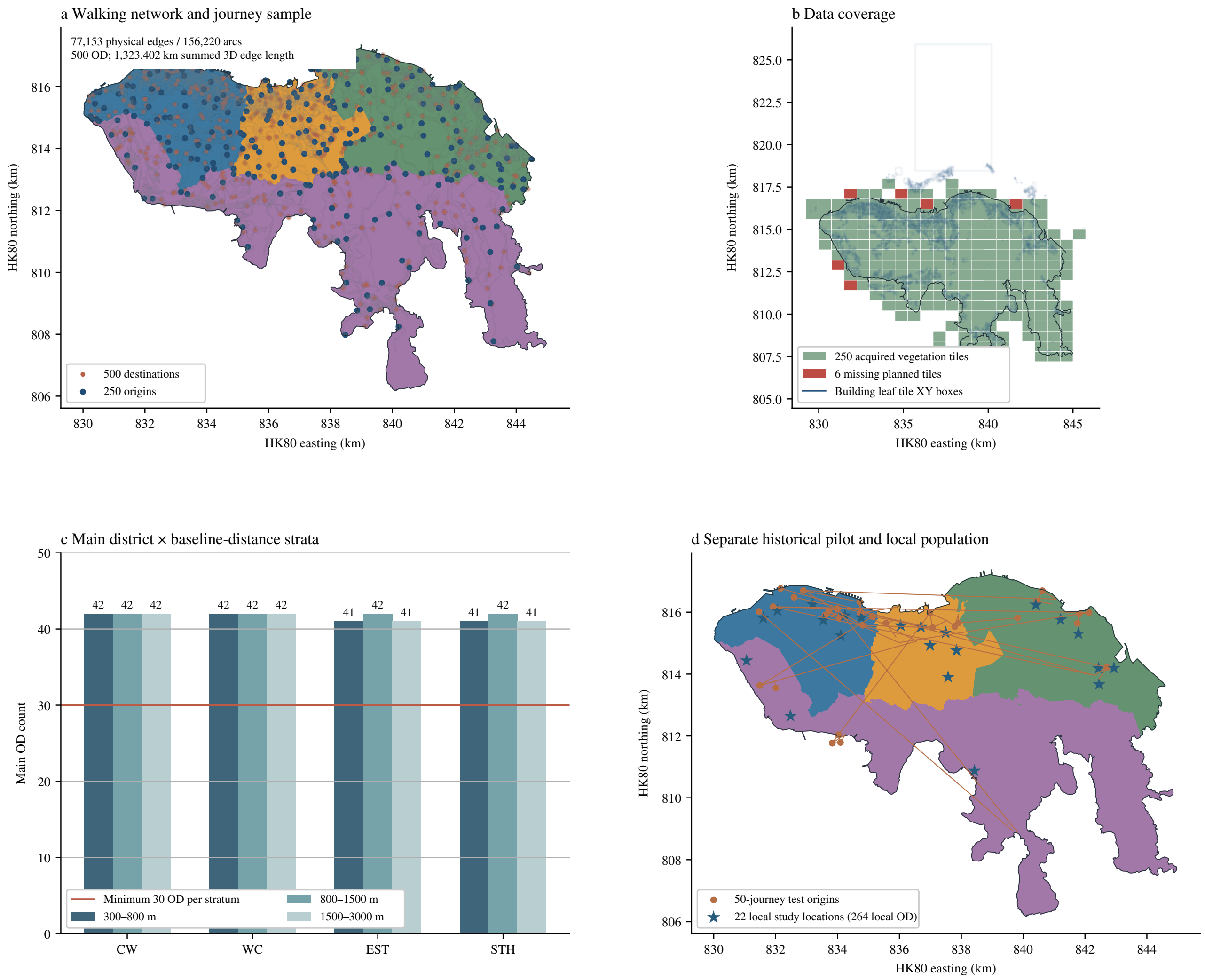

**Supplementary Fig. S1. Study area, data coverage and journey locations a**, Walking network, 250 origins and 500 destinations. **b**, Coverage of the vegetation and building datasets. **c**, Journey counts across administrative areas and shortest-route distance bands. **d**, Origins from the earlier 50-journey test and the 22 locations used for local analyses.

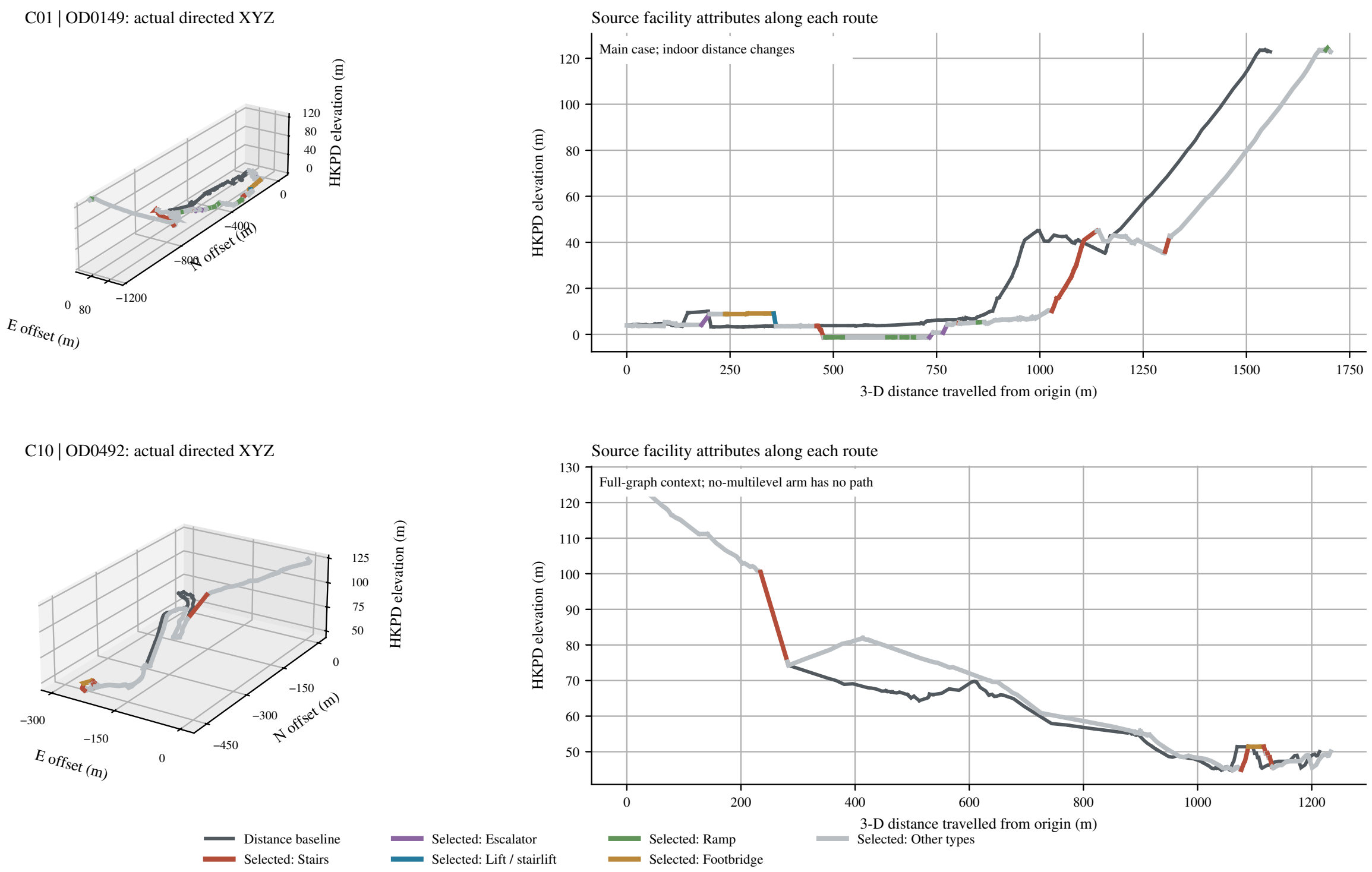

**Supplementary Fig. S2. Elevations along multilevel routes** Three-dimensional routes (left) and elevation profiles coloured by facility type (right) for two illustrative journeys. The three-dimensional views use twofold vertical exaggeration. Elevations are relative to the Hong Kong Principal Datum (HKPD).

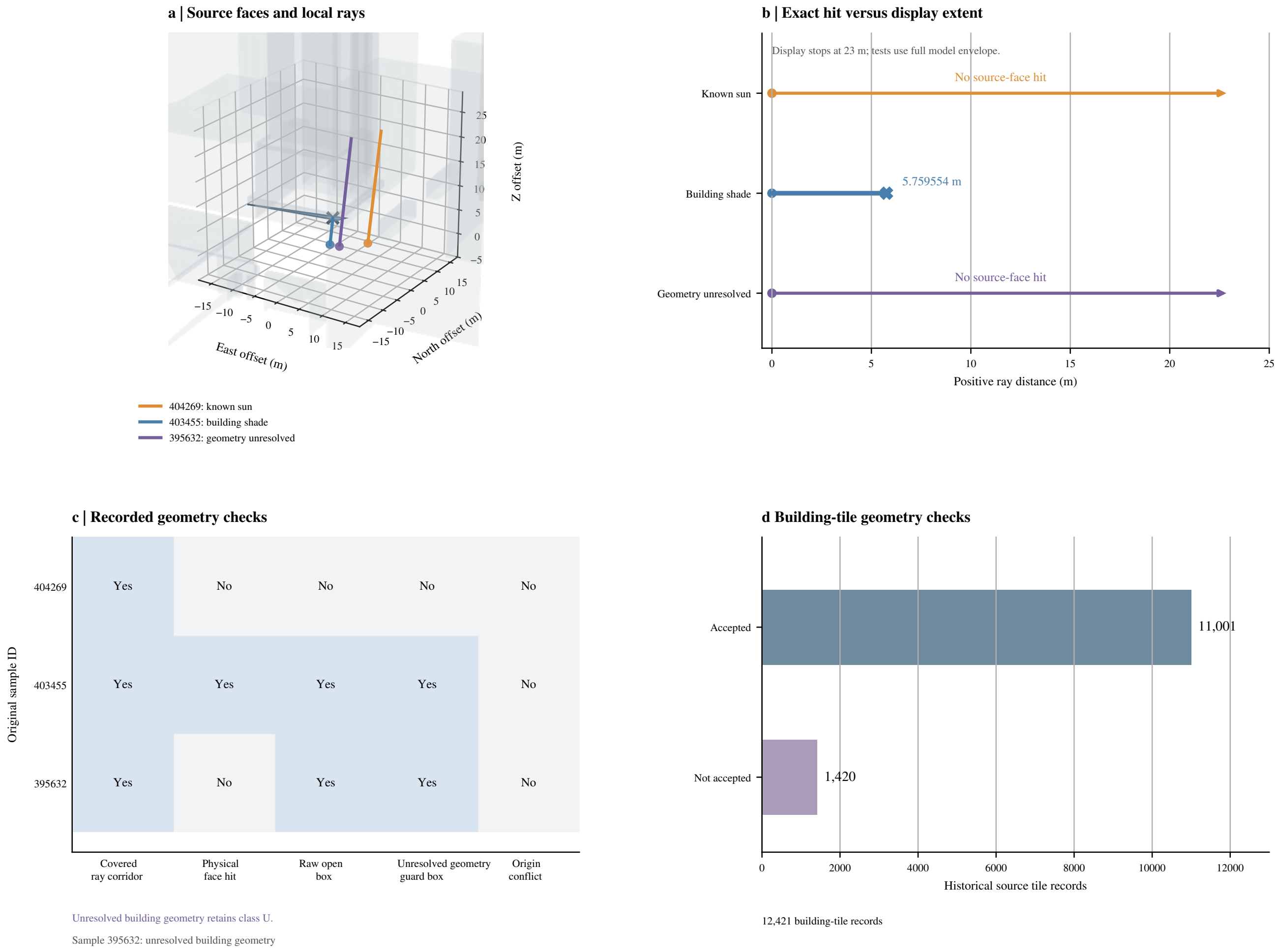

**Supplementary Fig. S3. Building shade classification a**, Building surfaces and solar rays for three sample positions classified as direct sun, building shade or unresolved exposure. **b**, Ray-intersection distances. **c**, Geometry and coverage checks for the three positions. **d**, Building-tile counts by geometry-check status. Examples use 12:00 HKT on 15 July 2026.

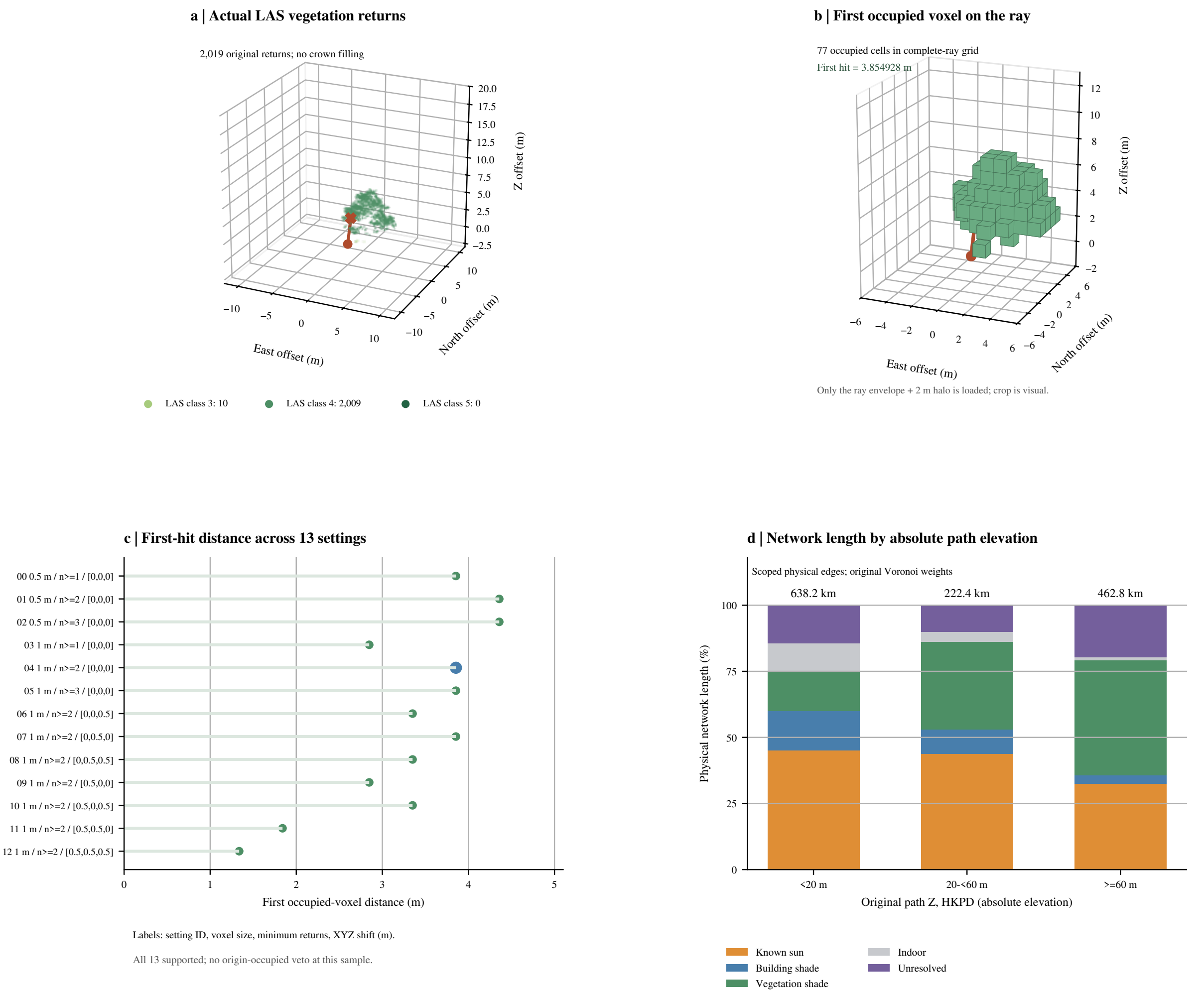

**Supplementary Fig. S4. Vegetation shade and path elevation a**, LiDAR vegetation returns around a sample position. **b**, Occupied voxels and the first ray intersection. **c**, First-intersection distances across 13 vegetation settings. **d**, Network length by exposure class and path-elevation band. Elevations are relative to the Hong Kong Principal Datum. The example uses noon exposure and a reference grid of 1 m voxels with at least two returns per occupied voxel.

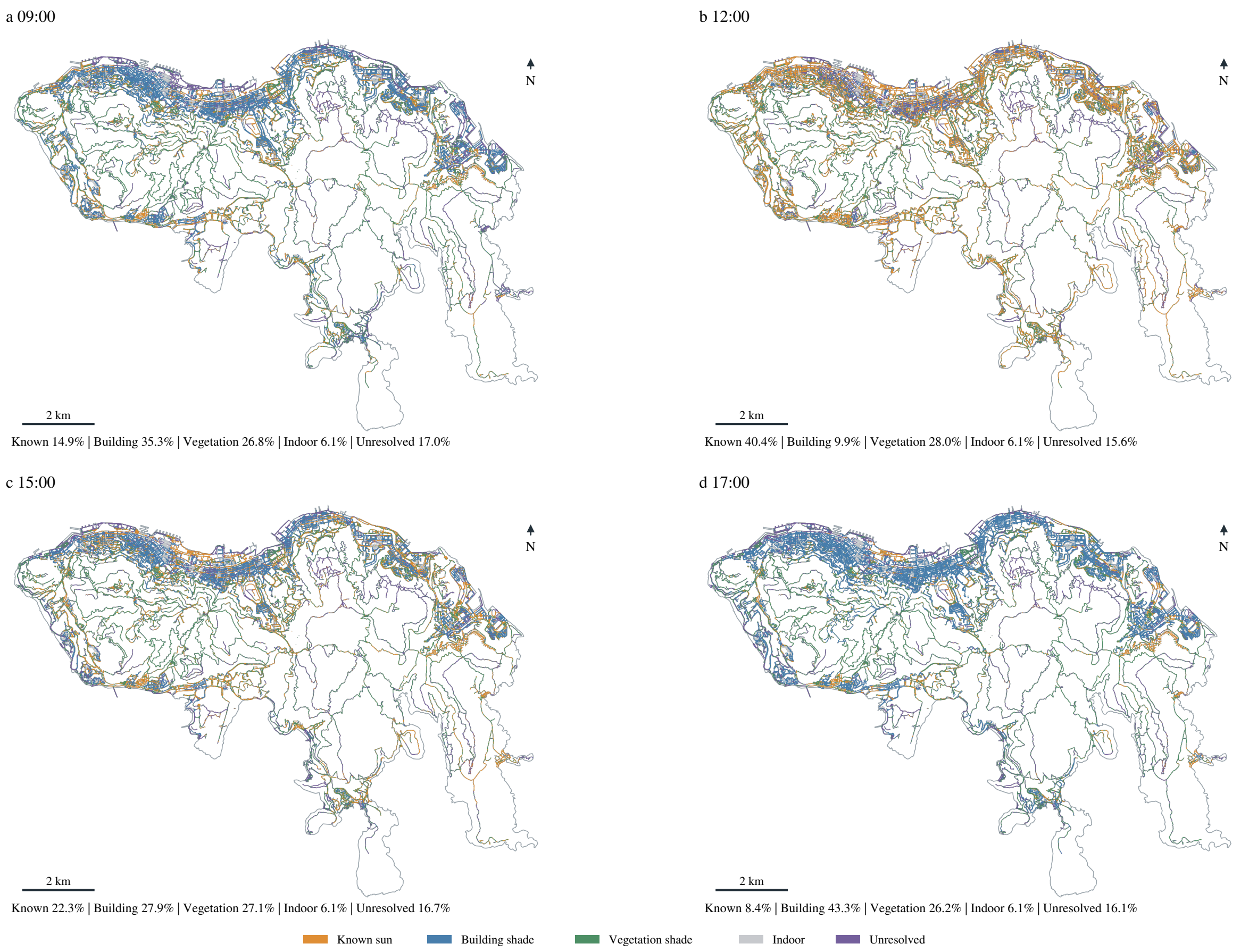

**Supplementary Fig. S5. Sun exposure and protection at four times** Exposure classes at 09:00, 12:00, 15:00 and 17:00 HKT in panels **a**, **b**, **c** and **d**, respectively, on 15 July 2026. Percentages are weighted by three-dimensional path length across the 1,323.4 km walking network.

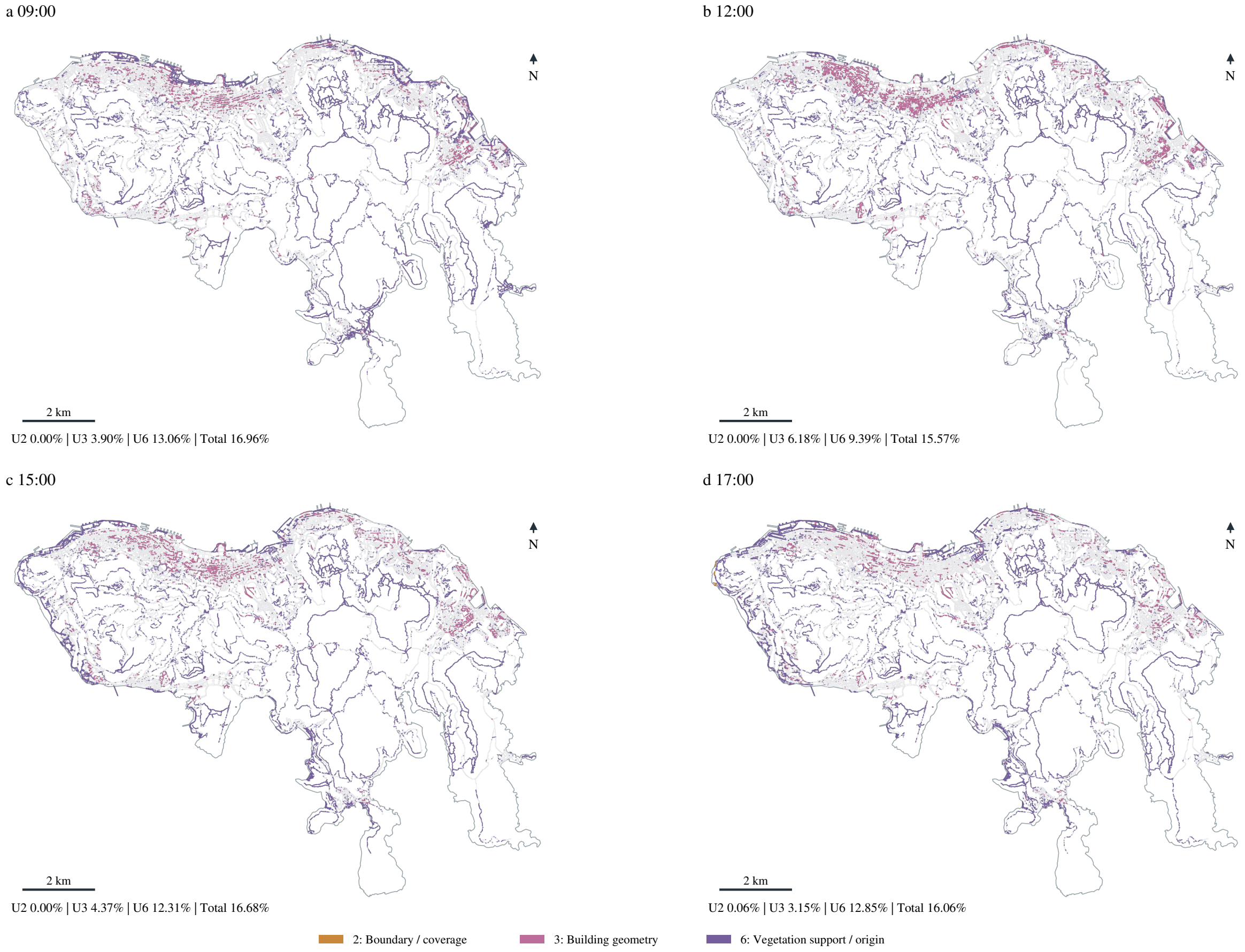

**Supplementary Fig. S6. Spatial distribution of unresolved exposure** Unresolved exposure at 09:00, 12:00, 15:00 and 17:00 HKT in panels **a**, **b**, **c** and **d**, respectively. Colours distinguish boundary or coverage gaps, unresolved building geometry, and unresolved vegetation coverage or occupied ray origins. Percentages use total three-dimensional network length.

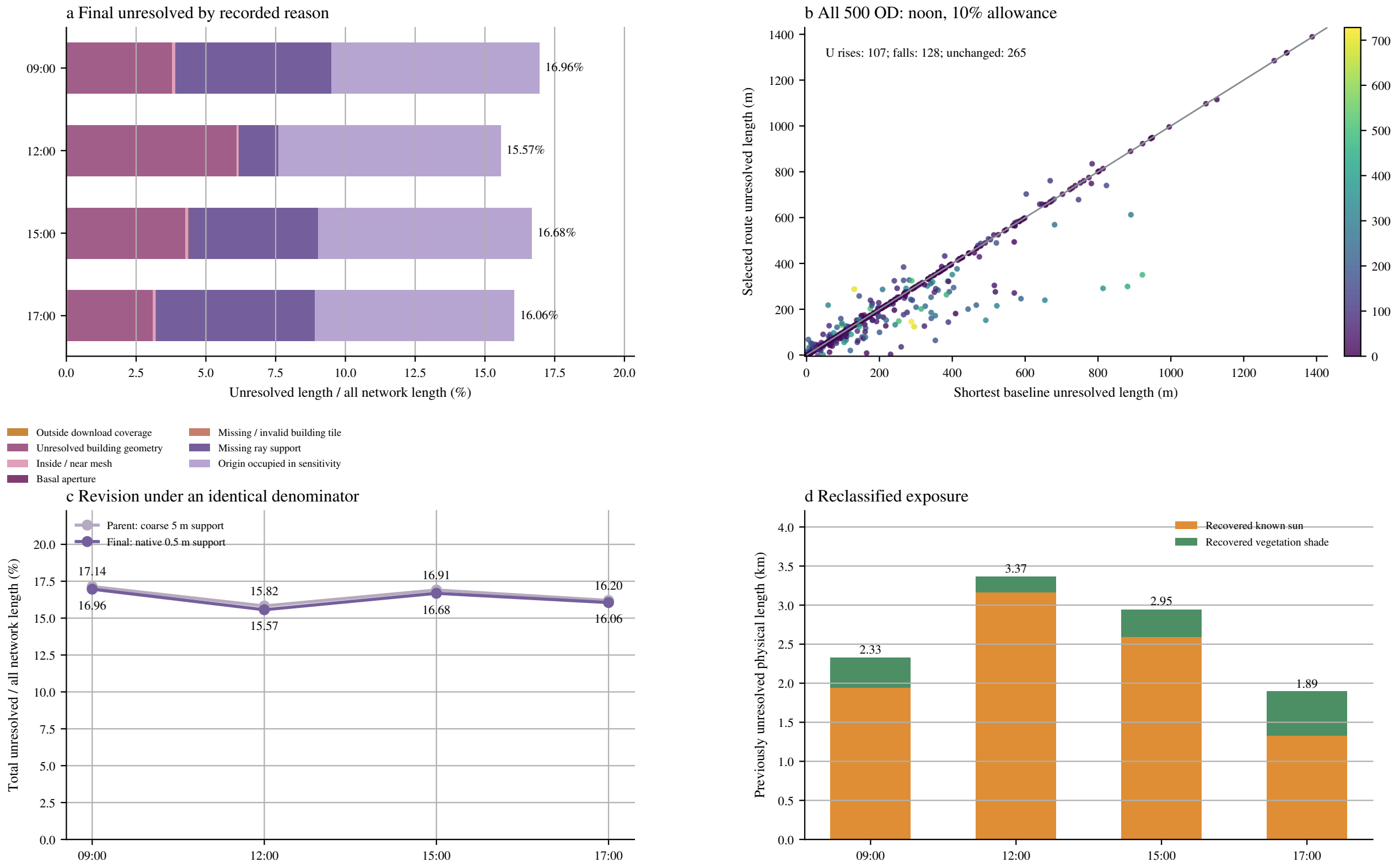

**Supplementary Fig. S7. Sources of unresolved exposure and route effects a**, Unresolved network length by classification reason at four times. **b**, Unresolved-exposure length on shortest and shade-informed routes for 500 journeys at noon with a 10% extra-distance limit. **c**, Unresolved network shares under 5 m and 0.5 m terrain-coverage checks. **d**, Previously unresolved network length classified as direct sun or vegetation shade using the 0.5 m coverage check.

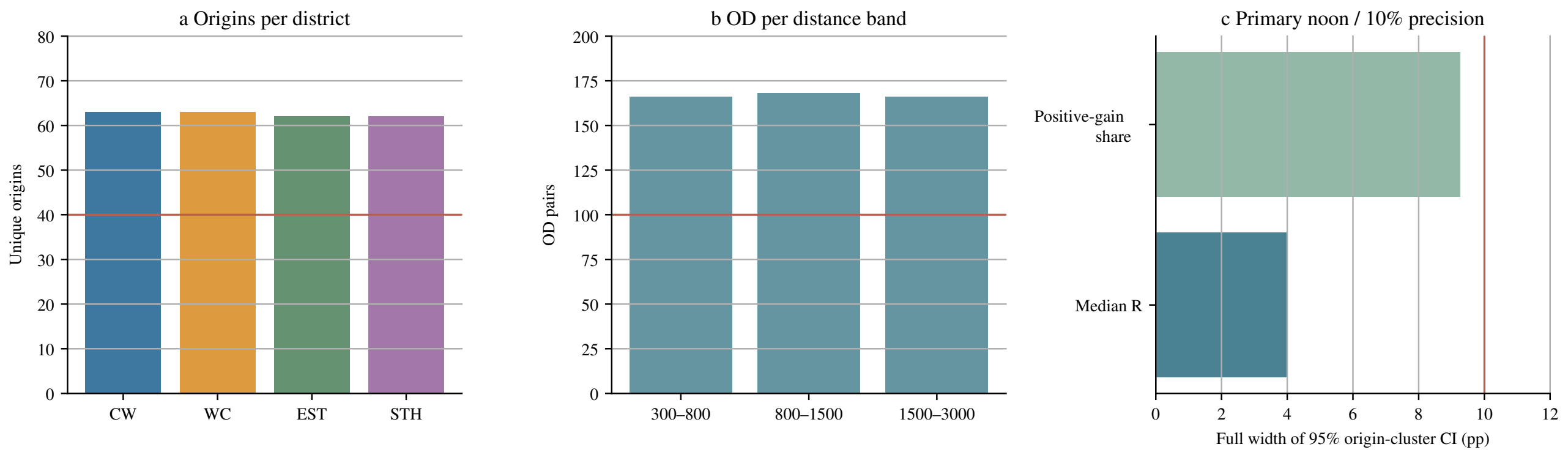

**Supplementary Fig. S8. Sample composition and estimation precision a**, Origin counts by administrative area. **b**, Journey counts by shortest-route distance band. **c**, Widths of the 95% origin-cluster confidence intervals for median proportional reduction in known sun and the share of journeys with reduced sun at noon with a 10% extra-distance limit. Dashed lines indicate the precision criteria.

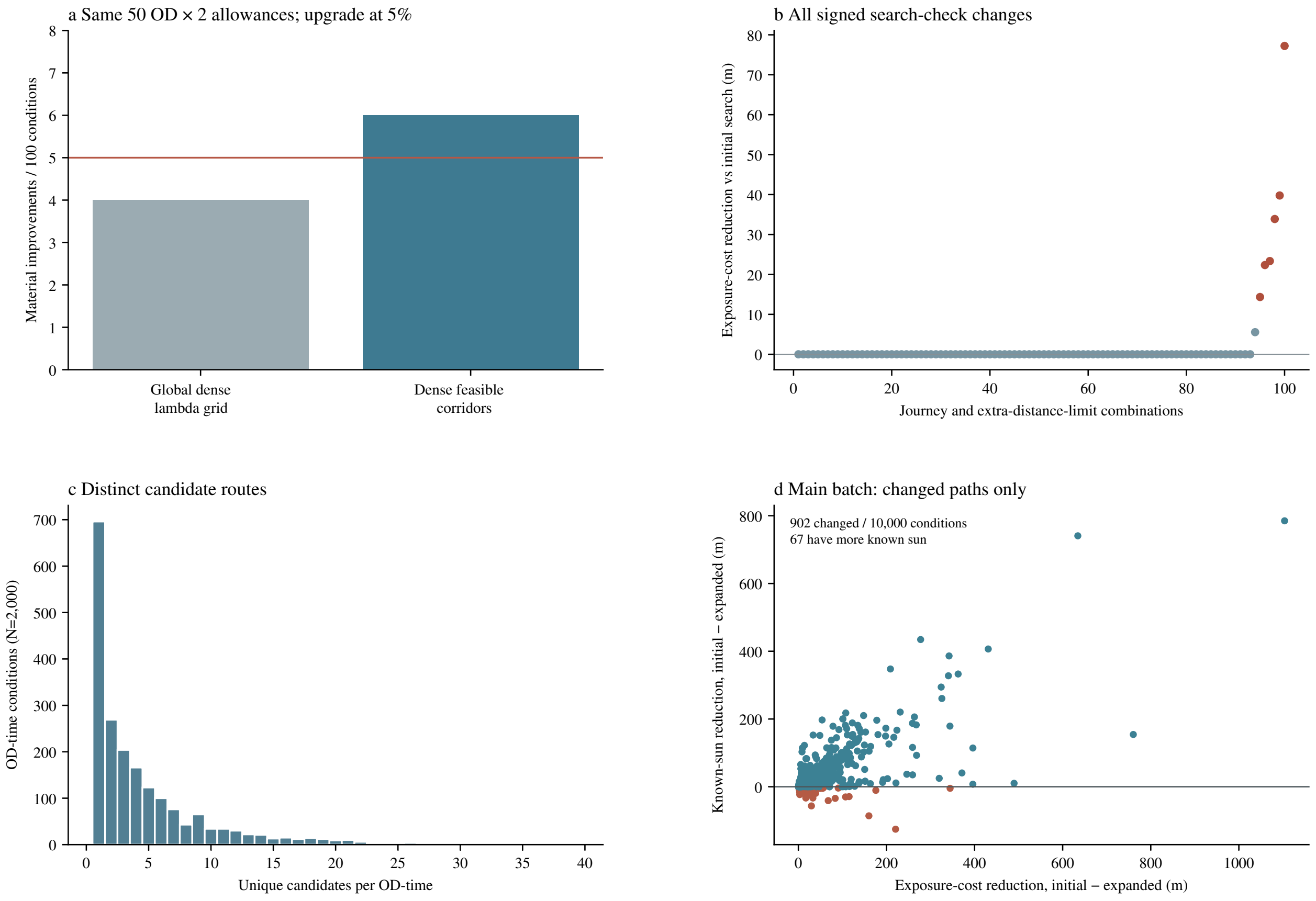

**Supplementary Fig. S9. Comparison of route-search strategies a**, Frequencies of substantial improvements from denser searches across distance and exposure weights and within feasible corridors for 50 journeys at two extra-distance limits. **b**, Changes in exposure cost relative to the initial search. **c**, Candidate-route counts for 2,000 journey-time combinations. **d**, Changes in known sun and exposure cost after search expansion. Exposure cost is the combined length of known sun and unresolved sections. Substantial improvements exceed the larger of 1 m or 1% of baseline exposure cost.

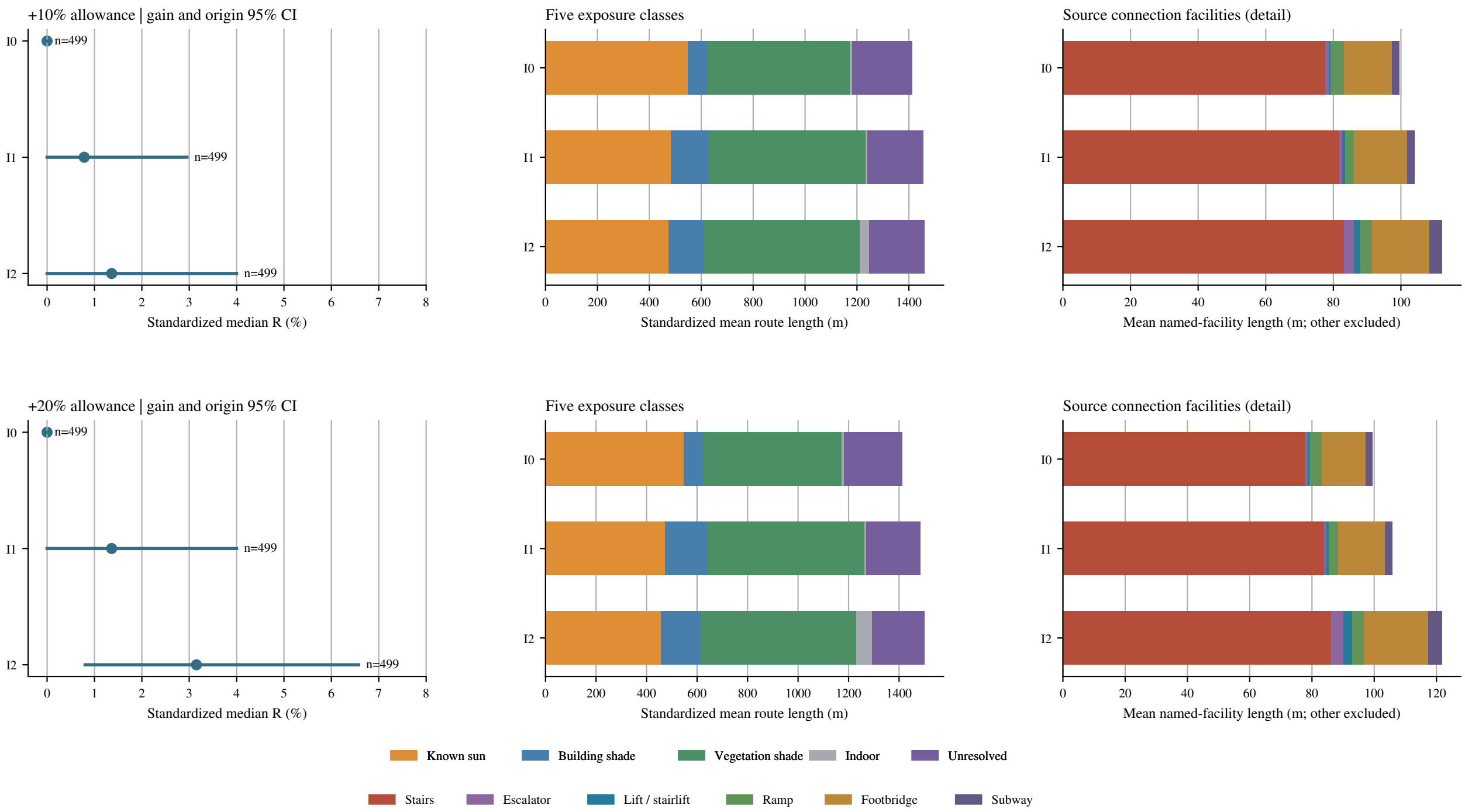

**Supplementary Fig. S10. Shade information and route composition** Proportional sun reductions (left), route composition by exposure class (middle) and facility type (right) under 10% and 20% extra-distance limits. I0 selects the shortest route. I1 uses outdoor shade information and I2 also recognises indoor protection. Each comparison uses a common candidate-route set with 500 journeys at noon. Proportional reductions use the 499 journeys with baseline sun. Error bars show 95% origin-cluster confidence intervals.

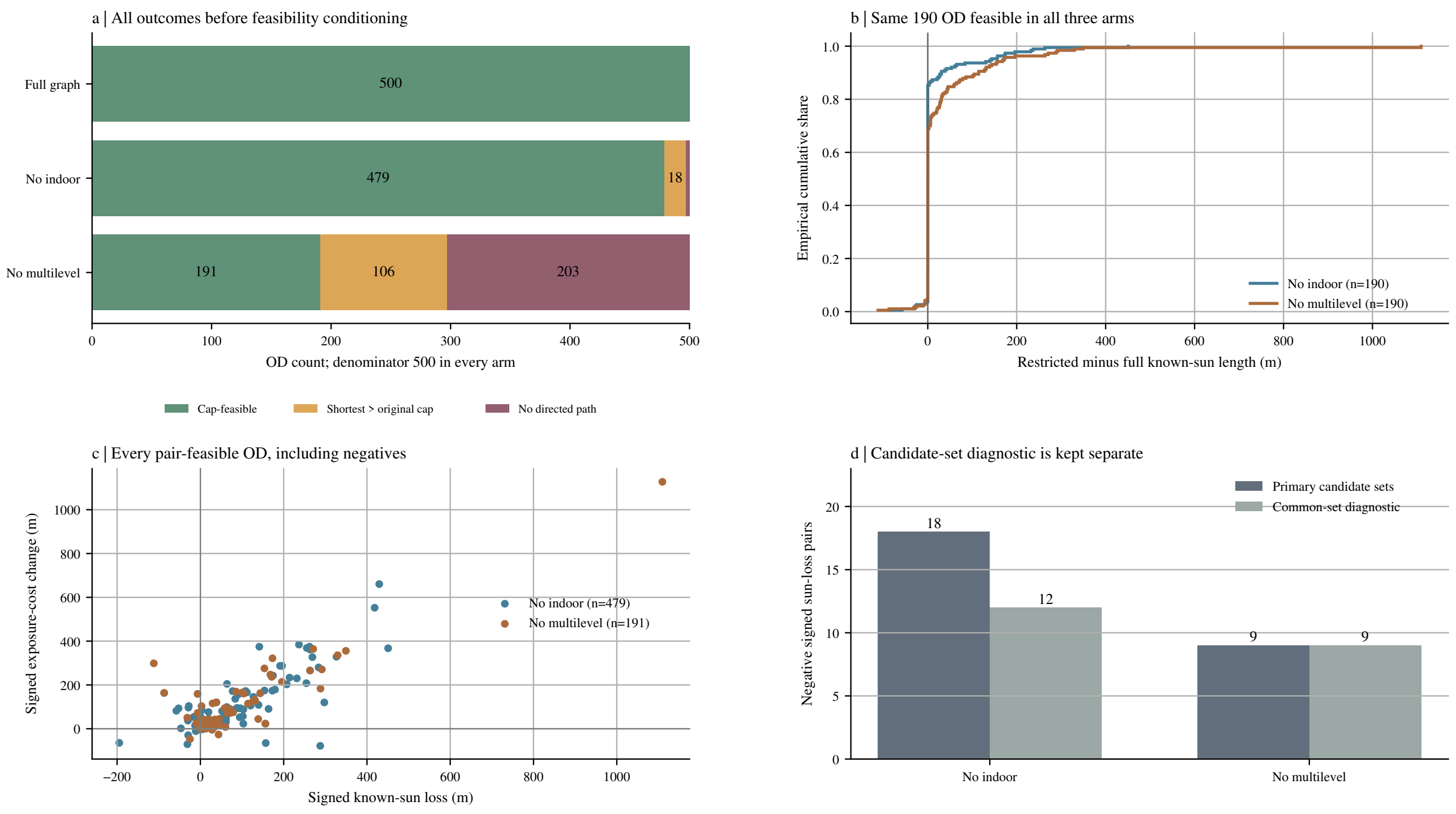

**Supplementary Fig. S11. Journey feasibility and sun exposure after removing connections a**, Feasibility counts for 500 journeys in the full network and after removing indoor or multilevel connections. **b**, Cumulative exposure changes for the 190 journeys feasible in all three networks. **c**, Exposure changes among journeys feasible in each restricted network. **d**, Counts of journeys with negative sun changes under the primary and common candidate-route sets. Counts and cumulative distributions are unweighted. Positive sun changes indicate greater exposure after removal. All panels use noon and the original 10% extra-distance limit.

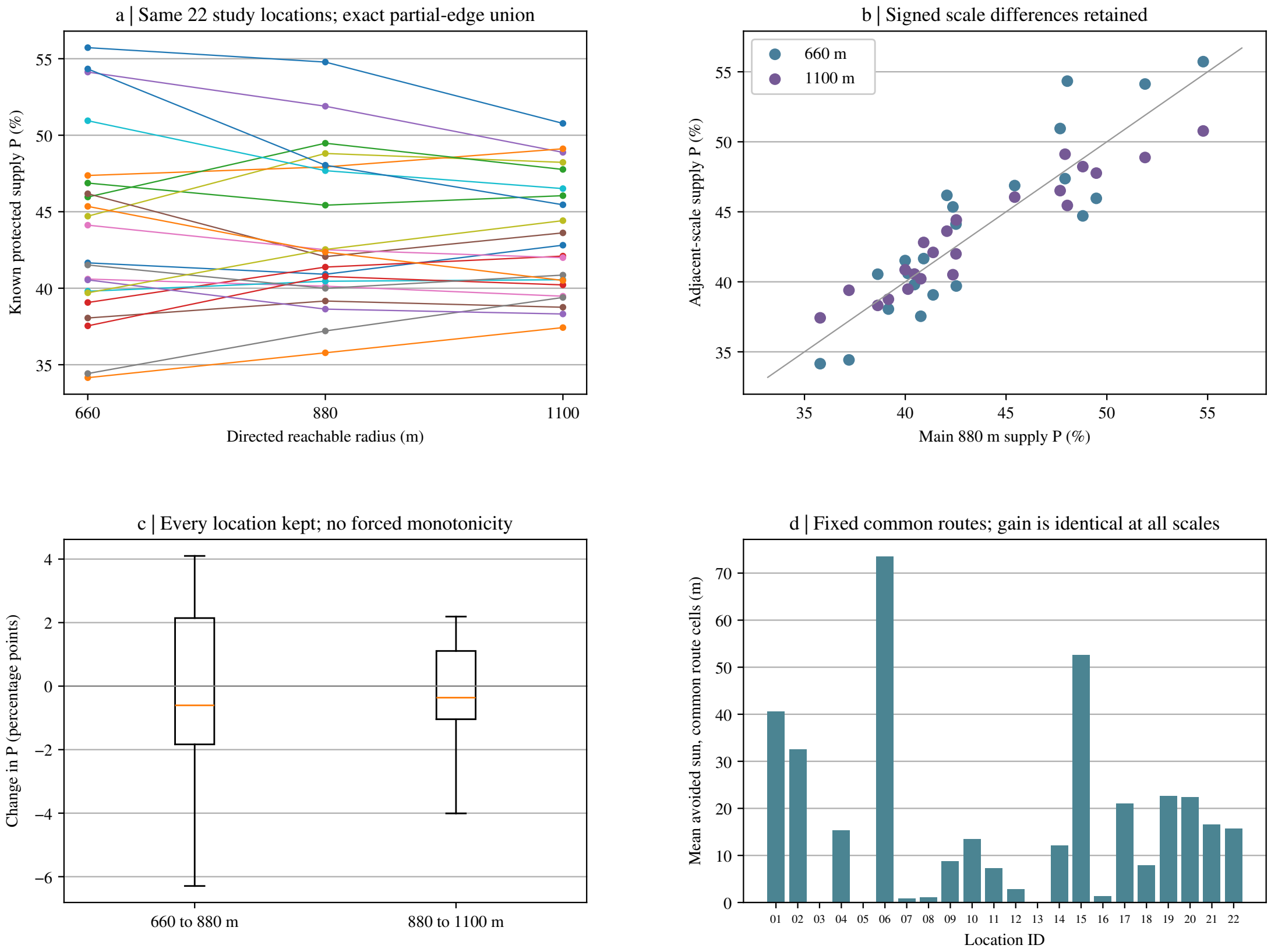

**Supplementary Fig. S12. Nearby protection across walking network radii a**, Protected-network shares at radii of 660, 880 and 1,100 m for 22 locations. **b**, Comparison with the 880 m estimates. **c**, Changes in protection share between adjacent radii. **d**, Mean avoided sun for the same 176 journeys used at every radius. Radii are distances along the walking network. Route geometry is held fixed as the walking network radius changes.

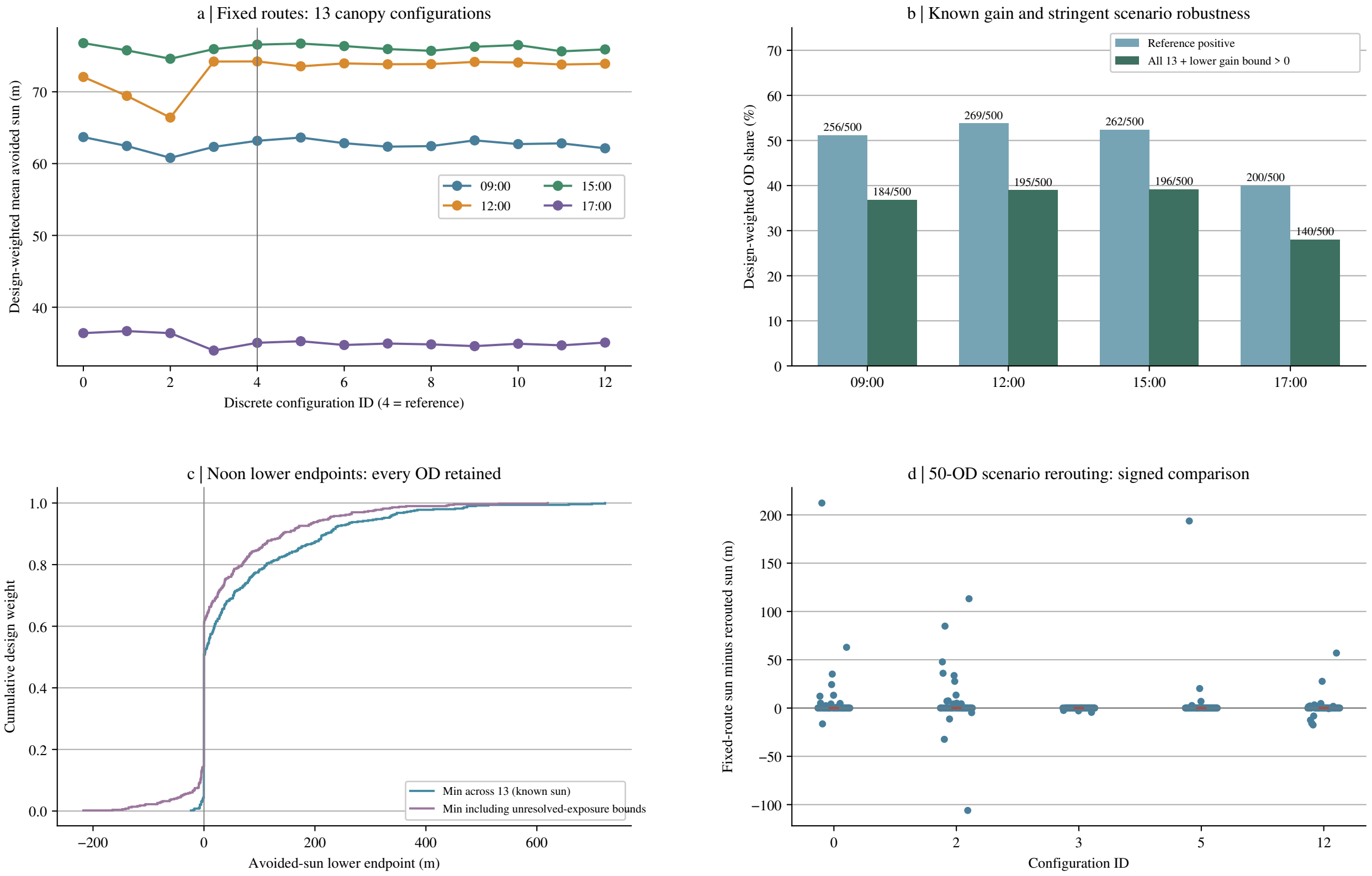

**Supplementary Fig. S13. Sensitivity to vegetation settings and unresolved exposure a**, Mean avoided sun for fixed routes across 13 vegetation settings. **b**, Shares and counts of journeys with reduced sun in the reference analysis and across all settings under the lower gain bound. **c**, Cumulative distributions of minimum gains across vegetation settings and assignments of unresolved exposure. Shared route sections receive consistent assignments. **d**, Changes in known sun after rerouting 50 journeys under alternative vegetation settings. Fixed-route comparisons use 500 journeys at four times with a 10% extra-distance limit.

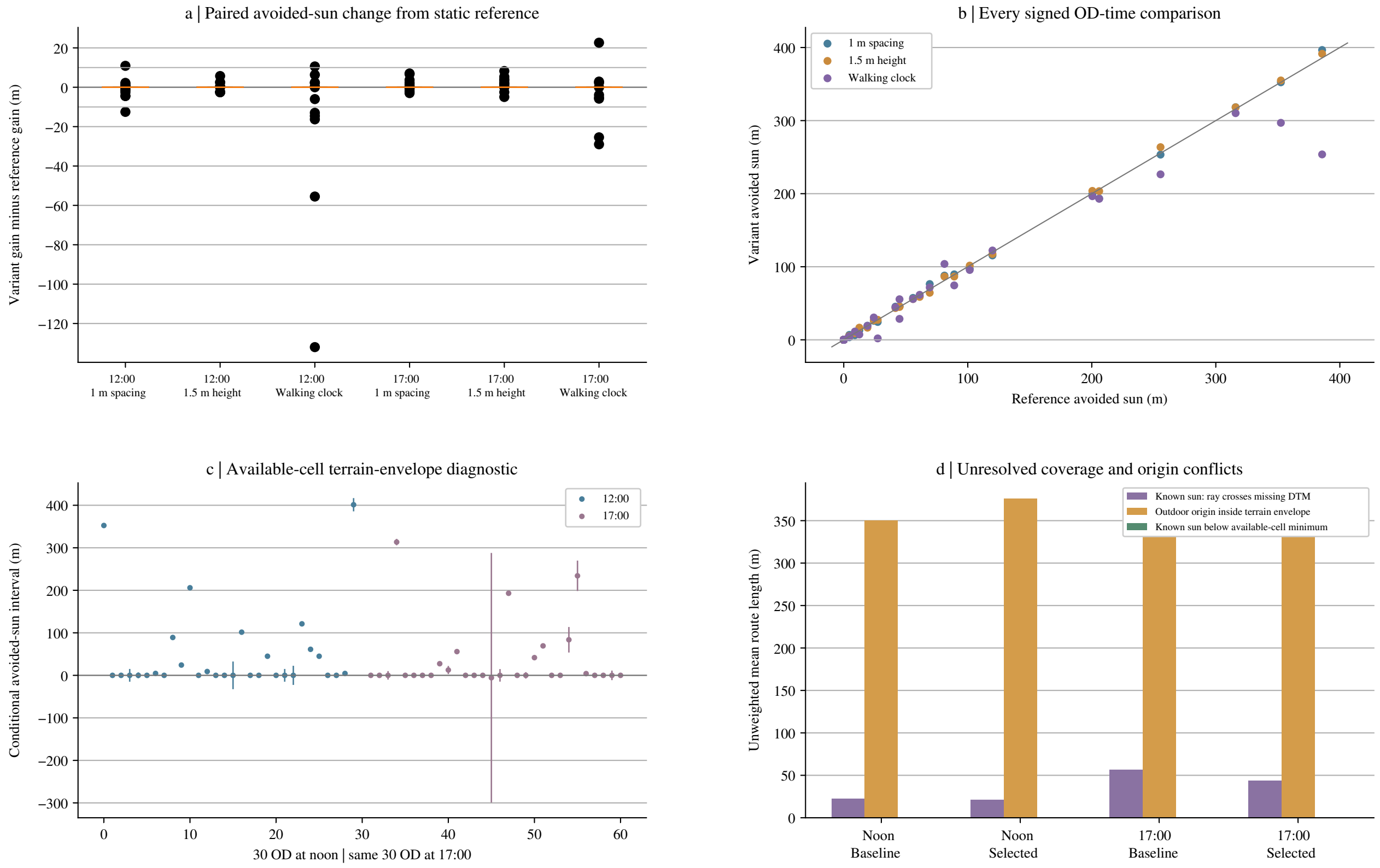

**Supplementary Fig. S14. Sensitivity to path sampling, solar position and terrain obstruction a**, Changes in avoided sun under 1 m sampling, a 1.5 m ray-origin height and solar-position updates during walking. **b**, Corresponding comparisons with the reference estimates. **c**, Terrain-based gain ranges for sections with usable elevation data. **d**, Route lengths affected by missing terrain data, ray origins within the terrain envelope and potential terrain obstruction. Comparisons use the same fixed route pairs for 30 journeys from 29 origins at 12:00 and 17:00 HKT.

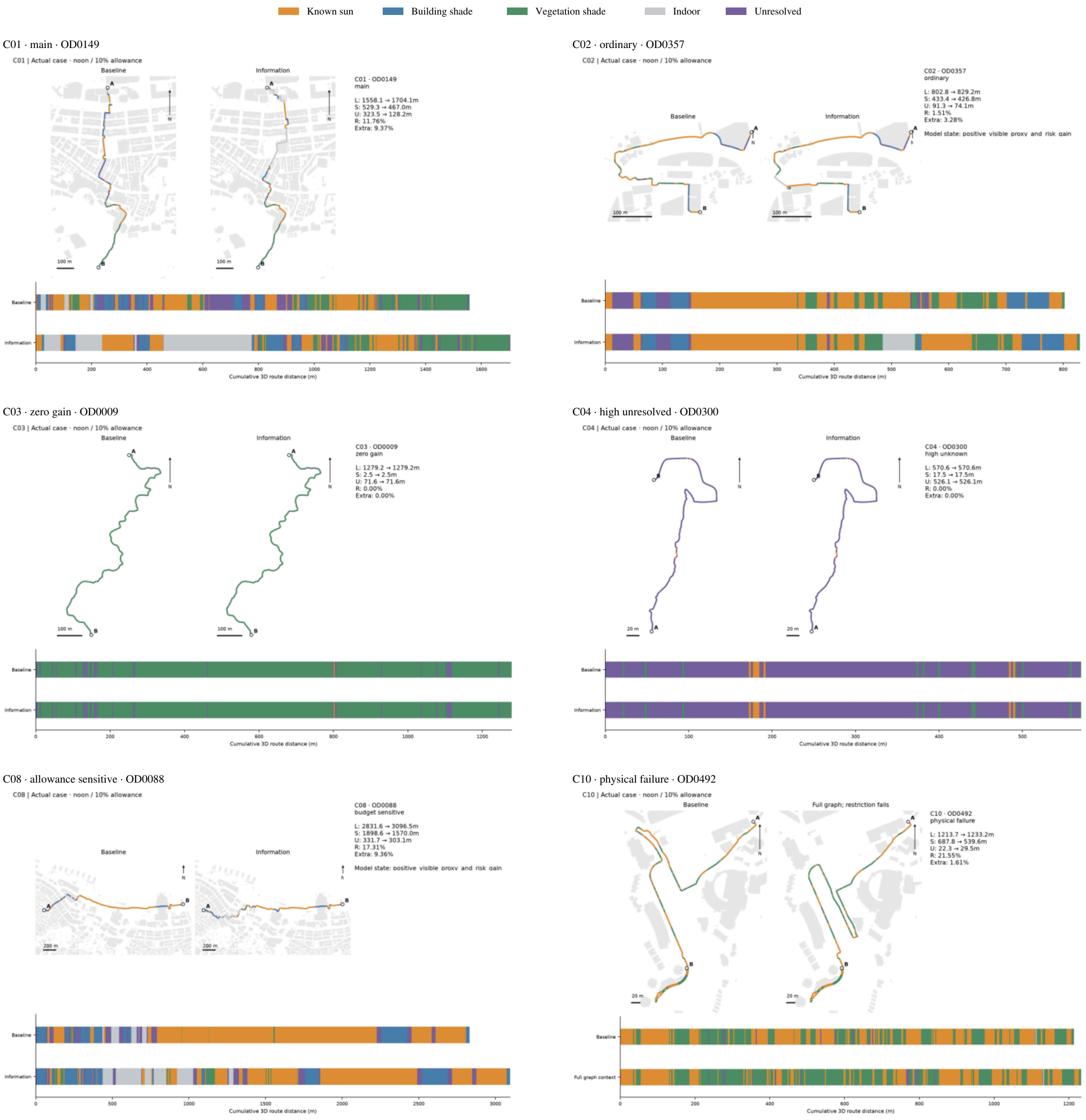

**Supplementary Fig. S15. Examples of shortest and shade-informed routes** Route maps and ordered exposure profiles for six cases. C01 is the main illustrative journey. C02 shows an ordinary gain, C03 zero gain, C04 extensive unresolved exposure and C08 sensitivity to the extra-distance limit. C10 is a journey disconnected by removing multilevel links. All cases use 12:00 HKT on 15 July 2026 and a 10% extra-distance limit.

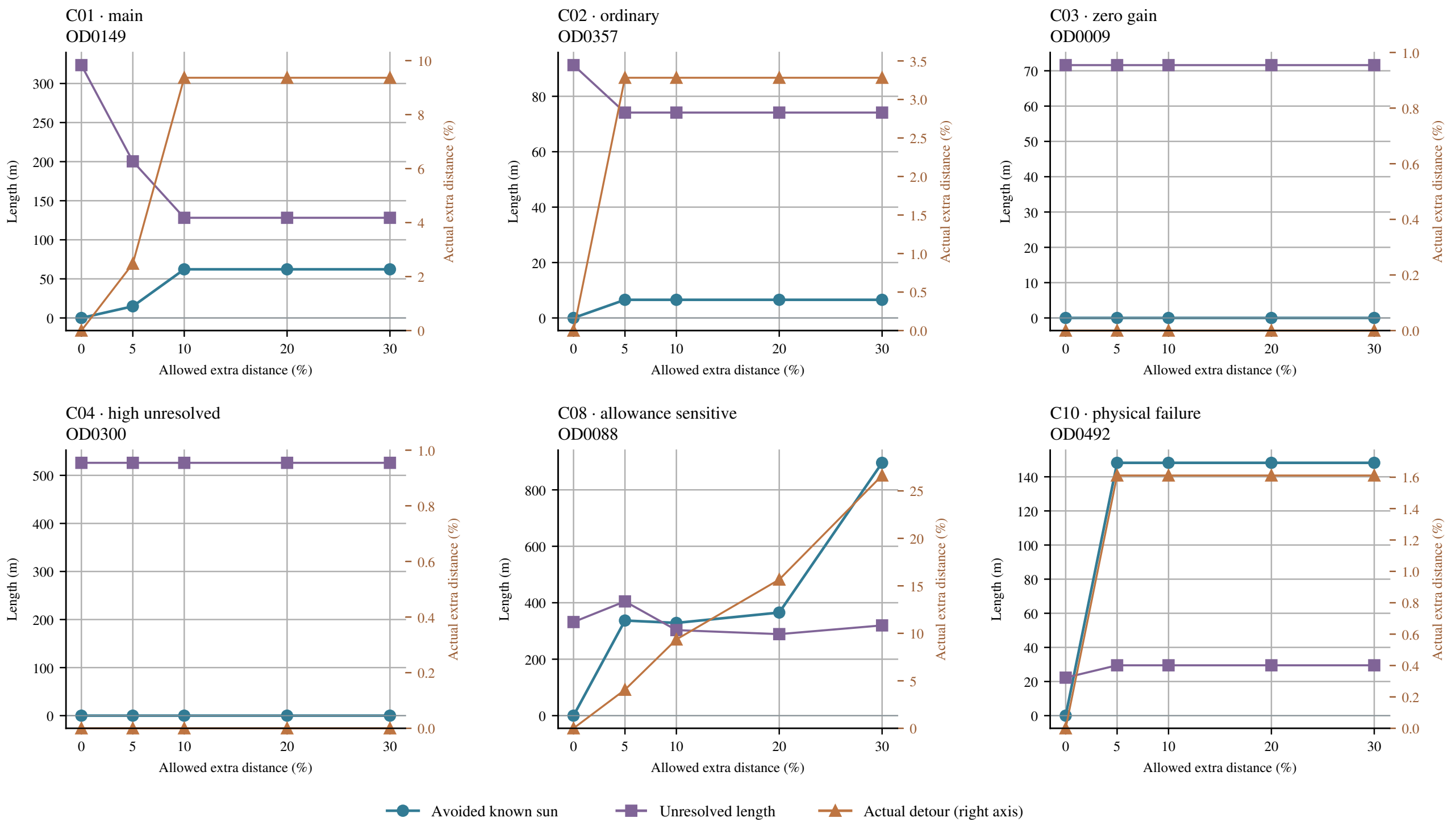

**Supplementary Fig. S16. Sun avoidance at different extra-distance limits** Known-sun reduction and unresolved route length (left axes), and actual additional walking as a percentage of shortest-route length (right axes), across the five extra-distance limits. The six cases match Supplementary Fig. S15. All results use noon exposure in the full walking network.

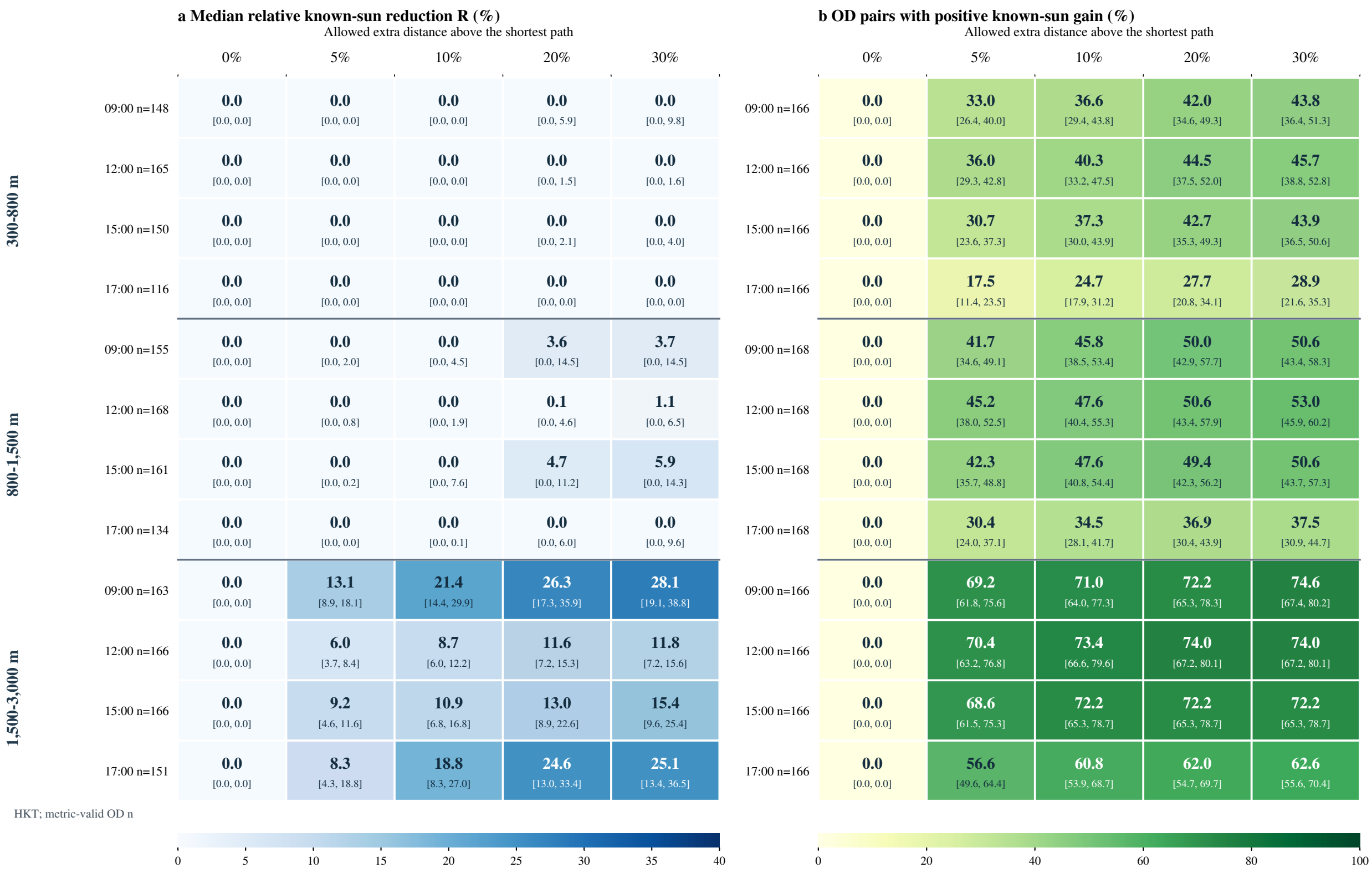

**Supplementary Fig. S17. Sun avoidance across times, journey distances and extra-distance limits a**, Median proportional reduction in known sun. **b**, Shares of journeys with reduced known sun. Columns show five extra-distance limits and rows combine four times with three shortest-route distance bands. Cells show design-weighted estimates and 95% origin-cluster confidence intervals from 1,000 bootstrap replicates. Proportional reductions use journeys with baseline sun. The full sample contains 500 journeys from 250 origins on 15 July 2026.

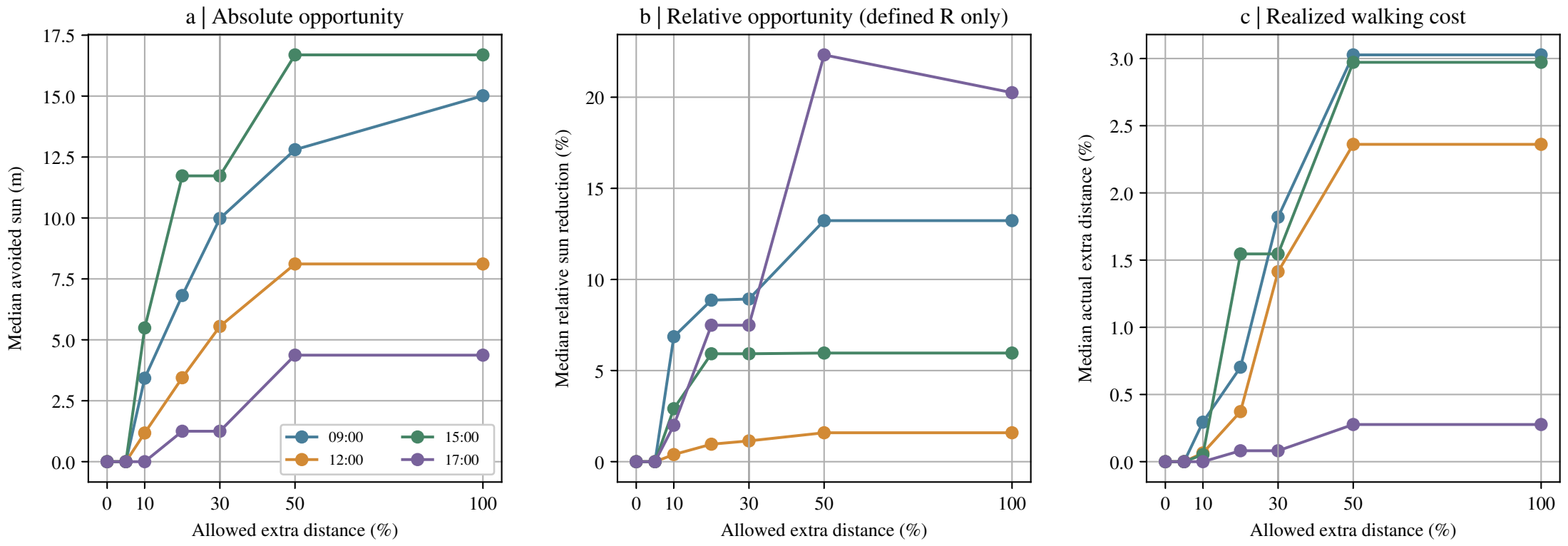

**Supplementary Fig. S18. Sun avoidance in extended route searches a**, Median avoided known-sun distance. **b**, Median proportional reduction. **c**, Median additional walking as a percentage of shortest-route length. Unweighted summaries use the same 50 journeys at four times. An expanded candidate-route set is used at 50% and 100% extra-distance limits. Dotted lines separate the primary and expanded searches.

# Supplementary Tables

Lengths are in metres. Percentages and dimensionless fractions are identified in the column headings. Estimates are displayed to two decimal places, with full precision retained in the source data. CI denotes confidence interval. Origin-cluster and spatial-block confidence intervals use the resampling units defined in Supplementary Methods 7. Changes in shares are expressed in percentage points (pp). Time IDs 0 to 3 correspond to 09:00, 12:00, 15:00 and 17:00 HKT.

## Supplementary Table S1. Data sources

The inventory identifies the source products, acquisition or archive dates, coordinate systems and official access links used to assemble the model. Geometry acquisition dates and the modelled date are recorded separately.

| Input | Product | Acquisition or archive | Coordinates | Official source |
|---|---|---|---|---|
| pedestrian_network | LandsD 3D Pedestrian Network | 2026-09-26 archived geometry snapshot; island_m1_network_v1/island_m3_routing_v1 | EPSG:2326 XY; EPSG:5738 HKPD Z where 3D | https://data.gov.hk/en-data/dataset/hk-landsd-openmap-3d-pedestrian-network |
| indoor_network | LandsD 3D Indoor Network | 2026-09-26 archived geometry snapshot | EPSG:2326 XY; EPSG:5738 HKPD Z where 3D | https://portal.csdi.gov.hk/server/rest/services/common/landsd_rcd_1742809197336_78665/MapServer |
| buildings | LandsD building-specific 3D Tiles | 2026-09-26 geometry; M2A refinement1 | EPSG:2326 XY; EPSG:5738 HKPD Z where 3D | https://portal.csdi.gov.hk/csdi-webpage/apidoc/3d-spatial-data-api |
| vegetation | CEDD LAS specified-class vegetation | Acquired 2019-12 through 2020-02; product 2020; research download 2026-09 | EPSG:2326 XY; EPSG:5738 HKPD Z where 3D | https://www.ginfo.cedd.gov.hk/geoopendata/eng/LIDAR.aspx |
| terrain | CEDD native DTM | 2020 product; native 0.5m; diagnostic 5m min/max cells | EPSG:2326 XY; EPSG:5738 HKPD Z where 3D | https://www.ginfo.cedd.gov.hk/geoopendata/Data/LiDAR/2020%20LiDAR.xml |
| coastline | LandsD HYDRLINE coast, CLASS SLN and TYPE COA | Metadata 2026-05-29; retrieved 2026-10-10 UTC; dictionary v1.2 | EPSG:2326 XY; EPSG:5738 HKPD Z where 3D | https://portal.csdi.gov.hk/csdi-webpage/metadata/landsd_rcd_1637221775627_85634/html |
| districts | HAD administrative boundaries | 2026-09-26 archived district snapshot; four polygons clipped to strict landmask | EPSG:2326 XY; EPSG:5738 HKPD Z where 3D | https://data.gov.hk/en-data/dataset/hk-had-json1-hong-kong-administrative-boundaries |

## Supplementary Table S2. Network sun exposure and protection

The four-time network classification was integrated over 1,323.4 km of retained physical segments. At noon, known sun accounted for 40.40%, building shade 9.95%, vegetation shade 28.02%, indoor protection 6.07% and unresolved exposure 15.57%. Part a reports every evaluation time, part b uses each administrative area's own network length as its denominator, and part c decomposes unresolved sources. Known protection combines building shade, vegetation shade and indoor protection. OD outcomes use their separately defined route and sampling weights. The spatial classifications and unresolved-source diagnostics appear in Supplementary Figs. S5–S7.

### Supplementary Table S2a. Exposure-class lengths at four evaluation times

| Time | Class | Length (m) | Share (%) |
|---|---|---|---|
| 09:00 | Known sun | 196,694.66 | 14.86 |
| 09:00 | Building shade | 467,730.81 | 35.34 |
| 09:00 | Unresolved exposure | 224,486.92 | 16.96 |
| 09:00 | Indoor protection | 80,274.27 | 6.07 |
| 09:00 | Vegetation shade | 354,215.43 | 26.77 |
| 12:00 | Known sun | 534,640.26 | 40.40 |
| 12:00 | Building shade | 131,672.37 | 9.95 |
| 12:00 | Unresolved exposure | 206,054.21 | 15.57 |
| 12:00 | Indoor protection | 80,274.27 | 6.07 |
| 12:00 | Vegetation shade | 370,760.97 | 28.02 |
| 15:00 | Known sun | 294,651.56 | 22.26 |
| 15:00 | Building shade | 368,882.40 | 27.87 |
| 15:00 | Unresolved exposure | 220,805.21 | 16.68 |
| 15:00 | Indoor protection | 80,274.27 | 6.07 |
| 15:00 | Vegetation shade | 358,788.64 | 27.11 |
| 17:00 | Known sun | 110,880.39 | 8.38 |
| 17:00 | Building shade | 573,614.90 | 43.34 |
| 17:00 | Unresolved exposure | 212,485.97 | 16.06 |
| 17:00 | Indoor protection | 80,274.27 | 6.07 |

| Time | Class | Length (m) | Share (%) |
|---|---|---|---|
| 17:00 | Vegetation shade | 346,146.56 | 26.16 |

## Supplementary Table S2b. Noon exposure-class shares within administrative areas

| Administrative area on Hong Kong Island | Network length (km) | Known sun (%) | Building shade (%) | Vegetation shade (%) | Indoor protection (%) | Unresolved (%) | Known protection total (%) |
|---|---|---|---|---|---|---|---|
| Central and Western | 341.25 | 37.30 | 12.10 | 26.97 | 9.89 | 13.74 | 48.96 |
| Eastern | 375.12 | 41.52 | 11.77 | 23.49 | 5.88 | 17.34 | 41.14 |
| Southern | 361.45 | 41.47 | 3.69 | 35.00 | 3.29 | 16.54 | 41.99 |
| Wan Chai | 245.58 | 41.41 | 13.39 | 26.10 | 5.12 | 13.98 | 44.61 |

## Supplementary Table S2c. Sources of unresolved exposure

| Time | Family | Reason | Samples | Length (m) | Network share (%) |
|---|---|---|---|---|---|
| 09:00 | inherited | Outside download coverage | 0 | 0.00 | 0.00 |
| 09:00 | inherited | Uncertified source object | 25252 | 50,124.90 | 3.79 |
| 09:00 | inherited | Inside / near mesh | 1176 | 1,520.50 | 0.11 |
| 09:00 | inherited | Basal aperture | 21 | 24.11 | 0.00 |
| 09:00 | inherited | Missing / invalid building tile | 0 | 0.00 | 0.00 |
| 09:00 | canopy | Missing ray support | 35288 | 74,060.51 | 5.60 |
| 09:00 | canopy | Origin occupied in sensitivity | 42427 | 98,756.90 | 7.46 |
| 12:00 | inherited | Outside download coverage | 0 | 0.00 | 0.00 |
| 12:00 | inherited | Uncertified source object | 40541 | 80,659.57 | 6.09 |
| 12:00 | inherited | Inside / near mesh | 646 | 1,039.43 | 0.08 |
| 12:00 | inherited | Basal aperture | 97 | 93.15 | 0.01 |
| 12:00 | inherited | Missing / invalid building tile | 0 | 0.00 | 0.00 |
| 12:00 | canopy | Missing ray support | 8725 | 18,685.20 | 1.41 |
| 12:00 | canopy | Origin occupied in sensitivity | 45661 | 105,576.86 | 7.98 |
| 15:00 | inherited | Outside download coverage | 0 | 0.00 | 0.00 |

| Time | Family | Reason | Samples | Length (m) | Network share (%) |
|---|---|---|---|---|---|
| 15:00 | inherited | Uncertified source object | 28410 | 56,404.21 | 4.26 |
| 15:00 | inherited | Inside / near mesh | 1076 | 1,441.65 | 0.11 |
| 15:00 | inherited | Basal aperture | 18 | 23.93 | 0.00 |
| 15:00 | inherited | Missing / invalid building tile | 0 | 0.00 | 0.00 |
| 15:00 | canopy | Missing ray support | 29165 | 61,671.04 | 4.66 |
| 15:00 | canopy | Origin occupied in sensitivity | 43561 | 101,264.39 | 7.65 |
| 17:00 | inherited | Outside download coverage | 381 | 762.26 | 0.06 |
| 17:00 | inherited | Uncertified source object | 19873 | 40,055.45 | 3.03 |
| 17:00 | inherited | Inside / near mesh | 1206 | 1,556.23 | 0.12 |
| 17:00 | inherited | Basal aperture | 14 | 27.76 | 0.00 |
| 17:00 | inherited | Missing / invalid building tile | 0 | 0.00 | 0.00 |
| 17:00 | canopy | Missing ray support | 35654 | 75,493.78 | 5.70 |
| 17:00 | canopy | Origin occupied in sensitivity | 40483 | 94,590.50 | 7.15 |

## Supplementary Table S3. Sampling design and estimation precision

Part a reports sample counts and stratum weights for 500 journeys from 250 origins. Part b reports all 33 precision checks, their thresholds, units and outcomes. All checks passed. Sampling and precision assessment are described in Supplementary Methods 3, with sampling balance and interval widths shown in Supplementary Fig. S8.

### Supplementary Table S3a. Sampling counts and stratum weights

| Group | Value | OD | Origins | Stratum weight |
|---|---|---|---|---|
| All journeys | ALL | 500 | 250 | 0.0833333333333 |
| Administrative area | CW | 126 | 63 | 0.333333333333 |
| Administrative area | WC | 126 | 63 | 0.333333333333 |
| Administrative area | EST | 124 | 62 | 0.333333333333 |
| Administrative area | STH | 124 | 62 | 0.333333333333 |
| Distance band | 300_800 | 166 | 166 | 0.25 |
| Distance band | 800_1500 | 168 | 168 | 0.25 |
| Distance band | 1500_3000 | 166 | 166 | 0.25 |
| district_distance | CW/300_800 | 42 | 42 | 0.0833333333333 |
| district_distance | CW/800_1500 | 42 | 42 | 0.0833333333333 |
| district_distance | CW/1500_3000 | 42 | 42 | 0.0833333333333 |
| district_distance | WC/300_800 | 42 | 42 | 0.0833333333333 |
| district_distance | WC/800_1500 | 42 | 42 | 0.0833333333333 |
| district_distance | WC/1500_3000 | 42 | 42 | 0.0833333333333 |
| district_distance | EST/300_800 | 41 | 41 | 0.0833333333333 |
| district_distance | EST/800_1500 | 42 | 42 | 0.0833333333333 |
| district_distance | EST/1500_3000 | 41 | 41 | 0.0833333333333 |
| district_distance | STH/300_800 | 41 | 41 | 0.0833333333333 |
| district_distance | STH/800_1500 | 42 | 42 | 0.0833333333333 |
| district_distance | STH/1500_3000 | 41 | 41 | 0.0833333333333 |

### Supplementary Table S3b. Estimation precision checks

| Criterion | Group | Observed | Operator | Threshold | Unit | Passed |
|---|---|---|---|---|---|---|
| origins_each_district | CW | 63 | >= | 40 | origins | True |
| origins_each_district | WC | 63 | >= | 40 | origins | True |
| origins_each_district | EST | 62 | >= | 40 | origins | True |
| origins_each_district | STH | 62 | >= | 40 | origins | True |
| OD_each_distance_band | 300_800 | 166 | >= | 100 | OD | True |
| OD_each_distance_band | 800_1500 | 168 | >= | 100 | OD | True |
| OD_each_distance_band | 1500_3000 | 166 | >= | 100 | OD | True |
| OD_each_district_distance | CW/300_800 | 42 | >= | 30 | OD | True |
| origins_each_district_distance | CW/300_800 | 42 | >= | 20 | origins | True |
| OD_each_district_distance | CW/800_1500 | 42 | >= | 30 | OD | True |
| origins_each_district_distance | CW/800_1500 | 42 | >= | 20 | origins | True |

| Criterion | Group | Observed | Operator | Threshold | Unit | Passed |
|---|---|---|---|---|---|---|
| OD_each_district_distance | CW/1500_3000 | 42 | >= | 30 | OD | True |
| origins_each_district_distance | CW/1500_3000 | 42 | >= | 20 | origins | True |
| OD_each_district_distance | WC/300_800 | 42 | >= | 30 | OD | True |
| origins_each_district_distance | WC/300_800 | 42 | >= | 20 | origins | True |
| OD_each_district_distance | WC/800_1500 | 42 | >= | 30 | OD | True |
| origins_each_district_distance | WC/800_1500 | 42 | >= | 20 | origins | True |
| OD_each_district_distance | WC/1500_3000 | 42 | >= | 30 | OD | True |
| origins_each_district_distance | WC/1500_3000 | 42 | >= | 20 | origins | True |
| OD_each_district_distance | EST/300_800 | 41 | >= | 30 | OD | True |
| origins_each_district_distance | EST/300_800 | 41 | >= | 20 | origins | True |
| OD_each_district_distance | EST/800_1500 | 42 | >= | 30 | OD | True |
| origins_each_district_distance | EST/800_1500 | 42 | >= | 20 | origins | True |
| OD_each_district_distance | EST/1500_3000 | 41 | >= | 30 | OD | True |
| origins_each_district_distance | EST/1500_3000 | 41 | >= | 20 | origins | True |
| OD_each_district_distance | STH/300_800 | 41 | >= | 30 | OD | True |
| origins_each_district_distance | STH/300_800 | 41 | >= | 20 | origins | True |
| OD_each_district_distance | STH/800_1500 | 42 | >= | 30 | OD | True |
| origins_each_district_distance | STH/800_1500 | 42 | >= | 20 | origins | True |
| OD_each_district_distance | STH/1500_3000 | 41 | >= | 30 | OD | True |
| origins_each_district_distance | STH/1500_3000 | 41 | >= | 20 | origins | True |
| r_median_95ci_total_width_pp_max | overall/noon/budget10 | 4.00014126185 | <= | 10 | percentage_points | True |
| positive_gain_share_95ci_total_width_pp_max | overall/noon/budget10 | 9.27228022717 | <= | 10 | percentage_points | True |

## Supplementary Table S4. Sun avoidance at different extra-distance limits

At noon, mean avoided known-sun distance was 56.87, 74.23, 89.58 and 98.43 m at 5%, 10%, 20% and 30% extra-distance limits, respectively. The 5% extra-distance limit delivered 77% of the mean avoidance at the 10% extra-distance limit. Dividing mean paired increments by the size of each successive extra-distance limit increase gave 11.4, 3.5, 1.5 and 0.9 m per percentage point, in ascending extra-distance limit order. Route-level gains decreased on 10, 10 and nine journeys across the three transitions above 5%, respectively. The exposure cost combined known sun and unresolved exposure, so changes in shade-informed routes could alter both categories. Of 499 journeys with known sun on the shortest route, 25 achieved at least a 50% reduction at a tested extra-distance limit up to 30%. The first successful extra-distance limits were 5%, 10%, 20% and 30% for four, two, 12 and seven journeys, respectively. Parts a, b and c report current attainment, first tested attainment and paired increments. Search comparisons are shown in Supplementary Fig. S9. The independent 50-journey extension is reported in Supplementary Table S11 and Supplementary Fig. S18.

### Supplementary Table S4a. Sun-avoidance target attainment at each extra-distance limit

| Time | Extra-distance limit (%) | Target | Metric | Valid n | Estimate (%) | Origin-cluster CI (%) | Spatial-block CI (%) |
|---|---|---|---|---|---|---|---|
| 09:00 | 0 | 0.5 | Attainment at this extra-distance limit | 466 | 0.00 | 0.00–0.00 | 0.00–0.00 |
| 09:00 | 0 | 0.5 | Attainment by this extra-distance limit | 466 | 0.00 | 0.00–0.00 | 0.00–0.00 |
| 09:00 | 5 | 0.5 | Attainment at this extra-distance limit | 466 | 12.08 | 9.00–15.35 | 7.33–17.46 |
| 09:00 | 5 | 0.5 | Attainment by this extra-distance limit | 466 | 12.08 | 9.00–15.35 | 7.33–17.46 |
| 09:00 | 10 | 0.5 | Attainment at this extra-distance limit | 466 | 17.88 | 14.26–21.27 | 11.61–24.17 |
| 09:00 | 10 | 0.5 | Attainment by this extra-distance limit | 466 | 18.29 | 14.65–21.66 | 12.08–24.45 |
| 09:00 | 20 | 0.5 | Attainment at this extra-distance limit | 466 | 22.55 | 18.17–26.61 | 15.64–30.39 |
| 09:00 | 20 | 0.5 | Attainment by this extra-distance limit | 466 | 22.96 | 18.71–26.98 | 16.01–30.81 |
| 09:00 | 30 | 0.5 | Attainment at this extra-distance limit | 466 | 25.77 | 21.39–30.18 | 17.56–34.49 |
| 09:00 | 30 | 0.5 | Attainment by this extra-distance limit | 466 | 26.18 | 21.61–30.55 | 18.07–34.92 |
| 12:00 | 0 | 0.5 | Attainment at this extra-distance limit | 499 | 0.00 | 0.00–0.00 | 0.00–0.00 |
| 12:00 | 0 | 0.5 | Attainment by this extra-distance limit | 499 | 0.00 | 0.00–0.00 | 0.00–0.00 |
| 12:00 | 5 | 0.5 | Attainment at this extra-distance limit | 499 | 0.80 | 0.18–1.71 | 0.16–1.41 |
| 12:00 | 5 | 0.5 | Attainment by this extra-distance limit | 499 | 0.80 | 0.18–1.71 | 0.16–1.41 |
| 12:00 | 10 | 0.5 | Attainment at this extra-distance limit | 499 | 1.20 | 0.39–2.20 | 0.31–2.19 |
| 12:00 | 10 | 0.5 | Attainment by this extra-distance limit | 499 | 1.20 | 0.39–2.20 | 0.31–2.19 |

| Time | Extra-distance limit (%) | Target | Metric | Valid n | Estimate (%) | Origin-cluster CI (%) | Spatial-block CI (%) |
|---|---|---|---|---|---|---|---|
| 12:00 | 20 | 0.5 | Attainment at this extra-distance limit | 499 | 3.40 | 1.85–5.07 | 1.61–5.33 |
| 12:00 | 20 | 0.5 | Attainment by this extra-distance limit | 499 | 3.60 | 2.04–5.28 | 1.89–5.54 |
| 12:00 | 30 | 0.5 | Attainment at this extra-distance limit | 499 | 4.41 | 2.61–6.29 | 2.39–6.88 |
| 12:00 | 30 | 0.5 | Attainment by this extra-distance limit | 499 | 5.01 | 3.13–6.95 | 3.04–7.21 |
| 15:00 | 0 | 0.5 | Attainment at this extra-distance limit | 477 | 0.00 | 0.00–0.00 | 0.00–0.00 |
| 15:00 | 0 | 0.5 | Attainment by this extra-distance limit | 477 | 0.00 | 0.00–0.00 | 0.00–0.00 |
| 15:00 | 5 | 0.5 | Attainment at this extra-distance limit | 477 | 8.64 | 6.14–11.33 | 4.51–12.65 |
| 15:00 | 5 | 0.5 | Attainment by this extra-distance limit | 477 | 8.64 | 6.14–11.33 | 4.51–12.65 |
| 15:00 | 10 | 0.5 | Attainment at this extra-distance limit | 477 | 12.52 | 9.66–15.60 | 7.62–17.76 |
| 15:00 | 10 | 0.5 | Attainment by this extra-distance limit | 477 | 12.52 | 9.66–15.60 | 7.62–17.76 |
| 15:00 | 20 | 0.5 | Attainment at this extra-distance limit | 477 | 16.86 | 13.48–20.58 | 10.49–24.01 |
| 15:00 | 20 | 0.5 | Attainment by this extra-distance limit | 477 | 17.27 | 13.72–20.91 | 10.74–24.73 |
| 15:00 | 30 | 0.5 | Attainment at this extra-distance limit | 477 | 18.68 | 15.30–22.47 | 12.01–25.76 |
| 15:00 | 30 | 0.5 | Attainment by this extra-distance limit | 477 | 19.29 | 15.86–23.17 | 12.45–26.48 |
| 17:00 | 0 | 0.5 | Attainment at this extra-distance limit | 401 | 0.00 | 0.00–0.00 | 0.00–0.00 |
| 17:00 | 0 | 0.5 | Attainment by this extra-distance limit | 401 | 0.00 | 0.00–0.00 | 0.00–0.00 |
| 17:00 | 5 | 0.5 | Attainment at this extra-distance limit | 401 | 10.57 | 7.52–13.67 | 6.14–14.26 |
| 17:00 | 5 | 0.5 | Attainment by this extra-distance limit | 401 | 10.57 | 7.52–13.67 | 6.14–14.26 |
| 17:00 | 10 | 0.5 | Attainment at this extra-distance limit | 401 | 16.70 | 13.15–20.47 | 10.43–22.76 |
| 17:00 | 10 | 0.5 | Attainment by this extra-distance limit | 401 | 17.21 | 13.61–21.04 | 11.26–22.95 |
| 17:00 | 20 | 0.5 | Attainment at this extra-distance limit | 401 | 21.19 | 16.63–25.35 | 13.56–28.56 |
| 17:00 | 20 | 0.5 | Attainment by this extra-distance limit | 401 | 21.92 | 17.39–26.11 | 14.65–29.19 |
| 17:00 | 30 | 0.5 | Attainment at this extra-distance limit | 401 | 24.01 | 19.55–28.34 | 15.68–32.41 |
| 17:00 | 30 | 0.5 | Attainment by this extra-distance limit | 401 | 25.23 | 20.38–29.45 | 17.22–33.34 |

## Supplementary Table S4b. First tested limit halving known sun

| Time | First extra-distance limit (fraction) | Valid n | Share (%) | Origin-cluster CI (%) | Spatial-block CI (%) |
|---|---|---|---|---|---|
| 09:00 | 0.0 | 466 | 0.00 | 0.00–0.00 | 0.00–0.00 |
| 09:00 | 0.05 | 466 | 12.08 | 9.00–15.35 | 7.33–17.46 |
| 09:00 | 0.1 | 466 | 6.21 | 4.13–8.35 | 3.92–8.45 |
| 09:00 | 0.2 | 466 | 4.67 | 2.85–6.83 | 2.44–7.77 |
| 09:00 | 0.3 | 466 | 3.22 | 1.76–4.81 | 1.49–5.32 |
| 09:00 | not_reached_by_30 | 466 | 73.82 | 69.45–78.39 | 65.08–81.93 |
| 12:00 | 0.0 | 499 | 0.00 | 0.00–0.00 | 0.00–0.00 |
| 12:00 | 0.05 | 499 | 0.80 | 0.18–1.71 | 0.16–1.41 |
| 12:00 | 0.1 | 499 | 0.40 | 0.00–1.00 | 0.00–1.06 |
| 12:00 | 0.2 | 499 | 2.40 | 1.06–3.78 | 1.32–3.76 |
| 12:00 | 0.3 | 499 | 1.41 | 0.41–2.49 | 0.58–2.48 |
| 12:00 | not_reached_by_30 | 499 | 94.99 | 93.05–96.87 | 92.79–96.96 |
| 15:00 | 0.0 | 477 | 0.00 | 0.00–0.00 | 0.00–0.00 |
| 15:00 | 0.05 | 477 | 8.64 | 6.14–11.33 | 4.51–12.65 |
| 15:00 | 0.1 | 477 | 3.88 | 2.30–5.75 | 2.30–6.00 |
| 15:00 | 0.2 | 477 | 4.75 | 2.95–6.65 | 2.46–8.18 |
| 15:00 | 0.3 | 477 | 2.02 | 0.85–3.35 | 1.01–3.57 |
| 15:00 | not_reached_by_30 | 477 | 80.71 | 76.83–84.14 | 73.52–87.55 |
| 17:00 | 0.0 | 401 | 0.00 | 0.00–0.00 | 0.00–0.00 |
| 17:00 | 0.05 | 401 | 10.57 | 7.52–13.67 | 6.14–14.26 |
| 17:00 | 0.1 | 401 | 6.64 | 4.45–9.18 | 3.88–9.95 |
| 17:00 | 0.2 | 401 | 4.71 | 2.68–6.86 | 2.82–6.73 |
| 17:00 | 0.3 | 401 | 3.30 | 1.70–4.99 | 1.25–5.90 |
| 17:00 | not_reached_by_30 | 401 | 74.77 | 70.55–79.62 | 66.66–82.78 |

## Supplementary Table S4c. Changes in sun avoidance between extra-distance limits

| Time | From extra-distance limit (%) | To extra-distance limit (%) | Statistic | Valid n | Change (m) | Origin-cluster CI (m) | Spatial-block CI (m) |
|---|---|---|---|---|---|---|---|
| 09:00 | 0 | 5 | median | 500 | 0.00 | 0.00–4.56 | 0.00–7.50 |
| 09:00 | 0 | 5 | mean | 500 | 48.14 | 39.89–56.51 | 35.95–60.53 |
| 09:00 | 5 | 10 | median | 500 | 0.00 | 0.00–0.00 | 0.00–0.00 |
| 09:00 | 5 | 10 | mean | 500 | 15.03 | 10.91–19.40 | 10.98–19.35 |
| 09:00 | 10 | 20 | median | 500 | 0.00 | 0.00–0.00 | 0.00–0.00 |
| 09:00 | 10 | 20 | mean | 500 | 11.74 | 8.42–15.62 | 7.87–17.01 |
| 09:00 | 20 | 30 | median | 500 | 0.00 | 0.00–0.00 | 0.00–0.00 |
| 09:00 | 20 | 30 | mean | 500 | 4.28 | 2.98–5.84 | 2.66–6.03 |
| 12:00 | 0 | 5 | median | 500 | 1.80 | 0.00–6.86 | 0.00–19.10 |
| 12:00 | 0 | 5 | mean | 500 | 56.87 | 48.42–65.41 | 41.34–73.60 |
| 12:00 | 5 | 10 | median | 500 | 0.00 | 0.00–0.00 | 0.00–0.00 |
| 12:00 | 5 | 10 | mean | 500 | 17.36 | 13.77–21.16 | 11.59–23.73 |
| 12:00 | 10 | 20 | median | 500 | 0.00 | 0.00–0.00 | 0.00–0.00 |
| 12:00 | 10 | 20 | mean | 500 | 15.35 | 11.01–19.94 | 10.59–23.40 |
| 12:00 | 20 | 30 | median | 500 | 0.00 | 0.00–0.00 | 0.00–0.00 |
| 12:00 | 20 | 30 | mean | 500 | 8.85 | 5.94–12.36 | 6.83–11.54 |
| 15:00 | 0 | 5 | median | 500 | 0.00 | 0.00–2.25 | 0.00–7.14 |
| 15:00 | 0 | 5 | mean | 500 | 57.11 | 47.77–67.85 | 41.67–71.43 |
| 15:00 | 5 | 10 | median | 500 | 0.00 | 0.00–0.00 | 0.00–0.00 |
| 15:00 | 5 | 10 | mean | 500 | 19.45 | 14.74–24.22 | 13.50–26.35 |
| 15:00 | 10 | 20 | median | 500 | 0.00 | 0.00–0.00 | 0.00–0.00 |
| 15:00 | 10 | 20 | mean | 500 | 16.73 | 12.24–21.76 | 11.33–22.36 |
| 15:00 | 20 | 30 | median | 500 | 0.00 | 0.00–0.00 | 0.00–0.00 |
| 15:00 | 20 | 30 | mean | 500 | 6.12 | 3.98–8.46 | 3.47–9.17 |
| 17:00 | 0 | 5 | median | 500 | 0.00 | 0.00–0.00 | 0.00–0.00 |

| Time | From extra-distance limit (%) | To extra-distance limit (%) | Statistic | Valid n | Change (m) | Origin-cluster CI (m) | Spatial-block CI (m) |
|---|---|---|---|---|---|---|---|
| 17:00 | 0 | 5 | mean | 500 | 26.43 | 21.40–31.76 | 18.81–33.57 |
| 17:00 | 5 | 10 | median | 500 | 0.00 | 0.00–0.00 | 0.00–0.00 |
| 17:00 | 5 | 10 | mean | 500 | 8.62 | 5.63–11.96 | 5.72–12.69 |
| 17:00 | 10 | 20 | median | 500 | 0.00 | 0.00–0.00 | 0.00–0.00 |
| 17:00 | 10 | 20 | mean | 500 | 7.99 | 4.62–12.38 | 4.40–11.00 |
| 17:00 | 20 | 30 | median | 500 | 0.00 | 0.00–0.00 | 0.00–0.00 |
| 17:00 | 20 | 30 | mean | 500 | 5.33 | 2.76–8.58 | 2.70–8.99 |

## Supplementary Table S5. Sun avoidance with outdoor shade and indoor protection information

At noon, when routes could be up to 10% longer than the shortest route, recognising outdoor shade reduced known sun by a mean of 64.37 m (95% origin-cluster confidence interval, 55.31 to 73.63 m). Recognising indoor protection added a further mean reduction of 10.06 m (6.53 to 13.82 m). These steps together reduced known sun by 74.43 m using the separately generated common candidate-route set. With routes up to 20% longer, the corresponding reductions were 75.52 m (64.64 to 86.52 m) and 14.89 m (10.72 to 19.37 m). In the table, I0 chooses the shortest route, I1 recognises outdoor shade, and I2 also recognises indoor protection. Both distance comparisons retain positive, zero and negative increments. Length statistics are in metres, and signed outcome shares are percentages. Information conditions and paired increments are illustrated in Supplementary Fig. S10.

| Extra-distance limit (%) | Kind | Arm | Contrast | Statistic | Unit | Valid n | Estimate | Origin-cluster CI | Spatial-block CI |
|---|---|---|---|---|---|---|---|---|---|
| 10 | Information condition | I0 | | median | m | 500 | 0.00 | 0.00–0.00 | 0.00–0.00 |
| 10 | Information condition | I0 | | mean | m | 500 | 0.00 | 0.00–0.00 | 0.00–0.00 |
| 10 | Information condition | I1 | | median | m | 500 | 4.20 | 0.00–10.50 | 0.00–21.25 |
| 10 | Information condition | I1 | | mean | m | 500 | 64.37 | 55.31–73.63 | 48.20–81.56 |
| 10 | Information condition | I2 | | median | m | 500 | 6.55 | 0.00–18.98 | 0.00–29.14 |
| 10 | Information condition | I2 | | mean | m | 500 | 74.43 | 64.26–84.70 | 54.94–96.79 |
| 20 | Information condition | I0 | | median | m | 500 | 0.00 | 0.00–0.00 | 0.00–0.00 |
| 20 | Information condition | I0 | | mean | m | 500 | 0.00 | 0.00–0.00 | 0.00–0.00 |
| 20 | Information condition | I1 | | median | m | 500 | 6.55 | 0.00–19.10 | 0.00–26.94 |
| 20 | Information condition | I1 | | mean | m | 500 | 75.52 | 64.64–86.52 | 57.23–95.17 |
| 20 | Information condition | I2 | | median | m | 500 | 9.66 | 4.20–28.79 | 0.00–41.11 |
| 20 | Information condition | I2 | | mean | m | 500 | 90.42 | 77.92–103.43 | 66.69–116.18 |
| 10 | Paired comparison | Paired comparison | I0→I1 | median | m | 500 | 4.20 | 0.00–10.50 | 0.00–21.25 |
| 10 | Paired comparison | Paired comparison | I0→I1 | mean | m | 500 | 64.37 | 55.31–73.63 | 48.20–81.56 |
| 20 | Paired comparison | Paired comparison | I0→I1 | median | m | 500 | 6.55 | 0.00–19.10 | 0.00–26.94 |
| 20 | Paired comparison | Paired comparison | I0→I1 | mean | m | 500 | 75.52 | 64.64–86.52 | 57.23–95.17 |
| 10 | Paired comparison | Paired comparison | I1→I2 | median | m | 500 | 0.00 | 0.00–0.00 | 0.00–0.00 |
| 10 | Paired comparison | Paired comparison | I1→I2 | mean | m | 500 | 10.06 | 6.53–13.82 | 4.86–17.02 |
| 20 | Paired comparison | Paired comparison | I1→I2 | median | m | 500 | 0.00 | 0.00–0.00 | 0.00–0.00 |

| Extra-distance limit (%) | Kind | Arm | Contrast | Statistic | Unit | Valid n | Estimate | Origin-cluster CI | Spatial-block CI |
|---|---|---|---|---|---|---|---|---|---|
| 20 | Paired comparison | Paired comparison | I1→I2 | mean | m | 500 | 14.89 | 10.72–19.37 | 8.23–23.78 |

## Supplementary Table S6. Journey feasibility and sun exposure after removing connections

Part a retains feasibility counts for all 500 journeys under each connection restriction. Part b reports signed changes among the 190 journeys feasible in the full network and both restricted networks, conditioning the original sampling weights on this subset. The separate common-set diagnostic is shown in Supplementary Fig. S11. Restriction definitions are given in Supplementary Methods 5.

### Supplementary Table S6a. Journey feasibility after connection removal

| Connection restriction | Disconnected | Feasible | Above original route-length limit |
|---|---|---|---|
| full | 0 | 500 | 0 |
| no_indoor | 3 | 479 | 18 |
| no_multilevel | 203 | 191 | 106 |

### Supplementary Table S6b. Exposure changes on routes feasible in all networks

| Intervention | Metric | Statistic | Valid n | Estimate (m) | Origin-cluster CI (m) | Spatial-block CI (m) |
|---|---|---|---|---|---|---|
| no_indoor | Change in known-sun distance | median | 190 | 0.00 | 0.00–0.00 | 0.00–0.00 |
| no_indoor | Change in known-sun distance | mean | 190 | 14.53 | 6.91–23.62 | 6.04–24.31 |
| no_indoor | Change in exposure cost | median | 190 | 0.00 | 0.00–0.00 | 0.00–0.00 |
| no_indoor | Change in exposure cost | mean | 190 | 18.13 | 9.87–27.87 | 8.39–30.65 |
| no_indoor | extra_length_m | median | 190 | 0.00 | 0.00–0.00 | 0.00–0.00 |
| no_indoor | extra_length_m | mean | 190 | -2.24 | -5.84–1.31 | -6.47–0.80 |
| no_indoor | unknown_change_m | median | 190 | 0.00 | 0.00–0.00 | 0.00–0.00 |
| no_indoor | unknown_change_m | mean | 190 | 3.61 | -1.29–9.07 | -1.26–10.70 |
| no_multilevel | Change in known-sun distance | median | 190 | 0.00 | 0.00–0.00 | 0.00–0.00 |
| no_multilevel | Change in known-sun distance | mean | 190 | 31.17 | 18.79–48.17 | 16.78–49.69 |
| no_multilevel | Change in exposure cost | median | 190 | 0.00 | 0.00–0.00 | 0.00–0.00 |
| no_multilevel | Change in exposure cost | mean | 190 | 40.30 | 27.28–59.37 | 22.32–64.05 |
| no_multilevel | extra_length_m | median | 190 | 0.00 | 0.00–0.00 | 0.00–0.00 |
| no_multilevel | extra_length_m | mean | 190 | 2.55 | -4.05–9.55 | -5.31–9.37 |
| no_multilevel | unknown_change_m | median | 190 | 0.00 | 0.00–0.00 | 0.00–0.00 |
| no_multilevel | unknown_change_m | mean | 190 | 9.13 | 2.87–17.10 | 1.98–19.23 |

## Supplementary Table S7. Nearby protection and local journey outcomes

Each of the 22 study locations had 12 local journeys, giving 264 journeys, including 118 with zero avoided sun. At the reference 880 m walking radius, known protection covered 35.78% to 54.78% of the reachable network. Mean avoided known sun ranged from 2.74 to 83.31 m. The correlation between nearby protection and gain was −0.343 when unresolved sections were excluded from protection and 0.222 when they were counted as protected (Spearman ρ). Rankings of nearby protection across 660 and 880 m radii had a Spearman correlation of 0.835, and ranks across 880 and 1,100 m had a correlation of 0.939. The corresponding mean changes in the nearby protected share were −0.218 and −0.214 percentage points. The 176-journey common subset retained fixed route outcomes while the walking network radius changed. Part a reports all locations, mean proportional reductions, gain-to-baseline-sun ratios and protection-share bounds including unresolved exposure. Part b reports radius changes. Locations A and B had protected shares of 41.38% and 40.77%, respectively, and mean avoided sun of 21.77 and 31.86 m. At B, all four journeys with 250 m target distances avoided no sun, while the four 750 m journeys avoided 37.56 to 141.01 m. At A, eastward journeys avoided no sun at all three target distances. Representative journeys were selected near each location's median avoided-sun distance. The two routes coincided at A, while the alternative at B avoided 28.53 m. The maps show local route geometry, and the exposure profiles cover the full journeys. The scale comparison is shown in Supplementary Fig. S12.

### Supplementary Table S7a. Nearby protection and journey outcomes by location

| Study location | Administrative area | Target cells | P (%) | P upper (%) | Unresolved share (%) | Mean ΔS (m) | Mean proportional reduction R (%) | Aggregate proportional reduction (%) | Mean extra (m) | Zero-gain target cells | Valid R |
|---|---|---|---|---|---|---|---|---|---|---|---|
| LC01 | WC | 12 | 40.91 | 57.41 | 16.50 | 65.71 | 22.25 | 21.54 | 34.53 | 3 | 12 |
| LC02 | EST | 12 | 35.78 | 56.36 | 20.58 | 36.44 | 18.70 | 18.70 | 28.48 | 3 | 11 |
| LC03 | STH | 12 | 45.43 | 49.50 | 4.08 | 8.90 | 2.87 | 3.89 | 3.04 | 9 | 12 |
| LC04 | STH | 12 | 41.38 | 48.83 | 7.45 | 21.77 | 5.89 | 6.75 | 9.20 | 6 | 12 |
| LC05 | EST | 12 | 51.90 | 58.13 | 6.24 | 25.49 | 7.51 | 14.27 | 10.29 | 10 | 9 |
| LC06 | CW | 12 | 39.16 | 54.87 | 15.71 | 83.31 | 34.93 | 31.56 | 26.72 | 1 | 12 |
| LC07 | STH | 12 | 40.13 | 51.77 | 11.64 | 2.74 | 1.28 | 1.15 | 3.35 | 8 | 12 |
| LC08 | WC | 12 | 37.21 | 42.62 | 5.41 | 13.59 | 2.85 | 5.41 | 11.96 | 8 | 12 |
| LC09 | CW | 12 | 48.81 | 67.50 | 18.70 | 17.26 | 5.54 | 5.38 | 20.60 | 0 | 12 |
| LC10 | EST | 12 | 40.46 | 43.91 | 3.46 | 18.56 | 5.57 | 5.96 | 10.72 | 2 | 12 |
| LC11 | CW | 12 | 54.78 | 58.64 | 3.85 | 16.11 | 6.22 | 8.57 | 18.12 | 7 | 12 |
| LC12 | WC | 12 | 47.93 | 56.86 | 8.93 | 3.21 | 1.39 | 1.88 | 3.31 | 10 | 12 |

| Study location | Administrative area | Target cells | P (%) | P upper (%) | Unresolved share (%) | Mean ΔS (m) | Mean proportional reduction R (%) | Aggregate proportional reduction (%) | Mean extra (m) | Zero-gain target cells | Valid R |
|---|---|---|---|---|---|---|---|---|---|---|---|
| LC13 | CW | 12 | 49.48 | 57.37 | 7.89 | 2.90 | 0.85 | 1.00 | 7.89 | 10 | 12 |
| LC14 | EST | 12 | 40.77 | 48.30 | 7.53 | 31.86 | 9.26 | 12.59 | 19.94 | 5 | 12 |
| LC15 | EST | 12 | 38.63 | 54.01 | 15.38 | 53.99 | 17.60 | 20.83 | 16.26 | 2 | 12 |
| LC16 | WC | 12 | 42.07 | 53.01 | 10.94 | 9.80 | 2.98 | 4.57 | 8.04 | 8 | 12 |
| LC17 | WC | 12 | 42.52 | 61.12 | 18.60 | 44.45 | 17.27 | 20.66 | 23.20 | 2 | 12 |
| LC18 | WC | 12 | 39.99 | 50.13 | 10.14 | 24.17 | 8.60 | 11.81 | 18.03 | 3 | 12 |
| LC19 | CW | 12 | 42.53 | 51.93 | 9.40 | 41.14 | 11.65 | 14.45 | 16.47 | 3 | 12 |
| LC20 | CW | 12 | 47.68 | 66.00 | 18.32 | 35.06 | 23.06 | 26.43 | 13.42 | 6 | 12 |
| LC21 | CW | 12 | 48.04 | 54.66 | 6.62 | 17.47 | 6.98 | 8.60 | 13.70 | 7 | 12 |
| LC22 | EST | 12 | 42.36 | 57.60 | 15.23 | 32.13 | 8.93 | 12.72 | 16.08 | 5 | 12 |

## Supplementary Table S7b. Nearby protection estimates across walking network radii

| From (m) | To (m) | Study locations | Supply rank ρ | Mean ΔP (pp) | Min ΔP (pp) | Max ΔP (pp) |
|---|---|---|---|---|---|---|
| 660.0 | 880.0 | 22 | 0.835 | -0.22 | -6.29 | 4.10 |
| 880.0 | 1100.0 | 22 | 0.939 | -0.21 | -4.01 | 2.19 |
| 660.0 | 1100.0 | 22 | 0.816 | -0.43 | -8.88 | 4.97 |

## Supplementary Table S8. Sun exposure and walking distance across journeys and times

At noon with a 10% extra-distance limit, 231 journeys had zero gain. One had no known sun on the shortest route. Among the remaining 230 zero-gain journeys, 212 had no alternative within the route-length limit, and 18 had alternatives that did not satisfy the two criteria for exposure improvement. This no-alternative category contained 91 of 166, 81 of 168 and 40 of 166 journeys in the three distance bands, respectively. The bands spanned 300 to 800 m, 800 to 1,500 m and 1,500 to 3,000 m, measured along the shortest route. Across the four times, 135 journeys had positive gains at every evaluation time, 175 remained at zero, and 190 alternated between positive and zero gains. Part a retains metric-specific valid denominators and both bootstrap interval types. Part b reports within-journey changes from noon, and part c reports route-class means by administrative area. Supplementary Fig. S17 shows variation across distance bands, extra-distance limits and evaluation times. The six illustrative cases and their extra-distance limit profiles appear in Supplementary Figs. S15–S16. At noon, routes could be up to 10% longer than the shortest route. Mean additional protected lengths were 61.34 m in building shade, 47.24 m in vegetation shade and 29.79 m indoors. Unresolved exposure decreased by 16.93 m. Together with 74.23 m less known sun, these changes corresponded to 47.22 m more walking.

### Supplementary Table S8a. Weighted journey outcomes

| Time | Extra-distance limit (%) | Group | Value | Metric | Statistic | Valid n | Estimate (m or %) | Origin-cluster CI | Spatial-block CI |
|---|---|---|---|---|---|---|---|---|---|
| 09:00 | 0 | All journeys | ALL | Avoided known-sun distance (ΔS) | median | 500 | 0.00 | 0.00–0.00 | 0.00–0.00 |
| 09:00 | 0 | All journeys | ALL | Avoided known-sun distance (ΔS) | mean | 500 | 0.00 | 0.00–0.00 | 0.00–0.00 |
| 09:00 | 0 | All journeys | ALL | R | median | 466 | 0.00 | 0.00–0.00 | 0.00–0.00 |
| 09:00 | 0 | All journeys | ALL | Actual detour fraction | mean | 500 | 0.00 | 0.00–0.00 | 0.00–0.00 |
| 09:00 | 0 | All journeys | ALL | Additional walking distance (ΔL) | median | 500 | 0.00 | 0.00–0.00 | 0.00–0.00 |
| 09:00 | 0 | All journeys | ALL | Additional walking distance (ΔL) | mean | 500 | 0.00 | 0.00–0.00 | 0.00–0.00 |
| 09:00 | 0 | All journeys | ALL | Positive-gain share | positive | 500 | 0.00 | 0.00–0.00 | 0.00–0.00 |
| 09:00 | 5 | All journeys | ALL | Avoided known-sun distance (ΔS) | median | 500 | 0.00 | 0.00–4.56 | 0.00–7.50 |
| 09:00 | 5 | All journeys | ALL | Avoided known-sun distance (ΔS) | mean | 500 | 48.14 | 39.89–56.51 | 35.95–60.53 |
| 09:00 | 5 | All journeys | ALL | R | median | 466 | 1.00 | 0.00–3.02 | 0.00–4.91 |
| 09:00 | 5 | All journeys | ALL | Actual detour fraction | mean | 500 | 1.33 | 1.18–1.48 | 1.03–1.59 |
| 09:00 | 5 | All journeys | ALL | Additional walking distance (ΔL) | median | 500 | 0.00 | 0.00–2.27 | 0.00–7.70 |
| 09:00 | 5 | All journeys | ALL | Additional walking distance (ΔL) | mean | 500 | 22.99 | 20.22–25.66 | 17.99–27.80 |
| 09:00 | 5 | All journeys | ALL | Positive-gain share | positive | 500 | 47.95 | 43.67–52.61 | 39.41–54.99 |

| Time | Extra-distance limit (%) | Group | Value | Metric | Statistic | Valid n | Estimate (m or %) | Origin-cluster CI | Spatial-block CI |
|---|---|---|---|---|---|---|---|---|---|
| 09:00 | 10 | All journeys | ALL | Avoided known-sun distance (ΔS) | median | 500 | 2.04 | 0.00–8.32 | 0.00–12.16 |
| 09:00 | 10 | All journeys | ALL | Avoided known-sun distance (ΔS) | mean | 500 | 63.18 | 52.72–73.74 | 47.92–78.30 |
| 09:00 | 10 | All journeys | ALL | R | median | 466 | 2.57 | 0.00–8.93 | 0.00–13.71 |
| 09:00 | 10 | All journeys | ALL | Actual detour fraction | mean | 500 | 2.96 | 2.62–3.26 | 2.33–3.48 |
| 09:00 | 10 | All journeys | ALL | Additional walking distance (ΔL) | median | 500 | 1.52 | 0.00–15.42 | 0.00–20.65 |
| 09:00 | 10 | All journeys | ALL | Additional walking distance (ΔL) | mean | 500 | 51.03 | 45.21–56.24 | 40.99–59.89 |
| 09:00 | 10 | All journeys | ALL | Positive-gain share | positive | 500 | 51.15 | 46.62–55.95 | 41.99–58.47 |
| 09:00 | 20 | All journeys | ALL | Avoided known-sun distance (ΔS) | median | 500 | 7.50 | 1.96–13.59 | 0.00–20.62 |
| 09:00 | 20 | All journeys | ALL | Avoided known-sun distance (ΔS) | mean | 500 | 74.91 | 63.04–86.76 | 57.34–92.55 |
| 09:00 | 20 | All journeys | ALL | R | median | 466 | 9.25 | 2.40–14.82 | 0.00–23.54 |
| 09:00 | 20 | All journeys | ALL | Actual detour fraction | mean | 500 | 5.58 | 4.97–6.24 | 4.50–6.65 |
| 09:00 | 20 | All journeys | ALL | Additional walking distance (ΔL) | median | 500 | 13.93 | 0.52–26.84 | 0.00–34.11 |
| 09:00 | 20 | All journeys | ALL | Additional walking distance (ΔL) | mean | 500 | 93.38 | 82.77–104.35 | 74.73–112.63 |
| 09:00 | 20 | All journeys | ALL | Positive-gain share | positive | 500 | 54.74 | 50.41–59.49 | 45.56–62.23 |
| 09:00 | 30 | All journeys | ALL | Avoided known-sun distance (ΔS) | median | 500 | 9.07 | 3.19–15.35 | 0.00–25.06 |
| 09:00 | 30 | All journeys | ALL | Avoided known-sun distance (ΔS) | mean | 500 | 79.20 | 67.42–91.29 | 60.91–97.74 |
| 09:00 | 30 | All journeys | ALL | R | median | 466 | 11.83 | 3.62–17.66 | 1.15–26.82 |
| 09:00 | 30 | All journeys | ALL | Actual detour fraction | mean | 500 | 8.13 | 7.20–9.13 | 6.62–9.62 |
| 09:00 | 30 | All journeys | ALL | Additional walking distance (ΔL) | median | 500 | 21.68 | 2.45–38.37 | 0.00–51.46 |
| 09:00 | 30 | All journeys | ALL | Additional walking distance (ΔL) | mean | 500 | 134.65 | 118.25–151.24 | 109.62–161.09 |
| 09:00 | 30 | All journeys | ALL | Positive-gain share | positive | 500 | 56.34 | 51.98–61.11 | 47.70–63.65 |
| 12:00 | 0 | All journeys | ALL | Avoided known-sun distance (ΔS) | median | 500 | 0.00 | 0.00–0.00 | 0.00–0.00 |
| 12:00 | 0 | All journeys | ALL | Avoided known-sun distance (ΔS) | mean | 500 | 0.00 | 0.00–0.00 | 0.00–0.00 |
| 12:00 | 0 | All journeys | ALL | R | median | 499 | 0.00 | 0.00–0.00 | 0.00–0.00 |
| 12:00 | 0 | All journeys | ALL | Actual detour fraction | mean | 500 | 0.00 | 0.00–0.00 | 0.00–0.00 |
| 12:00 | 0 | All journeys | ALL | Additional walking distance (ΔL) | median | 500 | 0.00 | 0.00–0.00 | 0.00–0.00 |

| Time | Extra-distance limit (%) | Group | Value | Metric | Statistic | Valid n | Estimate (m or %) | Origin-cluster CI | Spatial-block CI |
|---|---|---|---|---|---|---|---|---|---|
| 12:00 | 0 | All journeys | ALL | Additional walking distance (ΔL) | mean | 500 | 0.00 | 0.00–0.00 | 0.00–0.00 |
| 12:00 | 0 | All journeys | ALL | Positive-gain share | positive | 500 | 0.00 | 0.00–0.00 | 0.00–0.00 |
| 12:00 | 5 | All journeys | ALL | Avoided known-sun distance (ΔS) | median | 500 | 1.80 | 0.00–6.86 | 0.00–19.10 |
| 12:00 | 5 | All journeys | ALL | Avoided known-sun distance (ΔS) | mean | 500 | 56.87 | 48.42–65.41 | 41.34–73.60 |
| 12:00 | 5 | All journeys | ALL | R | median | 499 | 0.36 | 0.00–1.53 | 0.00–3.65 |
| 12:00 | 5 | All journeys | ALL | Actual detour fraction | mean | 500 | 1.30 | 1.14–1.46 | 1.00–1.60 |
| 12:00 | 5 | All journeys | ALL | Additional walking distance (ΔL) | median | 500 | 0.52 | 0.00–4.06 | 0.00–12.88 |
| 12:00 | 5 | All journeys | ALL | Additional walking distance (ΔL) | mean | 500 | 22.01 | 19.23–24.67 | 17.43–26.77 |
| 12:00 | 5 | All journeys | ALL | Positive-gain share | positive | 500 | 50.55 | 45.46–54.91 | 41.02–59.67 |
| 12:00 | 10 | All journeys | ALL | Avoided known-sun distance (ΔS) | median | 500 | 6.55 | 0.00–18.98 | 0.00–29.14 |
| 12:00 | 10 | All journeys | ALL | Avoided known-sun distance (ΔS) | mean | 500 | 74.23 | 63.87–84.79 | 54.62–96.80 |
| 12:00 | 10 | All journeys | ALL | R | median | 499 | 1.36 | 0.00–4.00 | 0.00–6.23 |
| 12:00 | 10 | All journeys | ALL | Actual detour fraction | mean | 500 | 2.85 | 2.51–3.17 | 2.20–3.48 |
| 12:00 | 10 | All journeys | ALL | Additional walking distance (ΔL) | median | 500 | 3.98 | 0.00–15.15 | 0.00–21.88 |
| 12:00 | 10 | All journeys | ALL | Additional walking distance (ΔL) | mean | 500 | 47.22 | 41.56–52.65 | 37.15–56.96 |
| 12:00 | 10 | All journeys | ALL | Positive-gain share | positive | 500 | 53.77 | 48.98–58.26 | 44.23–62.72 |
| 12:00 | 10 | Administrative area | CW | Avoided known-sun distance (ΔS) | median | 126 | 18.98 | 0.00–41.11 | 0.00–62.76 |
| 12:00 | 10 | Administrative area | CW | Avoided known-sun distance (ΔS) | mean | 126 | 90.12 | 70.79–111.54 | 56.14–130.39 |
| 12:00 | 10 | Administrative area | CW | R | median | 126 | 3.65 | 0.00–10.71 | 0.00–15.76 |
| 12:00 | 10 | Administrative area | CW | Actual detour fraction | mean | 126 | 3.24 | 2.53–3.92 | 2.10–4.62 |
| 12:00 | 10 | Administrative area | CW | Additional walking distance (ΔL) | median | 126 | 2.54 | 0.00–30.04 | 0.00–65.08 |
| 12:00 | 10 | Administrative area | CW | Additional walking distance (ΔL) | mean | 126 | 54.09 | 43.07–65.01 | 36.84–72.35 |
| 12:00 | 10 | Administrative area | CW | Positive-gain share | positive | 126 | 57.94 | 48.64–67.46 | 38.72–75.50 |
| 12:00 | 10 | Administrative area | WC | Avoided known-sun distance (ΔS) | median | 126 | 15.83 | 0.00–45.25 | 0.00–87.68 |
| 12:00 | 10 | Administrative area | WC | Avoided known-sun distance (ΔS) | mean | 126 | 84.94 | 62.72–107.51 | 37.68–120.65 |
| 12:00 | 10 | Administrative area | WC | R | median | 125 | 4.59 | 0.00–11.76 | 0.00–14.93 |

| Time | Extra-distance limit (%) | Group | Value | Metric | Statistic | Valid n | Estimate (m or %) | Origin-cluster CI | Spatial-block CI |
|---|---|---|---|---|---|---|---|---|---|
| 12:00 | 10 | Administrative area | WC | Actual detour fraction | mean | 126 | 3.40 | 2.65–4.13 | 1.81–4.63 |
| 12:00 | 10 | Administrative area | WC | Additional walking distance (ΔL) | median | 126 | 18.61 | 0.00–37.03 | 0.00–50.01 |
| 12:00 | 10 | Administrative area | WC | Additional walking distance (ΔL) | mean | 126 | 55.48 | 43.21–67.65 | 30.03–73.96 |
| 12:00 | 10 | Administrative area | WC | Positive-gain share | positive | 126 | 58.73 | 49.32–68.09 | 39.49–74.89 |
| 12:00 | 10 | Administrative area | EST | Avoided known-sun distance (ΔS) | median | 124 | 27.26 | 0.00–49.11 | 0.00–69.67 |
| 12:00 | 10 | Administrative area | EST | Avoided known-sun distance (ΔS) | mean | 124 | 84.58 | 60.50–109.09 | 42.77–134.69 |
| 12:00 | 10 | Administrative area | EST | R | median | 124 | 4.84 | 0.00–9.71 | 0.00–14.40 |
| 12:00 | 10 | Administrative area | EST | Actual detour fraction | mean | 124 | 3.39 | 2.68–4.05 | 2.06–4.81 |
| 12:00 | 10 | Administrative area | EST | Additional walking distance (ΔL) | median | 124 | 20.40 | 0.00–33.14 | 0.00–50.62 |
| 12:00 | 10 | Administrative area | EST | Additional walking distance (ΔL) | mean | 124 | 53.35 | 41.11–65.46 | 32.46–75.28 |
| 12:00 | 10 | Administrative area | EST | Positive-gain share | positive | 124 | 57.99 | 48.20–67.34 | 38.49–77.50 |
| 12:00 | 10 | Administrative area | STH | Avoided known-sun distance (ΔS) | median | 124 | 0.00 | 0.00–0.00 | 0.00–1.80 |
| 12:00 | 10 | Administrative area | STH | Avoided known-sun distance (ΔS) | mean | 124 | 37.26 | 23.15–52.61 | 19.75–55.20 |
| 12:00 | 10 | Administrative area | STH | R | median | 124 | 0.00 | 0.00–0.00 | 0.00–0.19 |
| 12:00 | 10 | Administrative area | STH | Actual detour fraction | mean | 124 | 1.38 | 0.97–1.83 | 0.79–1.97 |
| 12:00 | 10 | Administrative area | STH | Additional walking distance (ΔL) | median | 124 | 0.00 | 0.00–0.00 | 0.00–3.98 |
| 12:00 | 10 | Administrative area | STH | Additional walking distance (ΔL) | mean | 124 | 25.95 | 17.33–34.63 | 13.92–38.10 |
| 12:00 | 10 | Administrative area | STH | Positive-gain share | positive | 124 | 40.40 | 31.13–49.09 | 28.09–51.20 |
| 12:00 | 10 | Distance band | 300_800 | Avoided known-sun distance (ΔS) | median | 166 | 0.00 | 0.00–0.00 | 0.00–0.00 |
| 12:00 | 10 | Distance band | 300_800 | Avoided known-sun distance (ΔS) | mean | 166 | 26.46 | 19.45–33.78 | 16.59–35.23 |
| 12:00 | 10 | Distance band | 300_800 | R | median | 165 | 0.00 | 0.00–0.00 | 0.00–0.55 |
| 12:00 | 10 | Distance band | 300_800 | Actual detour fraction | mean | 166 | 1.99 | 1.54–2.46 | 1.39–2.59 |
| 12:00 | 10 | Distance band | 300_800 | Additional walking distance (ΔL) | median | 166 | 0.00 | 0.00–0.00 | 0.00–0.00 |
| 12:00 | 10 | Distance band | 300_800 | Additional walking distance (ΔL) | mean | 166 | 12.83 | 9.78–15.89 | 8.44–16.95 |
| 12:00 | 10 | Distance band | 300_800 | Positive-gain share | positive | 166 | 40.29 | 33.16–47.53 | 29.64–49.87 |
| 12:00 | 10 | Distance band | 800_1500 | Avoided known-sun distance (ΔS) | median | 168 | 0.00 | 0.00–9.16 | 0.00–21.58 |

| Time | Extra-distance limit (%) | Group | Value | Metric | Statistic | Valid n | Estimate (m or %) | Origin-cluster CI | Spatial-block CI |
|---|---|---|---|---|---|---|---|---|---|
| 12:00 | 10 | Distance band | 800_1500 | Avoided known-sun distance (ΔS) | mean | 168 | 50.41 | 38.47–63.32 | 30.75–72.36 |
| 12:00 | 10 | Distance band | 800_1500 | R | median | 168 | 0.00 | 0.00–1.95 | 0.00–4.96 |
| 12:00 | 10 | Distance band | 800_1500 | Actual detour fraction | mean | 168 | 2.59 | 2.14–3.13 | 1.78–3.47 |
| 12:00 | 10 | Distance band | 800_1500 | Additional walking distance (ΔL) | median | 168 | 0.00 | 0.00–14.51 | 0.00–20.59 |
| 12:00 | 10 | Distance band | 800_1500 | Additional walking distance (ΔL) | mean | 168 | 31.59 | 25.39–38.21 | 21.35–42.65 |
| 12:00 | 10 | Distance band | 800_1500 | Positive-gain share | positive | 168 | 47.62 | 40.36–55.28 | 35.78–59.05 |
| 12:00 | 10 | Distance band | 1500_3000 | Avoided known-sun distance (ΔS) | median | 166 | 80.67 | 41.52–113.42 | 35.74–150.49 |
| 12:00 | 10 | Distance band | 1500_3000 | Avoided known-sun distance (ΔS) | mean | 166 | 145.81 | 121.36–171.16 | 108.53–189.62 |
| 12:00 | 10 | Distance band | 1500_3000 | R | median | 166 | 8.74 | 6.01–12.17 | 4.24–14.20 |
| 12:00 | 10 | Distance band | 1500_3000 | Actual detour fraction | mean | 166 | 3.98 | 3.42–4.52 | 3.19–4.76 |
| 12:00 | 10 | Distance band | 1500_3000 | Additional walking distance (ΔL) | median | 166 | 75.66 | 38.51–113.85 | 31.33–126.73 |
| 12:00 | 10 | Distance band | 1500_3000 | Additional walking distance (ΔL) | mean | 166 | 97.24 | 83.14–111.09 | 78.37–116.74 |
| 12:00 | 10 | Distance band | 1500_3000 | Positive-gain share | positive | 166 | 73.39 | 66.59–79.59 | 63.96–83.18 |
| 12:00 | 20 | All journeys | ALL | Avoided known-sun distance (ΔS) | median | 500 | 10.00 | 4.21–29.05 | 0.00–41.52 |
| 12:00 | 20 | All journeys | ALL | Avoided known-sun distance (ΔS) | mean | 500 | 89.58 | 77.12–101.93 | 66.12–115.88 |
| 12:00 | 20 | All journeys | ALL | R | median | 499 | 3.51 | 0.82–6.52 | 0.00–8.47 |
| 12:00 | 20 | All journeys | ALL | Actual detour fraction | mean | 500 | 5.26 | 4.64–5.90 | 4.08–6.54 |
| 12:00 | 20 | All journeys | ALL | Additional walking distance (ΔL) | median | 500 | 15.15 | 2.45–26.35 | 0.00–37.50 |
| 12:00 | 20 | All journeys | ALL | Additional walking distance (ΔL) | mean | 500 | 87.75 | 77.21–98.46 | 68.08–107.50 |

| Time | Extra-distance limit (%) | Group | Value | Metric | Statistic | Valid n | Estimate (m or %) | Origin-cluster CI | Spatial-block CI |
|---|---|---|---|---|---|---|---|---|---|
| 12:00 | 20 | All journeys | ALL | Positive-gain share | positive | 500 | 56.36 | 51.61–60.74 | 46.84–65.22 |
| 12:00 | 30 | All journeys | ALL | Avoided known-sun distance (ΔS) | median | 500 | 18.98 | 4.53–30.28 | 0.00–42.68 |
| 12:00 | 30 | All journeys | ALL | Avoided known-sun distance (ΔS) | mean | 500 | 98.43 | 84.87–112.35 | 73.82–125.37 |
| 12:00 | 30 | All journeys | ALL | R | median | 499 | 4.09 | 1.15–6.72 | 0.00–9.87 |
| 12:00 | 30 | All journeys | ALL | Actual detour fraction | mean | 500 | 7.05 | 6.23–7.93 | 5.55–8.76 |
| 12:00 | 30 | All journeys | ALL | Additional walking distance (ΔL) | median | 500 | 18.22 | 8.49–31.73 | 0.00–42.61 |
| 12:00 | 30 | All journeys | ALL | Additional walking distance (ΔL) | mean | 500 | 115.87 | 101.12–130.39 | 90.10–143.07 |
| 12:00 | 30 | All journeys | ALL | Positive-gain share | positive | 500 | 57.56 | 52.85–62.07 | 48.58–66.25 |
| 15:00 | 0 | All journeys | ALL | Avoided known-sun distance (ΔS) | median | 500 | 0.00 | 0.00–0.00 | 0.00–0.00 |
| 15:00 | 0 | All journeys | ALL | Avoided known-sun distance (ΔS) | mean | 500 | 0.00 | 0.00–0.00 | 0.00–0.00 |
| 15:00 | 0 | All journeys | ALL | R | median | 477 | 0.00 | 0.00–0.00 | 0.00–0.00 |
| 15:00 | 0 | All journeys | ALL | Actual detour fraction | mean | 500 | 0.00 | 0.00–0.00 | 0.00–0.00 |
| 15:00 | 0 | All journeys | ALL | Additional walking distance (ΔL) | median | 500 | 0.00 | 0.00–0.00 | 0.00–0.00 |
| 15:00 | 0 | All journeys | ALL | Additional walking distance (ΔL) | mean | 500 | 0.00 | 0.00–0.00 | 0.00–0.00 |
| 15:00 | 0 | All journeys | ALL | Positive-gain share | positive | 500 | 0.00 | 0.00–0.00 | 0.00–0.00 |
| 15:00 | 5 | All journeys | ALL | Avoided known-sun distance (ΔS) | median | 500 | 0.00 | 0.00–2.25 | 0.00–7.14 |
| 15:00 | 5 | All journeys | ALL | Avoided known-sun distance (ΔS) | mean | 500 | 57.11 | 47.77–67.85 | 41.67–71.43 |
| 15:00 | 5 | All journeys | ALL | R | median | 477 | 0.00 | 0.00–2.13 | 0.00–3.93 |
| 15:00 | 5 | All journeys | ALL | Actual detour fraction | mean | 500 | 1.29 | 1.12–1.44 | 1.01–1.54 |
| 15:00 | 5 | All journeys | ALL | Additional walking distance (ΔL) | median | 500 | 0.00 | 0.00–1.33 | 0.00–4.13 |
| 15:00 | 5 | All journeys | ALL | Additional walking distance (ΔL) | mean | 500 | 21.87 | 19.13–24.54 | 17.63–25.79 |
| 15:00 | 5 | All journeys | ALL | Positive-gain share | positive | 500 | 47.17 | 42.98–51.36 | 39.63–54.09 |
| 15:00 | 10 | All journeys | ALL | Avoided known-sun distance (ΔS) | median | 500 | 3.90 | 0.00–12.40 | 0.00–21.58 |
| 15:00 | 10 | All journeys | ALL | Avoided known-sun distance (ΔS) | mean | 500 | 76.57 | 65.24–88.22 | 57.32–96.02 |
| 15:00 | 10 | All journeys | ALL | R | median | 477 | 3.68 | 0.00–6.35 | 0.00–9.72 |
| 15:00 | 10 | All journeys | ALL | Actual detour fraction | mean | 500 | 2.94 | 2.62–3.29 | 2.20–3.64 |

| Time | Extra-distance limit (%) | Group | Value | Metric | Statistic | Valid n | Estimate (m or %) | Origin-cluster CI | Spatial-block CI |
|---|---|---|---|---|---|---|---|---|---|
| 15:00 | 10 | All journeys | ALL | Additional walking distance (ΔL) | median | 500 | 2.45 | 0.00–15.60 | 0.00–27.56 |
| 15:00 | 10 | All journeys | ALL | Additional walking distance (ΔL) | mean | 500 | 49.52 | 43.82–55.46 | 38.28–59.45 |
| 15:00 | 10 | All journeys | ALL | Positive-gain share | positive | 500 | 52.37 | 47.89–56.49 | 43.67–60.47 |
| 15:00 | 20 | All journeys | ALL | Avoided known-sun distance (ΔS) | median | 500 | 7.50 | 2.01–19.86 | 0.00–27.31 |
| 15:00 | 20 | All journeys | ALL | Avoided known-sun distance (ΔS) | mean | 500 | 93.30 | 79.15–107.45 | 70.81–116.52 |
| 15:00 | 20 | All journeys | ALL | R | median | 477 | 5.89 | 2.01–10.78 | 0.00–15.39 |
| 15:00 | 20 | All journeys | ALL | Actual detour fraction | mean | 500 | 6.01 | 5.32–6.69 | 4.43–7.53 |
| 15:00 | 20 | All journeys | ALL | Additional walking distance (ΔL) | median | 500 | 11.77 | 0.34–27.56 | 0.00–48.21 |
| 15:00 | 20 | All journeys | ALL | Additional walking distance (ΔL) | mean | 500 | 98.62 | 86.92–111.07 | 74.48–121.62 |
| 15:00 | 20 | All journeys | ALL | Positive-gain share | positive | 500 | 54.77 | 50.28–58.84 | 46.40–62.24 |
| 15:00 | 30 | All journeys | ALL | Avoided known-sun distance (ΔS) | median | 500 | 11.14 | 2.50–22.00 | 0.00–33.17 |
| 15:00 | 30 | All journeys | ALL | Avoided known-sun distance (ΔS) | mean | 500 | 99.41 | 85.58–114.22 | 74.83–124.18 |
| 15:00 | 30 | All journeys | ALL | R | median | 477 | 6.90 | 3.00–12.08 | 0.00–19.58 |
| 15:00 | 30 | All journeys | ALL | Actual detour fraction | mean | 500 | 8.02 | 7.04–8.99 | 5.79–10.33 |
| 15:00 | 30 | All journeys | ALL | Additional walking distance (ΔL) | median | 500 | 13.84 | 0.92–33.66 | 0.00–61.32 |
| 15:00 | 30 | All journeys | ALL | Additional walking distance (ΔL) | mean | 500 | 131.62 | 115.08–147.94 | 96.83–164.11 |
| 15:00 | 30 | All journeys | ALL | Positive-gain share | positive | 500 | 55.57 | 51.00–59.73 | 46.83–63.49 |
| 17:00 | 0 | All journeys | ALL | Avoided known-sun distance (ΔS) | median | 500 | 0.00 | 0.00–0.00 | 0.00–0.00 |
| 17:00 | 0 | All journeys | ALL | Avoided known-sun distance (ΔS) | mean | 500 | 0.00 | 0.00–0.00 | 0.00–0.00 |
| 17:00 | 0 | All journeys | ALL | R | median | 401 | 0.00 | 0.00–0.00 | 0.00–0.00 |
| 17:00 | 0 | All journeys | ALL | Actual detour fraction | mean | 500 | 0.00 | 0.00–0.00 | 0.00–0.00 |
| 17:00 | 0 | All journeys | ALL | Additional walking distance (ΔL) | median | 500 | 0.00 | 0.00–0.00 | 0.00–0.00 |
| 17:00 | 0 | All journeys | ALL | Additional walking distance (ΔL) | mean | 500 | 0.00 | 0.00–0.00 | 0.00–0.00 |
| 17:00 | 0 | All journeys | ALL | Positive-gain share | positive | 500 | 0.00 | 0.00–0.00 | 0.00–0.00 |
| 17:00 | 5 | All journeys | ALL | Avoided known-sun distance (ΔS) | median | 500 | 0.00 | 0.00–0.00 | 0.00–0.00 |
| 17:00 | 5 | All journeys | ALL | Avoided known-sun distance (ΔS) | mean | 500 | 26.43 | 21.40–31.76 | 18.81–33.57 |

| Time | Extra-distance limit (%) | Group | Value | Metric | Statistic | Valid n | Estimate (m or %) | Origin-cluster CI | Spatial-block CI |
|---|---|---|---|---|---|---|---|---|---|
| 17:00 | 5 | All journeys | ALL | R | median | 401 | 0.00 | 0.00–0.00 | 0.00–0.00 |
| 17:00 | 5 | All journeys | ALL | Actual detour fraction | mean | 500 | 0.96 | 0.82–1.09 | 0.75–1.11 |
| 17:00 | 5 | All journeys | ALL | Additional walking distance (ΔL) | median | 500 | 0.00 | 0.00–0.00 | 0.00–0.00 |
| 17:00 | 5 | All journeys | ALL | Additional walking distance (ΔL) | mean | 500 | 17.59 | 15.12–20.14 | 13.82–20.31 |
| 17:00 | 5 | All journeys | ALL | Positive-gain share | positive | 500 | 34.80 | 30.54–39.00 | 27.58–40.11 |
| 17:00 | 10 | All journeys | ALL | Avoided known-sun distance (ΔS) | median | 500 | 0.00 | 0.00–0.00 | 0.00–0.00 |
| 17:00 | 10 | All journeys | ALL | Avoided known-sun distance (ΔS) | mean | 500 | 35.05 | 28.70–41.78 | 26.27–43.87 |
| 17:00 | 10 | All journeys | ALL | R | median | 401 | 0.00 | 0.00–3.04 | 0.00–3.98 |
| 17:00 | 10 | All journeys | ALL | Actual detour fraction | mean | 500 | 2.22 | 1.95–2.52 | 1.80–2.60 |
| 17:00 | 10 | All journeys | ALL | Additional walking distance (ΔL) | median | 500 | 0.00 | 0.00–0.00 | 0.00–0.00 |
| 17:00 | 10 | All journeys | ALL | Additional walking distance (ΔL) | mean | 500 | 40.33 | 35.09–45.89 | 33.22–46.13 |
| 17:00 | 10 | All journeys | ALL | Positive-gain share | positive | 500 | 40.00 | 35.75–44.34 | 32.27–45.65 |
| 17:00 | 20 | All journeys | ALL | Avoided known-sun distance (ΔS) | median | 500 | 0.00 | 0.00–0.00 | 0.00–0.00 |
| 17:00 | 20 | All journeys | ALL | Avoided known-sun distance (ΔS) | mean | 500 | 43.03 | 34.89–51.82 | 32.29–52.37 |
| 17:00 | 20 | All journeys | ALL | R | median | 401 | 1.87 | 0.00–7.31 | 0.00–13.11 |
| 17:00 | 20 | All journeys | ALL | Actual detour fraction | mean | 500 | 3.92 | 3.39–4.48 | 3.10–4.66 |
| 17:00 | 20 | All journeys | ALL | Additional walking distance (ΔL) | median | 500 | 0.00 | 0.00–0.00 | 0.00–0.00 |
| 17:00 | 20 | All journeys | ALL | Additional walking distance (ΔL) | mean | 500 | 70.61 | 60.92–81.22 | 58.06–82.31 |
| 17:00 | 20 | All journeys | ALL | Positive-gain share | positive | 500 | 42.19 | 37.80–46.83 | 34.51–48.06 |
| 17:00 | 30 | All journeys | ALL | Avoided known-sun distance (ΔS) | median | 500 | 0.00 | 0.00–0.00 | 0.00–0.00 |
| 17:00 | 30 | All journeys | ALL | Avoided known-sun distance (ΔS) | mean | 500 | 48.36 | 38.88–58.11 | 36.17–60.16 |
| 17:00 | 30 | All journeys | ALL | R | median | 401 | 2.88 | 0.00–10.38 | 0.00–14.97 |
| 17:00 | 30 | All journeys | ALL | Actual detour fraction | mean | 500 | 5.40 | 4.61–6.17 | 4.05–6.64 |
| 17:00 | 30 | All journeys | ALL | Additional walking distance (ΔL) | median | 500 | 0.00 | 0.00–0.00 | 0.00–0.00 |
| 17:00 | 30 | All journeys | ALL | Additional walking distance (ΔL) | mean | 500 | 94.99 | 80.92–108.80 | 76.47–113.82 |
| 17:00 | 30 | All journeys | ALL | Positive-gain share | positive | 500 | 42.99 | 38.60–47.60 | 35.09–49.13 |

### Supplementary Table S8b. Changes in sun avoidance relative to noon

| Time | Statistic | OD | Origins | Change (m) | Origin-cluster CI (m) | Spatial-block CI (m) |
|---|---|---|---|---|---|---|
| 09:00 | median | 500 | 250 | 0.00 | 0.00–0.00 | 0.00–0.00 |
| 09:00 | mean | 500 | 250 | -11.05 | -19.94–-2.81 | -23.61–-1.09 |
| 15:00 | median | 500 | 250 | 0.00 | 0.00–0.00 | 0.00–0.00 |
| 15:00 | mean | 500 | 250 | 2.34 | -7.29–12.02 | -6.16–10.27 |
| 17:00 | median | 500 | 250 | 0.00 | 0.00–0.00 | -3.28–0.00 |
| 17:00 | mean | 500 | 250 | -39.18 | -49.66–-28.96 | -55.88–-22.85 |

## Supplementary Table S8c. Route exposure by area

| Administrative area | Route | Exposure class | Mean class length (m) |
|---|---|---|---|
| CW | Baseline | Known sun | 476.98 |
| CW | Baseline | Building shade | 98.73 |
| CW | Baseline | Vegetation shade | 601.76 |
| CW | Baseline | Indoor protection | 12.56 |
| CW | Baseline | Unresolved exposure | 188.39 |
| CW | Information | Known sun | 386.86 |
| CW | Information | Building shade | 174.95 |
| CW | Information | Vegetation shade | 660.05 |
| CW | Information | Indoor protection | 47.69 |
| CW | Information | Unresolved exposure | 162.96 |
| WC | Baseline | Known sun | 581.03 |
| WC | Baseline | Building shade | 125.26 |
| WC | Baseline | Vegetation shade | 531.66 |
| WC | Baseline | Indoor protection | 10.83 |
| WC | Baseline | Unresolved exposure | 195.86 |
| WC | Information | Known sun | 496.10 |
| WC | Information | Building shade | 223.35 |
| WC | Information | Vegetation shade | 562.43 |
| WC | Information | Indoor protection | 46.27 |
| WC | Information | Unresolved exposure | 171.99 |
| EST | Baseline | Known sun | 586.95 |
| EST | Baseline | Building shade | 58.64 |
| EST | Baseline | Vegetation shade | 462.86 |
| EST | Baseline | Indoor protection | 8.87 |
| EST | Baseline | Unresolved exposure | 302.39 |
| EST | Information | Known sun | 502.37 |

| Administrative area | Route | Exposure class | Mean class length (m) |
|---|---|---|---|
| EST | Information | Building shade | 120.47 |
| EST | Information | Vegetation shade | 510.97 |
| EST | Information | Indoor protection | 53.57 |
| EST | Information | Unresolved exposure | 285.68 |
| STH | Baseline | Known sun | 547.13 |
| STH | Baseline | Building shade | 19.29 |
| STH | Baseline | Vegetation shade | 603.28 |
| STH | Baseline | Indoor protection | 3.24 |
| STH | Baseline | Unresolved exposure | 233.12 |
| STH | Information | Known sun | 509.87 |
| STH | Information | Building shade | 28.50 |
| STH | Information | Vegetation shade | 655.09 |
| STH | Information | Indoor protection | 7.13 |
| STH | Information | Unresolved exposure | 231.43 |

## Supplementary Table S9. Effects of exposure assumptions on route comparisons

At noon, 195 of the 269 route pairs with positive reference gains remained positive across all 13 vegetation settings and the lower gain bounds for unresolved exposure. Mean gains under the pair-specific lowest and highest settings were 63.25 and 79.14 m, compared with 74.23 m in the reference setting. Walking-clock updates were assessed for the same 30 journeys at 12:00 and 17:00 HKT. All 60 comparisons retained their positive, zero or negative gain classification under a 1 m tolerance. Changes exceeded 10 m in six noon comparisons and three 17:00 comparisons, with a maximum absolute change of 132.0 m. Parts a, b and c report configuration counts, ray and walking-clock variants, and terrain-screen summaries, respectively. Supplementary Fig. S13 displays configuration envelopes, and Supplementary Fig. S14 displays ray and terrain checks. The sensitivity design and consistent assignments on shared route sections are specified in Supplementary Methods 8.

### Supplementary Table S9a. Sun-avoidance gains across vegetation settings

| Time | OD | Reference positive | Positive across configs and bounds |
|---|---|---|---|
| 09:00 | 500 | 256 | 184 |
| 12:00 | 500 | 269 | 195 |
| 15:00 | 500 | 262 | 196 |
| 17:00 | 500 | 200 | 140 |

### Supplementary Table S9b. Sensitivity to path sampling, ray height and solar position

| Time ID | Variant | OD | Mean change (m) | Median change (m) | P95 \|change\| (m) | Max \|change\| (m) | Changes >10 m | Practical sign changes |
|---|---|---|---|---|---|---|---|---|
| 1 | height_1p5m | 30 | 0.04 | 0.00 | 2.50 | 5.70 | 0 | 0 |
| 1 | reference | 30 | 0.00 | 0.00 | 0.00 | 0.00 | 0 | 0 |
| 1 | spacing_1m | 30 | -0.33 | 0.00 | 7.99 | 12.51 | 2 | 0 |
| 1 | walking_clock | 30 | -7.18 | 0.00 | 37.86 | 131.99 | 6 | 0 |
| 3 | height_1p5m | 30 | 0.61 | 0.00 | 5.16 | 8.24 | 0 | 0 |
| 3 | reference | 30 | 0.00 | 0.00 | 0.00 | 0.00 | 0 | 0 |
| 3 | spacing_1m | 30 | 0.53 | 0.00 | 5.36 | 7.00 | 0 | 0 |
| 3 | walking_clock | 30 | -1.39 | 0.00 | 24.17 | 28.97 | 3 | 0 |

## Supplementary Table S9c. Effects of terrain obstruction

| Time ID | OD | Median lower gain (m) | Median upper gain (m) | Positive lower bound count |
|---|---|---|---|---|
| 1 | 30 | 0.00 | 9.27 | 12 |
| 3 | 30 | 0.00 | 4.56 | 10 |

## Supplementary Table S10. Journey exposure, walking cost and route access

This table reports continuous sun exposure, proportional reduction in known sun, walking efficiency, baseline-exposure and endpoint groups, connection restrictions and classification bounds. Definitions and resampling are given in Supplementary Methods 10.

| Group | Summary | OD | Estimate | 95% CI | Unit |
|---|---|---|---|---|---|
| Continuity (noon, 10%) | Longest continuous known sun, shortest route | 269 | 155.47 | 143.92–168.93 | m |
| Continuity (noon, 10%) | Longest continuous known sun, shade-informed route | 269 | 108.58 | 99.11–118.59 | m |
| Continuity (noon, 10%) | Change in longest continuous known sun | 269 | -46.89 | -55.72–-39.00 | m |
| Continuity (noon, 10%) | Share with a shortened longest section (156 shortened, 96 unchanged, 17 lengthened) | 269 | 57.97 | 52.23–64.02 | % |
| Baseline sun (10%) | Mean known sun on shortest routes, 09:00 | 500 | 226.50 | 204.37–244.72 | m |
| Baseline sun (10%) | Mean known sun on shortest routes, 12:00 | 500 | 548.02 | 513.16–581.33 | m |
| Baseline sun (10%) | Mean known sun on shortest routes, 15:00 | 500 | 334.31 | 308.30–360.31 | m |
| Baseline sun (10%) | Mean known sun on shortest routes, 17:00 | 500 | 132.21 | 116.63–148.04 | m |
| Aggregate reduction (10%) | Mean ΔS divided by mean $S_0$, 09:00 | 500 | 27.89 | 24.54–31.45 | % |
| Aggregate reduction (10%) | Mean ΔS divided by mean $S_0$, 12:00 | 500 | 13.54 | 11.98–15.02 | % |
| Aggregate reduction (10%) | Mean ΔS divided by mean $S_0$, 15:00 | 500 | 22.90 | 20.09–25.88 | % |
| Aggregate reduction (10%) | Mean ΔS divided by mean $S_0$, 17:00 | 500 | 26.51 | 22.78–30.36 | % |
| Efficiency (noon) | Avoided known sun per metre of extra walking, 5% extra-distance limit | 500 | 2.58 | 2.30–2.87 | m/m |
| Efficiency (noon) | Avoided known sun per metre of extra walking, 10% extra-distance limit | 500 | 1.57 | 1.44–1.71 | m/m |
| Efficiency (noon) | Avoided known sun per metre of extra walking, 20% extra-distance limit | 500 | 1.02 | 0.93–1.11 | m/m |
| Efficiency (noon) | Avoided known sun per metre of extra walking, 30% extra-distance limit | 500 | 0.85 | 0.77–0.92 | m/m |
| Baseline exposure (noon, 10%) | Median ΔS, shortest-route known-sun share at least 30% | 335 | 44.43 | 30.23–62.24 | m |
| Baseline exposure (noon, 10%) | Positive-gain share, shortest-route known-sun share at least 30% | 335 | 69.20 | 63.72–74.13 | % |
| Baseline exposure (noon, 10%) | Median ΔS, shortest-route known-sun share below 30% | 165 | 0.00 | 0.00–0.00 | m |
| Baseline exposure (noon, 10%) | Positive-gain share, shortest-route known-sun share below 30% | 165 | 22.39 | 15.98–29.38 | % |
| Baseline exposure (noon, 10%) | Journeys with at least 80% of the shortest route protected and positive gains | 18 | 0 | — | count |
| Endpoint path type (noon, 10%) | Mean ΔS, both ends on footways or other street-network links | 381 | 84.13 | 71.61–95.61 | m |

| Group | Summary | OD | Estimate | 95% CI | Unit |
|---|---|---|---|---|---|
| Endpoint path type (noon, 10%) | Positive-gain share, both ends on footways or other street-network links | 381 | 59.53 | 54.07–64.56 | % |
| Endpoint path type (noon, 10%) | Mean ΔS, an end on a footpath or park walkway | 119 | 42.59 | 26.08–61.76 | m |
| Endpoint path type (noon, 10%) | Positive-gain share, an end on a footpath or park walkway | 119 | 35.34 | 25.61–45.04 | % |
| Connection restriction (noon, 10%) | Known-sun increase without indoor paths, own feasible set | 479 | 19.93 | 14.09–25.68 | m |
| Connection restriction (noon, 10%) | Known-sun increase without multilevel links, own feasible set | 191 | 31.01 | 18.16–47.89 | m |
| Classification bounds (noon, 10%) | Mean of pair-specific lowest gains across 13 settings and unresolved assignments | 500 | 63.25 | 53.57–72.41 | m |
| Classification bounds (noon, 10%) | Mean of pair-specific highest gains across 13 settings and unresolved assignments | 500 | 79.14 | 67.93–89.67 | m |

## Supplementary Table S11. Sun avoidance in extended route searches

This preselected 50-journey subset was evaluated with an expanded candidate-route set at extra-distance limits up to 50% and 100%. At noon, mean avoided known sun was 97.39 m at the 30% extra-distance limit in the primary analysis. The independent extension yielded 116.22 m at 50% and 117.23 m at 100%. The rows label the candidate-route set used for each result. The design is described in Supplementary Methods 8 and displayed in Supplementary Fig. S18.

| Time ID | Extra-distance limit (%) | Candidate-route set version | OD | Mean ΔS (m) | Median ΔS (m) | Median R (%) | Median actual detour (%) | Positive (%) |
|---|---|---|---|---|---|---|---|---|
| 0 | 0 | frozen primary v2 | 50 | 0.00 | 0.00 | 0.00 | 0.00 | 0.00 |
| 0 | 5 | frozen primary v2 | 50 | 44.88 | 0.00 | 0.00 | 0.00 | 44.00 |
| 0 | 10 | frozen primary v2 | 50 | 53.69 | 3.43 | 6.86 | 0.29 | 52.00 |
| 0 | 20 | frozen primary v2 | 50 | 65.70 | 6.82 | 8.86 | 0.70 | 54.00 |
| 0 | 30 | frozen primary v2 | 50 | 70.37 | 9.98 | 8.93 | 1.82 | 56.00 |
| 0 | 50 | independent extension | 50 | 79.03 | 12.80 | 13.23 | 3.03 | 58.00 |
| 0 | 100 | independent extension | 50 | 79.38 | 15.01 | 13.23 | 3.03 | 58.00 |
| 1 | 0 | frozen primary v2 | 50 | 0.00 | 0.00 | 0.00 | 0.00 | 0.00 |
| 1 | 5 | frozen primary v2 | 50 | 56.58 | 0.00 | 0.00 | 0.00 | 48.00 |
| 1 | 10 | frozen primary v2 | 50 | 75.37 | 1.18 | 0.39 | 0.07 | 50.00 |
| 1 | 20 | frozen primary v2 | 50 | 93.77 | 3.44 | 0.96 | 0.37 | 52.00 |
| 1 | 30 | frozen primary v2 | 50 | 97.39 | 5.55 | 1.14 | 1.41 | 54.00 |
| 1 | 50 | independent extension | 50 | 116.22 | 8.11 | 1.59 | 2.36 | 56.00 |
| 1 | 100 | independent extension | 50 | 117.23 | 8.11 | 1.59 | 2.36 | 56.00 |
| 2 | 0 | frozen primary v2 | 50 | 0.00 | 0.00 | 0.00 | 0.00 | 0.00 |
| 2 | 5 | frozen primary v2 | 50 | 64.20 | 0.00 | 0.00 | 0.00 | 46.00 |
| 2 | 10 | frozen primary v2 | 50 | 75.59 | 5.49 | 2.91 | 0.05 | 52.00 |
| 2 | 20 | frozen primary v2 | 50 | 98.83 | 11.73 | 5.92 | 1.55 | 56.00 |
| 2 | 30 | frozen primary v2 | 50 | 104.18 | 11.73 | 5.92 | 1.55 | 56.00 |
| 2 | 50 | independent extension | 50 | 122.93 | 16.69 | 5.96 | 2.97 | 56.00 |
| 2 | 100 | independent extension | 50 | 127.79 | 16.69 | 5.96 | 2.97 | 56.00 |

| Time ID | Extra-distance limit (%) | Candidate-route set version | OD | Mean ΔS (m) | Median ΔS (m) | Median R (%) | Median actual detour (%) | Positive (%) |
|---|---|---|---|---|---|---|---|---|
| 3 | 0 | frozen primary v2 | 50 | 0.00 | 0.00 | 0.00 | 0.00 | 0.00 |
| 3 | 5 | frozen primary v2 | 50 | 30.66 | 0.00 | 0.00 | 0.00 | 42.00 |
| 3 | 10 | frozen primary v2 | 50 | 32.84 | 0.00 | 2.00 | 0.00 | 46.00 |
| 3 | 20 | frozen primary v2 | 50 | 36.26 | 1.25 | 7.49 | 0.08 | 50.00 |
| 3 | 30 | frozen primary v2 | 50 | 44.89 | 1.25 | 7.49 | 0.08 | 50.00 |
| 3 | 50 | independent extension | 50 | 56.60 | 4.37 | 22.32 | 0.28 | 52.00 |
| 3 | 100 | independent extension | 50 | 61.62 | 4.37 | 20.25 | 0.28 | 52.00 |